\documentclass[aps,prb,10pt,floatfix,reprint,superscriptaddress]{revtex4-2}
\usepackage{times}
\usepackage{float}
\usepackage{multirow}
\usepackage[dvipsnames]{xcolor}
\usepackage{amsmath}
\usepackage{amsthm, mathrsfs}
\usepackage{amssymb}
\usepackage{pifont}
\usepackage{amsbsy}
\usepackage{enumitem}
\usepackage{wasysym}
\usepackage[english]{babel}
\usepackage[T1]{fontenc}
\usepackage[utf8]{inputenc} 
\usepackage{graphicx}
\usepackage[colorlinks,bookmarks=false,citecolor=blue,linkcolor=red,urlcolor=blue]{hyperref}
\usepackage{ulem}
\usepackage{pstricks}
\usepackage{rotating}			       % creates rotated page (landscape) 
\usepackage{tabularx,hhline}	
\usepackage[caption=false]{subfig}	
\usepackage{dsfont}
\usepackage{bbm}	
\newcolumntype{P}[1]{>{\centering\arraybackslash}p{#1}}
\usepackage{xcolor}
\makeatletter
\AtBeginDocument{\let\LS@rot\@undefined}
\makeatother

\newcommand{\tetrahedron}{
  \mathchoice
    {\includegraphics[trim=0cm 2.cm 0cm -2.5cm, height=3ex]{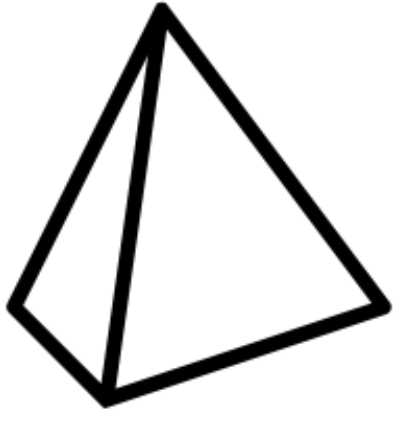}} % \displaystyle
    {\includegraphics[height=1ex]{tetrahedron}} % \textstyle
    {\includegraphics[height=2ex]{tetrahedron}} % \scriptstyle
    {\includegraphics[height=2ex]{tetrahedron}} % \scriptscriptstyle
}

\newcommand{\NiF}{$\mathrm{NaCaNi}_2\mathrm{F}_7$}
\newcommand{\RPU}{\mbox{$\mathbb{RP}^2 \times U(1)$}}

\begin{document}
%%%%%%%%%%%%%%%%%%%%%%%%%%%%%%%%%%%%%

%%%%%%%%%%%%%%%%%%%%%%%%%%%%%%%%%%%%%
\title{
Nematic spin liquid in a \mbox{spin--1} pyrochlore magnet and its realization 
in $\mathrm{NaCaNi}_2\mathrm{F}_7$
}
%%%%%%%%%%%%%%%%%%%%%%%%%%%%%%%%%%%%%

%%%%%%%%%%%%%%%%%%%%%%%%%%%%%%%%%%%%%
\author{Rico Pohle}
%%%%%%%%%%%%%%%%%%%%%%%%%%%%%%%%%%%%%
\affiliation{Creative Science Unit, Faculty of Science, 
Shizuoka University, Shizuoka 422-8529, Japan}
\affiliation{Institute for Materials Research,
Tohoku University, Sendai, Miyagi 980-8577, Japan}

%%%%%%%%%%%%%%%%%%%%%%%%%%%%%%%%%%%%%
\author{Nic Shannon}
%%%%%%%%%%%%%%%%%%%%%%%%%%%%%%%%%%%%%
\affiliation{Theory of Quantum Matter Unit, Okinawa Institute of Science and 
Technology Graduate University, Onna-son, Okinawa 904-0412, Japan} 
%%%%%%%%%%%%%%%%%%%%%%%%%%%%%%%%%%%%%

%%%%%%%%%%%%%%%%%%%%%%%%%%%%%%%%%%%%%
\date{\today}
%%%%%%%%%%%%%%%%%%%%%%%%%%%%%%%%%%%%%

%%%%%%%%%%%%%%%%%%%%%%%%%%%%%%%%%%%%%
\begin{abstract}
%%%%%%%%%%%%%%%%%%%%%%%%%%%%%%%%%%%%%

The search for spin liquids, magnetic phases which lie outside the Landau paradigm, 
remains one of the central challenges for modern condensed matter physics.
For a long time, the prime candidates were thought to be spin--1/2 magnets, but recently 
examples have been identified in many spin--1 materials, including the pyrochlore NaCaNi$_2$F$_7$.
Here we use numerical simulation to explore the spin liquid phases which arise in a minimal 
model of a spin--1 magnet on the pyrochlore lattice.
We find this model supports seven distinct spin liquid phases, 
including one with nematic correlations.
Through quantitative comparison with inelastic neutron scattering, we show that 
this nematic spin liquid provides a compelling scenario for NaCaNi$_2$F$_7$. 
These results suggest that the behaviour of spin liquids found in spin--1 pyrochlore 
magnets may be even richer than in materials with spin--1/2 moments.

%%%%%%%%%%%%%%%%%%%%%%%%%%%%%%%%%%%%%
\end{abstract}
%%%%%%%%%%%%%%%%%%%%%%%%%%%%%%%%%%%%%

%%%%%%%%%%%%%%%%%%%%%%%%%%%%%%%%%%%%%%%%%%
\maketitle
%%%%%%%%%%%%%%%%%%%%%%%%%%%%%%%%%%%%%%%%%%

\section{Introduction}

In the rush of excitement to understand cuprate high--temperature superconductivity, the   
possibility of finding a quantum spin liquid (QSL) rapidly became a {\it cause c\'el\`ebre} \cite{Anderson1987}.   
The abiding fascination of QSL, highly--entangled states with fractionalized ``spinon'' excitations, 
stems from the fact that their properties cannot be explained within the 
Landau paradigm of phases distinguished by symmetry.
At first, these exotic 
phases were assumed to be the exclusive preserve of \mbox{spin--1/2} magnets \cite{Anderson1973}. 
However, as our understanding of quantum spin systems has grown, so has the appreciation 
that spin liquids can occur in many different forms \cite{Wen2002}, and in many different systems, 
including those with highly anisotropic exchange \cite{Savary2017,Knolle2019}, and higher--spin 
moments \cite{Jin2022}.
Consistent with this, a growing number of spin--liquid candidates have been discussed in 
\mbox{spin--1} magnets, including 
the triangular--lattice antiferromagnet Ba$_3$NiSb$_2$O$_9$~\cite{Cheng2011}, 
the diamond--lattice system NiRh$_2$O$_4$~\cite{Chamorro2018},  
and the pyrochlore \mbox{antiferromagnet} $\mathrm{NaCaNi}_2\mathrm{F}_7$~\cite{Krizan2015,Plumb2019,Zhang2019}.

%%%%%%%%%%%%%%%%%%%%%%%%%%%%%%%%%%%%%%%%%%

%
Recent theoretical studies of  the \mbox{spin--1} Heisenberg model on the pyrochlore 
lattice are consistent with the existence of a spin liquid at low temperatures \cite{Mueller2019,Iqbal2019,Gao2020,Hagymasi2022,Hagymasi-arXiv}, 
and an extended Heisenberg model has also been found to offer a good account of the 
the equal time--spin correlations observed in $\mathrm{NaCaNi}_2\mathrm{F}_7$~\cite{Plumb2019,Zhang2019}. 
However, this model does not explain the suppression of spectral weight observed at low energy in 
$\mathrm{NaCaNi}_2\mathrm{F}_7$~\cite{Zhang2019}, or the spin glass transition it 
undergoes at \mbox{$T_{\sf g} \approx 3.6\ \text{K}$}~\cite{Krizan2015,Plumb2019}.
Moreover, published analysis rests on the approximation of ignoring \mbox{$S^z=0$} 
states of the Ni$^{2+}$ ions, which have an associated magnetic quadrupole, and so 
sharply distinguish \mbox{spin--1} from \mbox{spin--1/2} moments.  
Among the consequences of these neglected quadrupoles are  ``biquadratic'' interactions, 
forbidden in spin--1/2 magnets
\cite{Papanicolaou1988,Fazekas1999-WorldScientific,Penc2011-Springer}.
In classical spin models, these terms are known to have a 
singular effect, dramatically changing the character of  spin--liquid 
phases \cite{Shannon2010,Wan2016}.
However the impact of biquadratic interactions on spin--1 pyrochlore magnets,  
and in particular $\mathrm{NaCaNi}_2\mathrm{F}_7$, remains an open question.

%%%%%%%%%%%%%%%%%%%%%%%%%%%%%%%%%%%%%%
%%  Fig. 1 - finite temperatuer phase diagram
%%%%%%%%%%%%%%%%%%%%%%%%%%%%%%%%%%%%%%

\begin{figure*}[t]
	\centering
	 \includegraphics[width=0.9\textwidth]{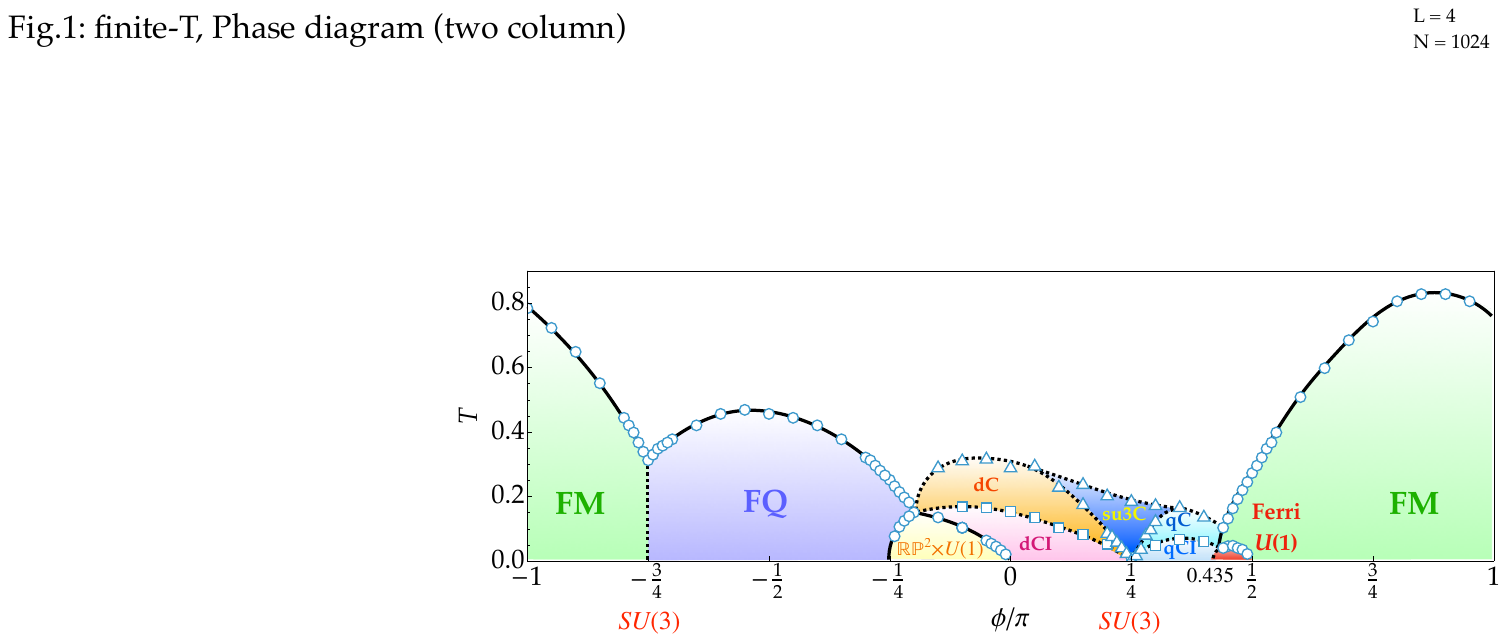}
	\caption{ 
	Finite--temperature phase diagram of the \mbox{$S=1$} bilinear-biquadratic (BBQ) model on the 
	pyrochlore lattice, showing an abundance of spin liquid phases.  
	In addition to ferromagnetic (FM), and ferroquadrupolar (FQ) order, 
	the model supports seven distinct Coulombic spin liquids, listed in 
	Table~\ref{table:spin.liquids}.  
	Results are taken from semi--classical Monte--Carlo (MC) simulation of 
	Eq.~(\ref{eq:H.BBQ}), with parametrization 
	Eq.~(\ref{eq:H.BBQ.parametrization}), 
	as described in the text.  
	Characteristic spin configurations are illustrated in Fig.~\ref{fig:zero.T.phases}.
	}
         \label{fig:finite.T.phase.diagram}
\end{figure*}

%%%%%%%%%%%%%%%%%%%%%%%%%%%%%%%%%%%%%
% In this Article we...
%%%%%%%%%%%%%%%%%%%%%%%%%%%%%%%%%%%%%

In this Article, we explore the new possibilities which arise for spin liquids 
in \mbox{spin--1} pyrochlore magnets.
We focus on the simplest model capable of describing the new features  
associated with \mbox{spin--1} moments, the bilinear biquadratic (BBQ) 
model.
Working within an approach which allows us to treat \mbox{spin--1} 
moments exactly at the level of a single site \cite{Remund2022}, we
establish a phase diagram for this model, finding an abundance of spin liquids.   
All of these spin liquids are ``Coulombic'' in character, showing a 
power--law decay of correlations, visible as ``pinch points'' in equal--time 
structure factors.
The majority of these phases do not have analogues in either 
classical spin models, or \mbox{spin--1/2} magnets, setting them apart 
from spin liquids already under discussion~\cite{Wen2002,Yan2024b}.
And because they blend dipole and quadrupole moments of spin, 
some of these spin liquids offer intriguing parallels with the 
colour physics of quarks.
We further use these results to establish a concrete scenario for both the 
spin liquid, and the spin--glass transition observed in 
$\mathrm{NaCaNi}_2\mathrm{F}_7$, 
in terms of a spin liquid with nematic correlations.
From this, we find good agreement with the low--energy 
dynamics measured in experiment, and identify the spin glass as 
intrinsic in nature, rather than a bi--product of disorder.

%%%%%%%%%%%%%%%%%%%%%%%%%%%%%%%%%%%%%

Our analysis proceeds from model to explicit comparison with experiment.
Solving variationaly for $T=0$, we find that the BBQ model exhibits five distinct spin liquid 
ground states, including one with an underlying $SU(3)$ symmetry, along with 
ferromagnetic (FM) and ferroquadrupolar (FQ) ordered phases. 
We use Monte Carlo (MC) simulation to extend this analysis to finite temperatures, 
uncovering two additional classical spin liquids stabilized by thermal fluctuations.   
These results are summarised in Fig.~\ref{fig:finite.T.phase.diagram}. 

%%%%%%%%%%%%%%%%%%%%%%%%%%%%%%%%%%%%%

Next, we use molecular dynamics (MD) simulation to explore the dynamics of the 
different spin liquid phases found in the vicinity of the Heisenberg point, highlighting 
the differences between positive and negative biquadratic interactions.   
We find that the nematic spin liquid stabilized by negative biquadratic interactions, 
and the ``colour ice'' phase stabilized by positive biquadratic interactions, 
have very different low--energy excitations, 
and can therefore be distinguished by inelastic neutron scattering.   
Once exchange anisotropy and further neighbour interactions are 
taken into account, the nematic spin liquid provides an  
excellent account of experiment on $\mathrm{NaCaNi}_2\mathrm{F}_7$,
reproducing both the suppression of spectral weight observed at low energies 
in inelastic neutron scattering, and providing an explanation for the spin glass 
transition observed at low temperatures. 

%%%%%%%%%%%%%%%%%%%%%%%%%%%%%%%%%%%%%%%% 
\section{Results}  
%%%%%%%%%%%%%%%%%%%%%%%%%%%%%%%%%%%%%%%% 
%
The model we consider is the \mbox{spin--1} bilinear biquadratic (BBQ) model 
on the pyrochlore lattice 
\begin{eqnarray}
	{\mathscr H}_{\sf BBQ}  &=&  J \sum_{\langle ij \rangle}  {\bf S}_i \cdot {\bf S}_j 
						+ \ K \ \sum_{\langle ij \rangle} ( {\bf S}_i \cdot {\bf S}_j )^2  \; .
	\label{eq:H.BBQ}
\end{eqnarray}
This Hamltonian exhibits the most general form of interactions between \mbox{spin--1} 
moments on first--neighbour bonds $\langle ij \rangle$ 
consistent with $SU(2)$ symmetry.   
Except where otherwise stated, we parametrize interactions as 
\begin{equation}
	(J, K) = (\cos{\phi}, \sin{\phi} ) 	\; ,
\label{eq:H.BBQ.parametrization}  
\end{equation}
and measure temperature relative 
to the magnitude of interactions, $\sqrt{J^2 + K^2}$, setting the Boltzmann 
constant $k_B = 1$.   

%%%%%%%%%%%%%%%%%%%%%%%%%%%%%%%%%%%%%%%% 

Biquadratic interactions can originate in multi--orbital exchange, where we generally expect 
\mbox{$K > 0$}, \cite{Fazekas1999-WorldScientific, Tchernyshyov2011}, and through coupling to 
phonons, where \mbox{$K < 0$} \cite{Yoshimori1981,Yamashita2000,Stoudenmire2009,Hoffmann2020, Soni2022}.
Physically, these interactions correspond to a coupling 
between quadrupole moments, made possible by the existence of 
a $S^z = 0$ state of the spin--1 moment.
It follows that the BBQ model can be expressed as 
\begin{eqnarray}				
	{\mathscr H}_{\sf BBQ}  
		&=& \sum_{\langle ij \rangle} 
		\left[\left( J - \frac{K}{2} \right) {\bf S}_i \cdot {\bf S}_j
		+ \frac{K}{2} \left. \bf{Q}_i \cdot \bf{Q}_j \right.  +  \frac{4}{3} K 
		\right]
		\; , \nonumber\\ 
	\label{eq:BBQ.model.Q}
\end{eqnarray}
where 
\begin{equation}
	{\textbf {\textit Q}}_i = \begin{pmatrix}
	Q_i^{x^2-y^2}	\\
	Q_i^{3z^2-r^2}	\\
	Q_i^{xy}		\\
	Q_i^{xz}		\\
	Q_i^{yz}				
\end{pmatrix}  =  \begin{pmatrix}
	(S^x_i)^2 - (S^y_i)^2  	\\
	\frac{2}{\sqrt{3}} \left( 
	   S^z_i S^z_i - \frac{1}{2} \left(S^x_i S^x_i + S^y_i S^y_i \right) 
	 \right) 	\\
	S^x_i S^y_i  + S^y_i S^x_i\\
	S^x_i S^z_i  + S^z_i S^x_i\\
	S^y_i S^z_i  + S^z_i S^y_i	\\
\end{pmatrix}  \, .
\label{eq:quadrupole}
\end{equation}
Because biquadratic interactions reflect on--site quadrupoles,  
they are forbidden for spin--1/2 magnets, and have generally been 
neglected in discussions of spin liquid phases.  

%%%%%%%%%%%%%%%%%%%%%%%%%%%%%%%%%%%%%.  

For purposes of numerical simulation, we work within a representation of
the BBQ model in terms of the group $U(3)$~\cite{Papanicolaou1988,Remund2022}. 
This allows us to faithfully represents both dipole-- and quadrupole moments 
at the level of a single site, and so explore the phases stabilized by biquadratic 
interactions.
Information about this approach is provided in the Methods Section 
and Supplementary Materials~\cite{SupplementaryMaterials}.

%%%%%%%%%%%%%%%%%%%%%%%%%%%%%%%%%%%%%.  
% Table 1 - list of spin liquid phases
%%%%%%%%%%%%%%%%%%%%%%%%%%%%%%%%%%%%%

\begin{table}[t]
\centering
\begin{tabular}[t]{|c|c|c|c|}
\hline
 full name & 
 abbreviation & 
 moments & 
 $T=0$ 
\\
\hline
 nematic &  
 $\mathbb{RP}^2 \times U(1)$ & 
 dipolar  &
 ~\checkmark\ [Fig.~\ref{fig:RP2U1}] 
\\
 dipolar Coulombic &
 dC  & 
 dipolar &
 \phantom\checkmark 
 \\
 dipolar color ice & 
 dCI & 
 dipolar &  
 ~\checkmark\ [Fig.~\ref{fig:dCI}] 
\\
 $SU(3)$ coulombic &  
 su3C & 
 mixed &  
 ~\checkmark\ [Fig.~\ref{fig:su3C}]
 %&
\\
 quadrupolar color ice & 
 qCI & 
 quadrupolar &  
 ~\checkmark\ [Fig.~\ref{fig:qCI}]
\\
 quadrupolar Coulombic &
 qC  & 
 quadrupolar & 
 \phantom\checkmark 
 \\
 ferrimagnetic &
 Ferri $U(1)$ & 
 mixed &
 ~\checkmark\ [Fig.~\ref{fig:FerriU1}]
 \\
\hline
\end{tabular}
\caption{
Spin liquids found in the spin--1 BBQ model on the pyrochlore lattice, 
as shown in the phase diagrams Fig.~\ref{fig:finite.T.phase.diagram}
and Fig.~\ref{fig:zero.T.phase.diagram}.
All spin liquids have Coulombic character, with corresponding predictions 
for the dipolar and quadrupolar structure factors shown in Fig.~\ref{fig:PinchPoints}.
Corresponding spin configurations at $T=0$ are 
illustrated in Fig.~\ref{fig:RP2U1}--\ref{fig:FerriU1}.
}
\label{table:spin.liquids}
\end{table}

%%%%%%%%%%%%%%%%%%%%%%%%%%%%%%%%%%%%%%
%  Fig. 2 -ground state phase diagram + real-space configs
%%%%%%%%%%%%%%%%%%%%%%%%%%%%%%%%%%%%%%

\begin{figure*}[th!]
	\captionsetup[subfigure]{farskip=2pt,captionskip=1pt}
	\centering	
	\subfloat[ Ground-state phase diagram found in variational calculations \label{fig:zero.T.phase.diagram}]{
  		\includegraphics[width=0.80\textwidth]{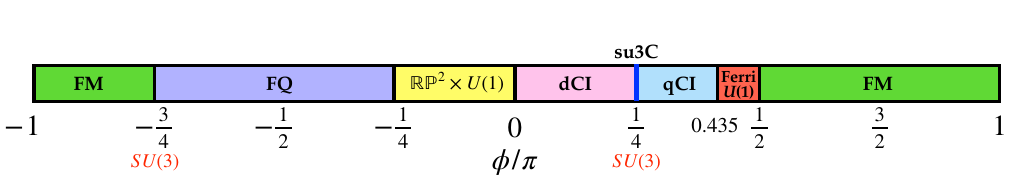}
		}
	\\[1ex]
	\subfloat[ FM \text{[}ordered\text{]} \label{fig:FM}]{ 
  		\includegraphics[width=0.45\columnwidth]{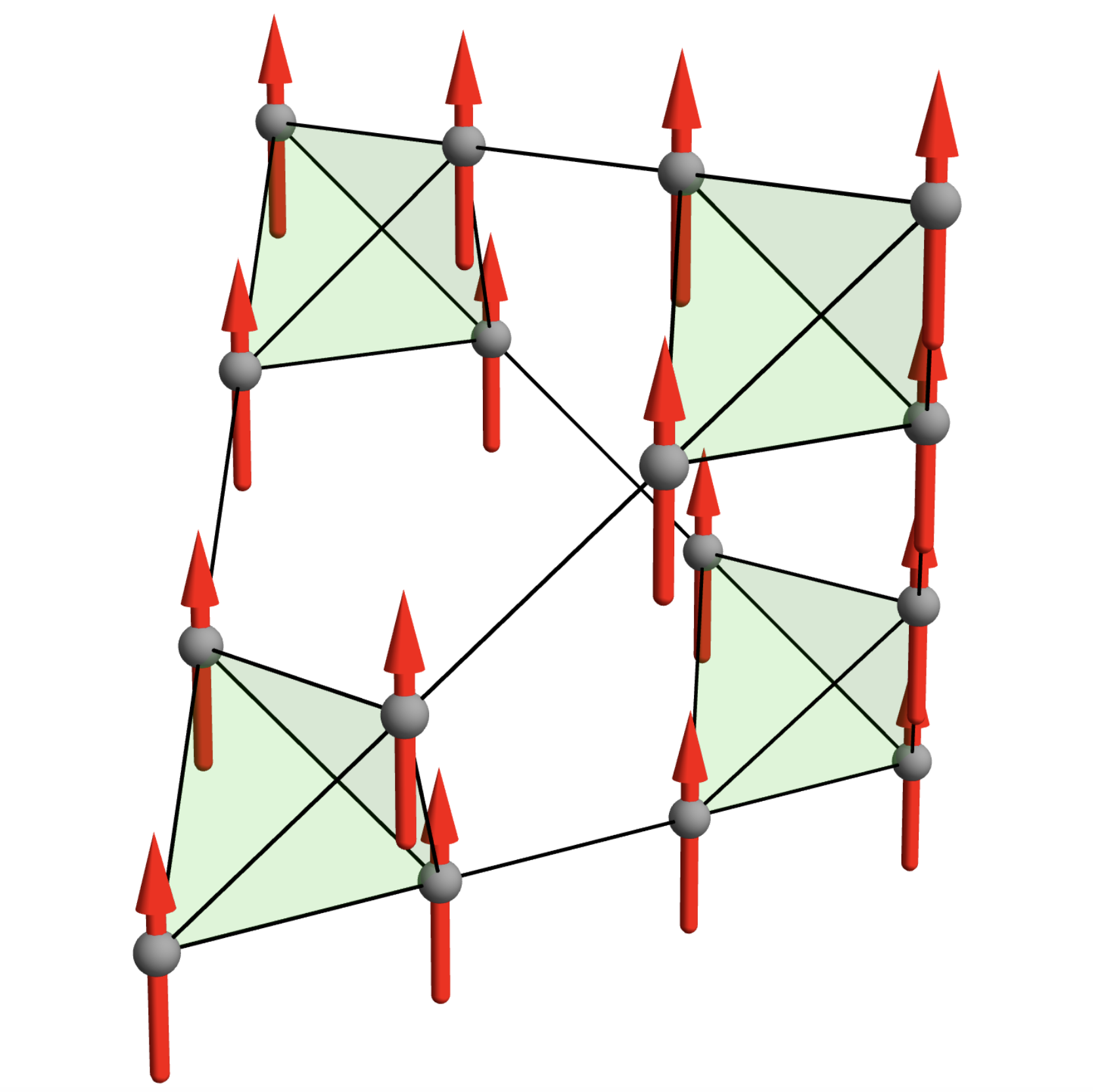}
		}
	\hspace{0.5cm}
	\subfloat[ FQ \text{[}ordered\text{]}  \label{fig:FQ}]{ 
  		\includegraphics[width=0.43\columnwidth]{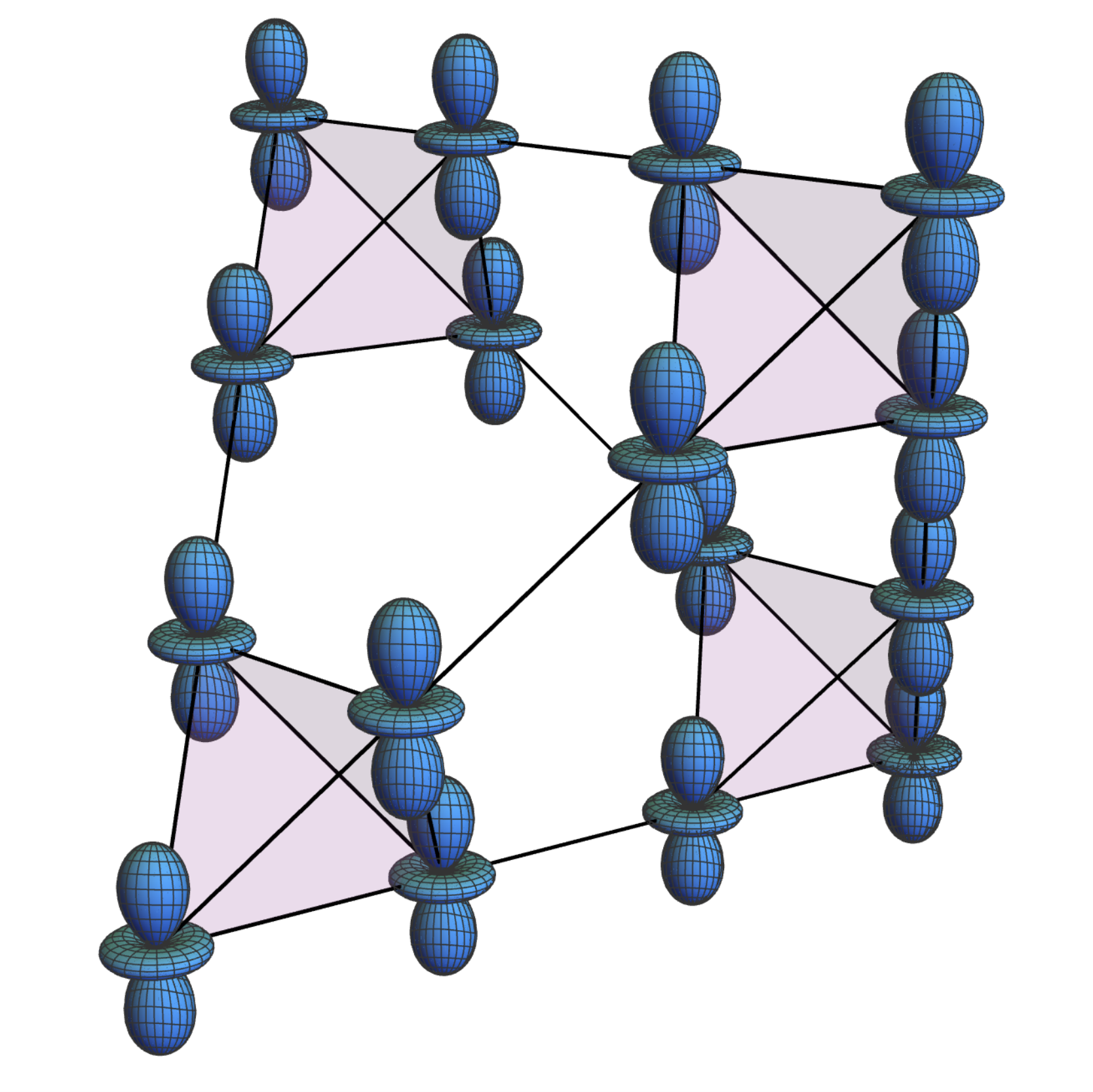}
		}
	\hspace{0.5cm}
	\subfloat[ $\mathbb{RP}^2 \times U(1)$ \text{[}spin liquid\text{]}  \label{fig:RP2U1}]{
  		\includegraphics[width=0.43\columnwidth]{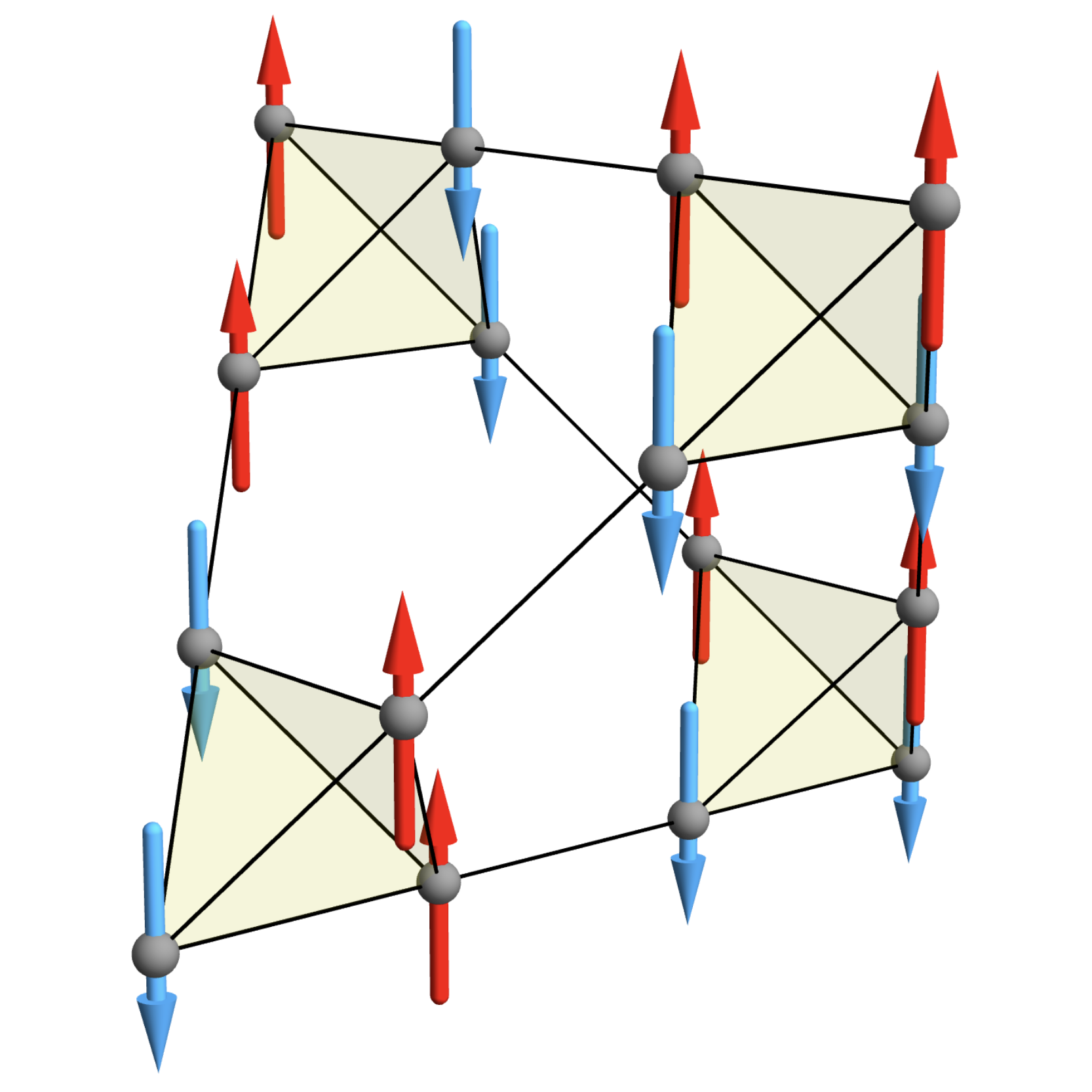}
		}	
	\hspace{0.5cm}
	\subfloat[ dCI \text{[}spin liquid\text{]} \label{fig:dCI}]{
  	\includegraphics[width=0.43\columnwidth]{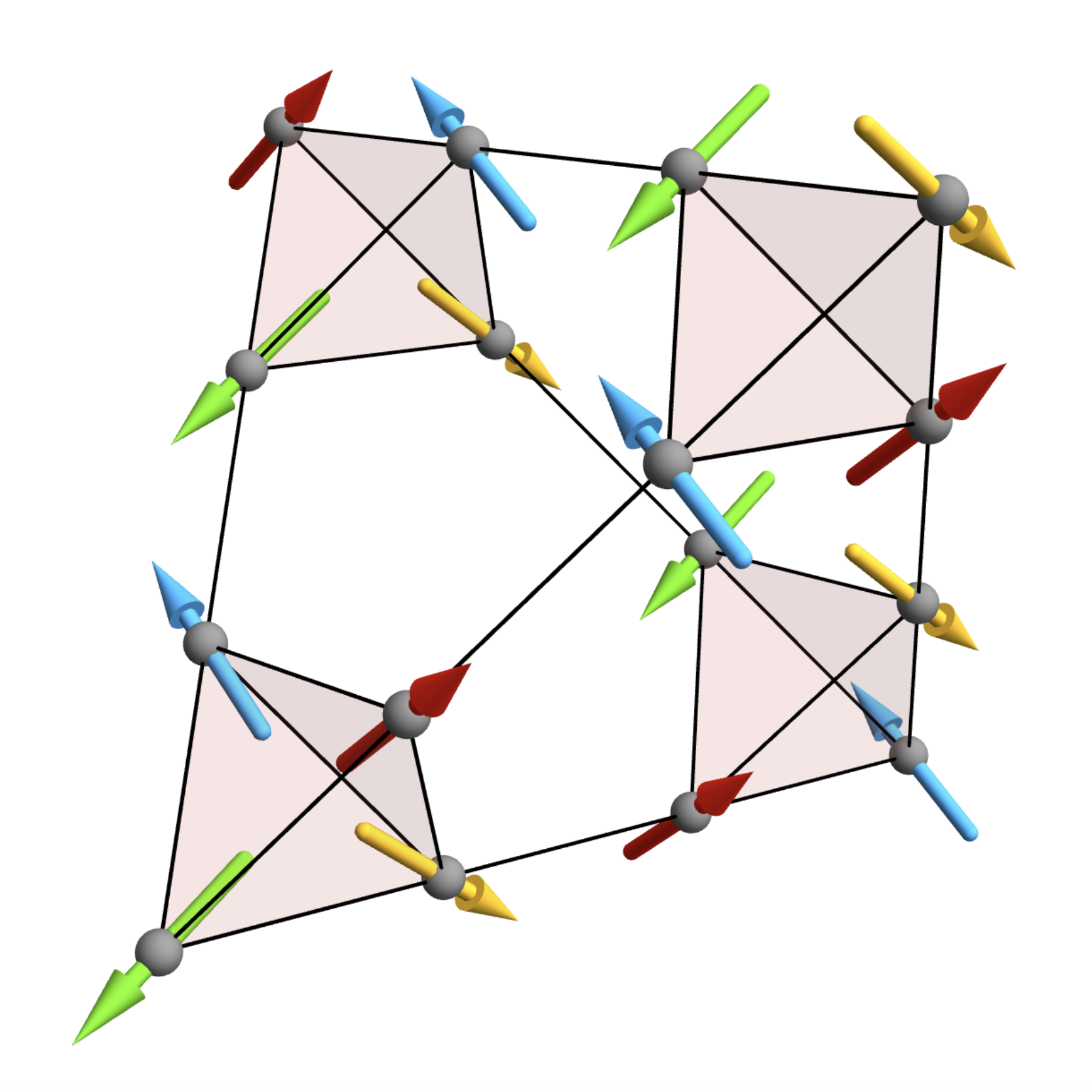}
	}
	\\[1ex]
	\subfloat[ su3C \text{[}spin liquid\text{]} \label{fig:su3C}]{ 
   		\includegraphics[width=0.43\columnwidth]{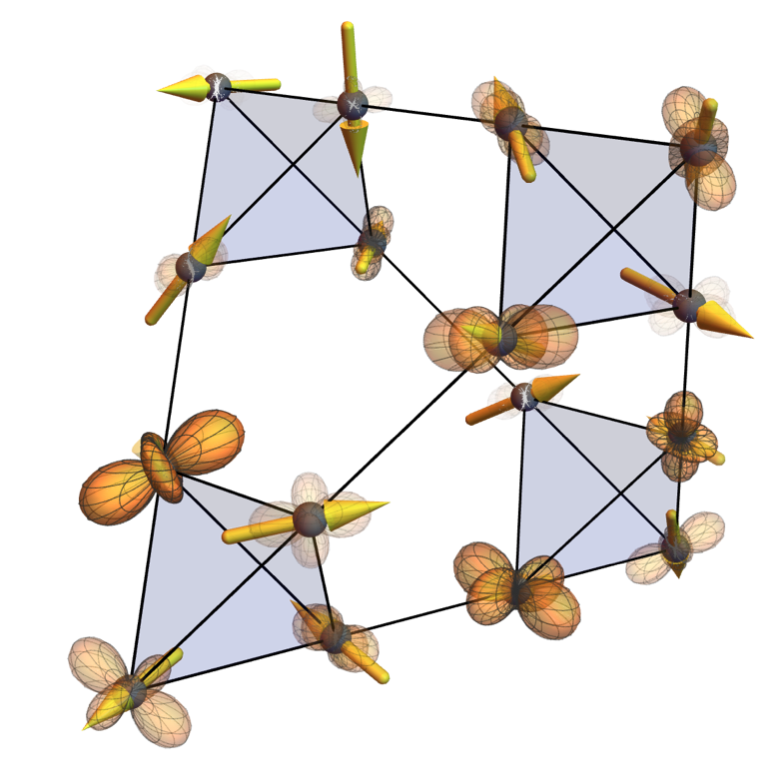}
		}
	\hspace{0.5cm}
	\subfloat[qCI \text{[}spin liquid\text{]}  \label{fig:qCI}]{ 
   		\includegraphics[width=0.43\columnwidth]{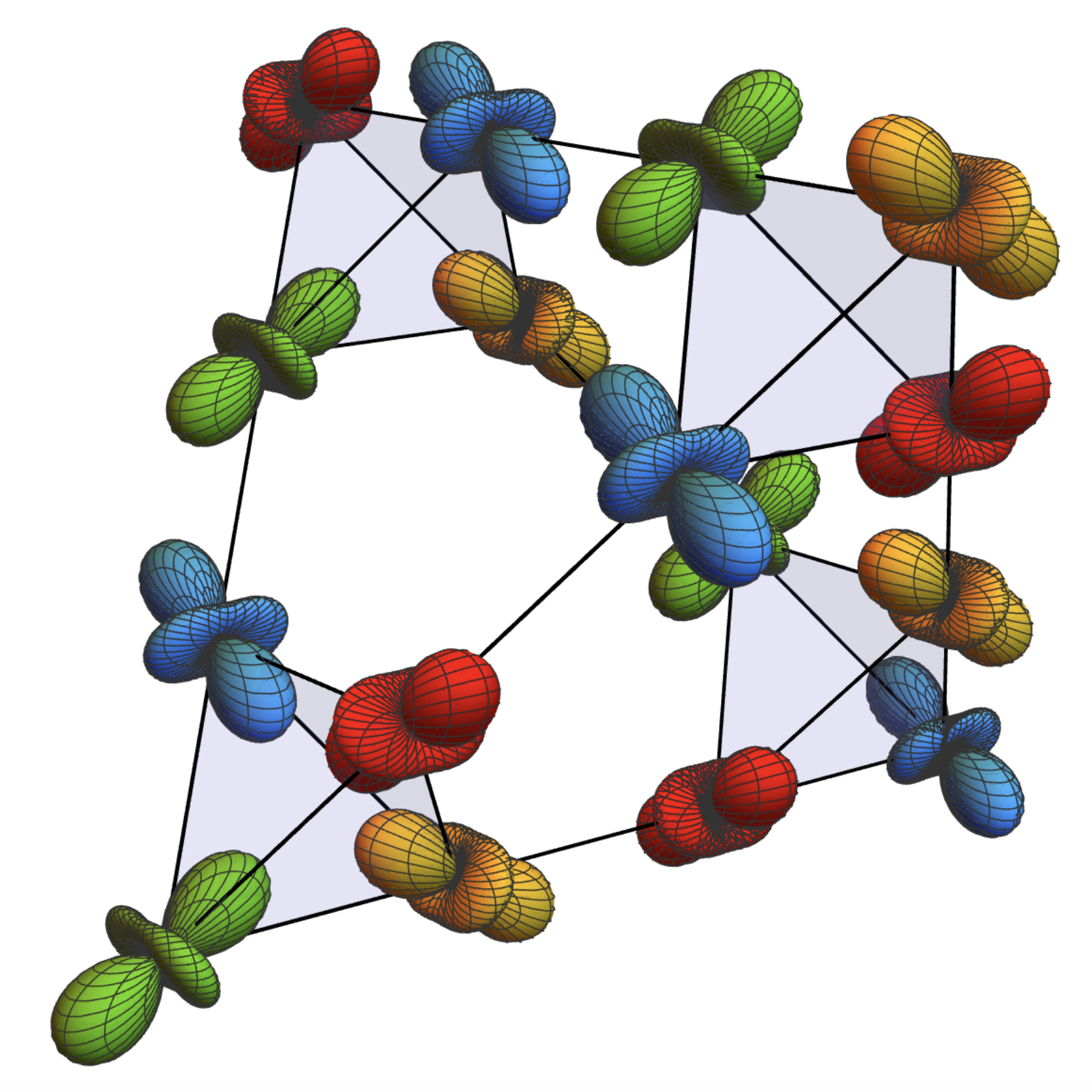}
		} 
	\hspace{0.5cm}
	\subfloat[ Ferri $U(1)$ \text{[}spin liquid\text{]} \label{fig:FerriU1}]{ 
  		\includegraphics[width=0.43\columnwidth]{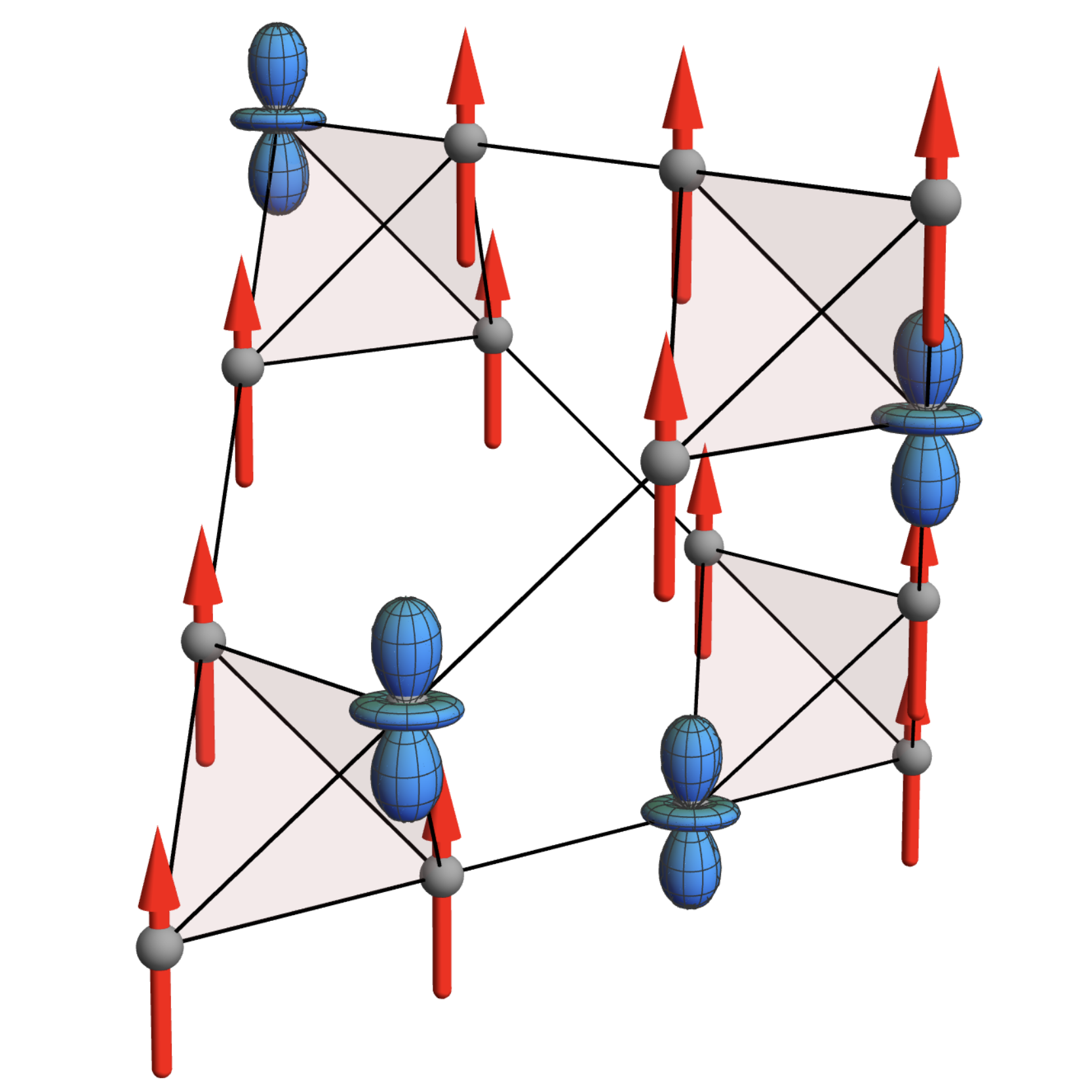}
	}
	\caption{ 
	Ground-state phase diagram of the \mbox{$S=1$} bilinear-biquadratic (BBQ) model on the 
	pyrochlore lattice, and associated zero--temperature phases.
	(a) Phase diagram found in semi--classical variational calculations for \mbox{$T=0$}.   
	In addition to two ordered ground states, the model supports ground-state manifolds 
	associated with five distinct Coulombic spin liquids, as listed in Table~\ref{table:spin.liquids}.
	(b) Example of spin configuration within ferromagnetic (FM) ordered state.
	Dipole moments are indicated with arrows.
	(c) Spin configuration within ferroquadrupolar (FQ) ordered state.
	Quadrupole moments are shown through surfaces of constant probability.
	(d) Spin configuration within nematic spin liquid [\RPU].
	(e) Spin configuration within dipolar colour ice (dCI).
	(f) Spin configuration within $SU(3)$ Coulombic spin liquid  (su3C).
	Spins have mixed dipolar and quadrupolar character. 
	(g) Spin configuration within quadrupolar colour ice (qCI).
	(h) Spin configuration within ferrimagnetic spin liquid [Ferri $U(1)$].
	In all cases spin configurations are shown for a single 16--site cubic unit cell, with 
	results taken from variational simulations of Eq.~(\ref{eq:H.BBQ}), using the
	parametrization Eq.~(\ref{eq:H.BBQ.parametrization}), as described in the text.
         }
        \label{fig:zero.T.phases} 
\end{figure*}

%%%%%%%%%%%%%%%%%%%%%%%%%%%%%%%%%%%%%
\subsection{Ground states} 
%%%%%%%%%%%%%%%%%%%%%%%%%%%%%%%%%%%%%
%
We first consider the zero--temperature properties of the BBQ model, 
Eq.~(\ref{eq:H.BBQ}), within the semi--classical approximation where 
the ground state is written as a product wavefunction of spin--1 moments.
Variational minimization of these product wave functions with respect to energy 
was carried out using libraries developed for machine learning, following \cite{Pohle2023}, 
as described in Methods.
Results are sumarised in Fig.~\ref{fig:zero.T.phases}, 
with phases listed in Table~\ref{table:spin.liquids}.
Corresponding predictions for the equal--time structure factors 
\begin{subequations}
\begin{eqnarray}
   S_{\sf S}({\bf q}) &=& \langle | {\bf S}({\bf q}) |^2 \rangle \; , \\ 
   S_{\sf Q}({\bf q}) &=& \langle | {\bf Q}({\bf q}) |^2 \rangle \; ,
\end{eqnarray}
\end{subequations}
calculated at finite temperature within Monte Carlo simulation (described below),  
are shown in Fig.~\ref{fig:PinchPoints}.    

%%%%%%%%%%%%%%%%%%%%%%%%%%%%%%%%%%%%%

{\it Heisenberg point}.  
The Heisenberg point, \mbox{$K=0$} [\mbox{$\phi =0$}], supports an extensive 
manifold of states in which spins have dipolar character, and satisfy the 
constraint that the total dipole moment on each tetrahedron vanishes 
\cite{Anderson1956,Henley2005}. 
Where spins are treated as classical $O(3)$ vectors, this gives rise to a 
spin liquid of ``Coulombic'' character \cite{Henley2010}, 
with algebraic correlation of dipole moments, as revealed by ``pinch point'' 
singularities in $S_{\sf S}({\bf q})$.
This well--studied classical spin liquid is known to be unstable against arbitrarily 
small biquadratic interaction $K$ \cite{Shannon2010,Wan2016}.

%%%%%%%%%%%%%%%%%%%%%%%%%%%

 {\it Nematic spin liquid}.  
For \mbox{$K \lesssim 0$} [\mbox{$-1/4 < \phi/\pi < 0$}], variational calculations
reveal a manifold of ground states in which spins have dipolar character, 
and select a common axis, giving rise to spin--nematic ($\mathbb{RP}^2$) 
order, while retaining an ice--like $U(1)$--gauge degeneracy [Fig.~\ref{fig:RP2U1}].
The Coulombic structure of this spin liquid is manifest in pinch--points in $S_{\sf S}({\bf q})$ 
[Fig.~\ref{fig:PinchPoints}(a)], matching the phenomenology found in the classical 
$O(3)$ BBQ model \cite{Shannon2010}.
Meanwhile $S_{\sf Q}({\bf q})$ exhibits zone--center Bragg peaks, consistent 
with spin--nematic order [Fig.~\ref{fig:PinchPoints}(b)]. 
We refer to this phase as the  {\it ``nematic spin liquid''} [ \RPU ].

%%%%%%%%%%%%%%%%%%%%%%%%%%%%%%%%%%%%%

{\it Dipolar color ice}.  
For  \mbox{$K \gtrsim 0$} [\mbox{$0 < \phi/\pi < 1/4$}], we find a state in which dipole moments 
align tetrahedrally [Fig.~\ref{fig:dCI}], giving rise to an extensive manifold of states previously dubbed 
``color ice'' \cite{Wan2016,Cerkauskas-Thesis}.
This spin liquid also exhibits pinch points in $S_{\sf S}({\bf q})$ [Fig.~\ref{fig:PinchPoints}(e)], 
accompanied by diffuse zone--center scattering (but no Bragg peaks) in $S_{\sf Q}({\bf q})$  
[Fig.~\ref{fig:PinchPoints}(f)].
We refer to this  phase as ``{\it dipolar color ice}'' (dCI).

%%%%%%%%%%%%%%%%%%%%%%%%%%%%%%%%%%%%%

{\it SU(3) Coulombic spin liquid}.  
For $J=K$ [\mbox{$\phi/\pi = 1/4$}] the BBQ model has $SU(3)$ symmetry.
Here variational calculations reveal a spin liquid in which both dipole 
and quadrupole moments exhibit algebraic correlations [Fig.~\ref{fig:su3C}], 
visible as pinch points in $S_{\sf S}({\bf q})$ [Fig.~\ref{fig:PinchPoints}(g)] 
and $S_{\sf Q}({\bf q})$ [Fig.~\ref{fig:PinchPoints}(h)].
We refer to the phase found at this point as the 
``{\it SU(3) Coulombic}'' (su3C) spin liquid.

%%%%%%%%%%%%%%%%%%%%%%%%%%%%%%%%%%%%%

{\it Quadrupolar color ice}.  
For \mbox{$J \gtrsim K$} [\mbox{$1/4 < \phi/\pi < 0.435(5)$}], we find a 
spin liquid analagogous to dipolar colour ice, 
but composed of quadrupole moments [Fig.~\ref{fig:qCI}].   
This exhibits pinch points in $S_{\sf Q}({\bf q})$ [Fig.~\ref{fig:PinchPoints}(j)], while 
$S_{\sf S}({\bf q})$ is essentially structureless [Fig.~\ref{fig:PinchPoints}(i)].
We refer to this spin liquid as ``{\it quadrupolar color ice}'' (qCI)

%%%%%%%%%%%%%%%%%%%%%%%%%%%%%%%%%%%%%

{\it Ferrimagnetic~$U(1)$~spin liquid}.   
The remaining zero--temperature spin liquid occurs for 
\mbox{$K \gg J$} [\mbox{$ 0.435(5) < \phi/\pi < 1/2$}], bordering FM order.
This has ferrimagnetic character, such that one spin in each 
tetrahedron supports a quadrupole moment, while the others 
exhibit dipole moments aligned with a common axis [Fig.~\ref{fig:FerriU1}].   
The sites occupied by the quadrupole moments are subject to 
the constraint that quadrupole moments never occupy 
neighboring sites, an extensive manifold of states
equivalent to hard--core dimer coverings 
of a diamond lattice, with an associated $U(1)$ gauge 
degeneracy \cite{Shannon2010}.  
As a consequence the ferrimagnetic state displays pinch points 
in $S_{\sf Q}({\bf q})$ [Fig.~\ref{fig:PinchPoints}(n)], accompanied 
by Bragg peaks in both $S_{\sf S}({\bf q})$ and $S_{\sf Q}({\bf q})$ 
[Fig.~\ref{fig:PinchPoints}(m), Fig.~\ref{fig:PinchPoints}(n)].
We refer to this state as ``{\it ferrimagnetic $U(1)$ spin liquid}'' [Ferri $U(1)$].

%%%%%%%%%%%%%%%%%%%%%%%%%%%%%%%%%%%%%
\subsection{Finite--temperature phases}
%%%%%%%%%%%%%%%%%%%%%%%%%%%%%%%%%%%%%
%
We have extended our analysis of the phase diagram to finite 
temperature, at 
a semi--classical level of approximation, 
using the $U(3)$ Monte Carlo (u3MC) method for spin--1 moments 
introduced in Ref.~\onlinecite{Remund2022}.  
Determining the extent of spin liquid phases at finite temperature 
presents a challenge, since the onset of spin liquids is not usually 
associated with a sharp anomaly in specific heat, or amenable  
to description in terms of a local order parameter.   
Here, we follow the approach taken in~\cite{Shannon2010,Greitemann2019}, 
identifying the crossovers into different spin liquids  through generalized 
susceptibilities of the form
\begin{eqnarray}
	\chi[\Lambda] 
	= \frac{N}{T} \left[ \left \langle \Lambda^{2} \right\rangle 
						 - \left\langle \Lambda \right\rangle^2 \right]	\, ,
\label{eq:generalized.susceptibility}
\end{eqnarray}
where $\Lambda$ is an operator characterizing the local correlations 
(constraint) within that spin liquid. 
Further details of this analysis are given  in \cite{SupplementaryMaterials}.

%%%%%%%%%%%%%%%%%%%%%%%%%%%%%%%%%%%%%

In Fig.~\ref{fig:finite.T.phase.diagram} we show the phase diagram 
found for a cubic cluster of \mbox{$N=1024$} spins, with linear 
dimension \mbox{$L=4$}.  
The location of anomalies in heat capacity associated with symmetry--breaking 
phase transitions are indicated through solid lines, while crossovers associated 
with spin liquids are shown with dashed lines.
In addition to the five spin--liquid phases identified at \mbox{$T=0$}, we 
find two further spin liquids stabilized by thermal fluctuations.

%%%%%%%%%%%%%%%%%%%%%%%%%%%%%%%%%%%%%

\begin{figure}
	\centering
	\includegraphics[width=0.49\textwidth]{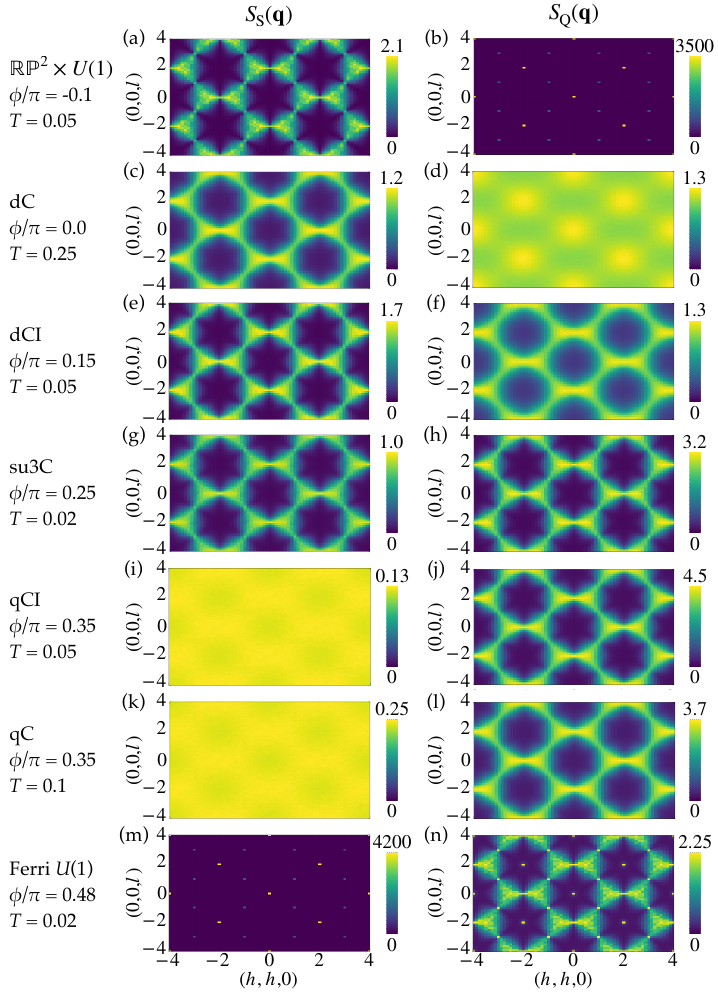}	
	\caption{ 
	Correlations characteristic of the seven spin liquid phases found in 
	the spin--1 bilinear biquadratic model on the pyrochlore lattice 
	[Table~\ref{table:spin.liquids}].
	{\it Left column}: equal--time structure factor for 
	dipole moments, $S_S({\bf q})$.
	{\it Right column}: equal--time structure factor for 
	quadrupole moments, $S_Q({\bf q})$.
	All seven spin liquids are Coulombic in character, displaying 
	pinch points in $S_S({\bf q})$ or $S_Q({\bf q})$.
	In addition to this, the spin--nematic and ferrimagnetic spin liquids 
	break spin--rotation symmetry, leading to Bragg peaks in the relevant 
	structure factor.
	Results are taken from semi--classical Monte--Carlo (MC) simulation of Eq.~(\ref{eq:H.BBQ}), 
	with parametrization Eq.~(\ref{eq:H.BBQ.parametrization}), as described in the text.  	
        }
	\label{fig:PinchPoints}
\end{figure}

{\it Dipolar Coulombic spin liquid}.  
In the vicinity of the Heisenberg point, \mbox{$\phi =0$} [\mbox{$K=0$}],  
thermal fluctuations at low temperature favour dipolar color ice over the 
competing nematic spin liquid.      
However, at higher temperatures we find a region where dipole moments  
exhibit the Coulombic correlations predicted for the classical $O(3)$ Heisenberg model 
\cite{Moessner1998-PRB58,Henley2005,Conlon2010}. 
The Coulombic character of this phase is manifest in pinch points in $S_{\sf S}({\bf q})$, 
broadened by the effects of finite temperature [Fig.~\ref{fig:PinchPoints}(c)].
Meanwhile, $S_{\sf Q}({\bf q})$ [Fig.~\ref{fig:PinchPoints}(d)] displays diffuse peaks 
in the same set of zone centers where the nematic spin liquid develops 
Bragg peaks [Fig.~\ref{fig:PinchPoints}(b)].
We refer to this state as a ``{\it dipolar Coulombic}" (dC) spin liquid.

%%%%%%%%%%%%%%%%%%%%%%%%%%%%%%%%%%%%%%
%%  Fig. 3 - pinch points
%%%%%%%%%%%%%%%%%%%%%%%%%%%%%%%%%%%%%%

%%%%%%%%%%%%%%%%%%%%%%%%%%%%%%%%%%%%%

{\it Quadrupolar Coulombic spin liquid}.
The dipolar Coulombic spin liquid also has a direct analogue for quadrupole 
moments, occurring for \mbox{$\phi/\pi \approx 0.4$}, and separating 
quadrupolar color ice from the high-temperature paramagnet.
This exhibits pinch points in $S_{\sf Q}({\bf q})$, broadened
by the effects of finite temperature, as shown in Fig.~\ref{fig:PinchPoints}(l).
We refer to this phase as a ``{\it quadrupolar Coulombic}" (qC) spin liquid.
%

%%%%%%%%%%%%%%%%%%%%%%%%%%%%%%%%%%%%%%
%  Fig. 4 - dynamics of RP2xU1 and dCI 
%%%%%%%%%%%%%%%%%%%%%%%%%%%%%%%%%%%%%%

\begin{figure*}[t]
	\captionsetup[subfigure]{labelformat=empty, farskip=2pt,captionskip=1pt}
	\centering	
	\subfloat[ (a) \RPU{} (dipoles) \label{fig:4a}]{
  		\includegraphics[width=0.5\columnwidth]{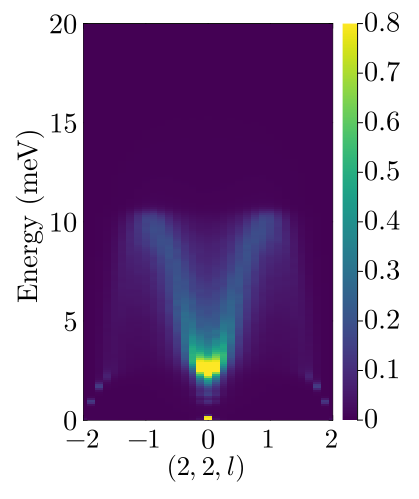}
		\llap{ \parbox[b]{4.5cm}{\textcolor{white}{$\frac{\omega}{k_B T} S_{\sf S}({\bf q}, \omega)$} \\\rule{0ex}{4.2cm} }}
		}
	\subfloat[ (b)  \RPU{} (quadrupoles)  \label{fig:4b}]{ 
  		\includegraphics[width=0.5\columnwidth]{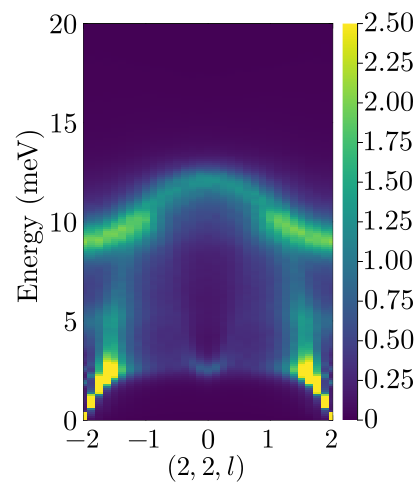}
		\llap{ \parbox[b]{4.5cm}{\textcolor{white}{$\frac{\omega}{k_B T} S_{\sf Q}({\bf q}, \omega)$} \\\rule{0ex}{4.2cm} }}
		}
	\subfloat[ (c) \RPU{} (dipoles) \label{fig:4c}]{
  		\includegraphics[width=0.5\columnwidth]{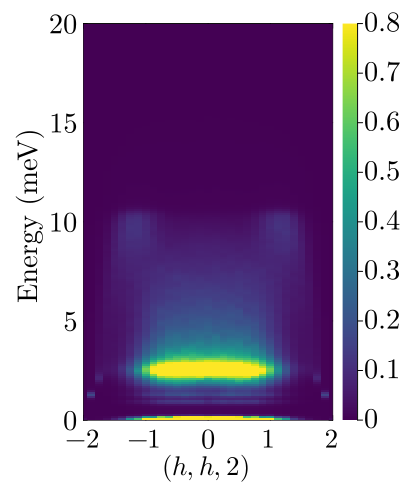}
		\llap{ \parbox[b]{4.5cm}{\textcolor{white}{$\frac{\omega}{k_B T} S_{\sf S}({\bf q}, \omega)$} \\\rule{0ex}{4.2cm} }}
		}
	\subfloat[ (d)  \RPU{} (quadrupoles)  \label{fig:4d}]{ 
  		\includegraphics[width=0.5\columnwidth]{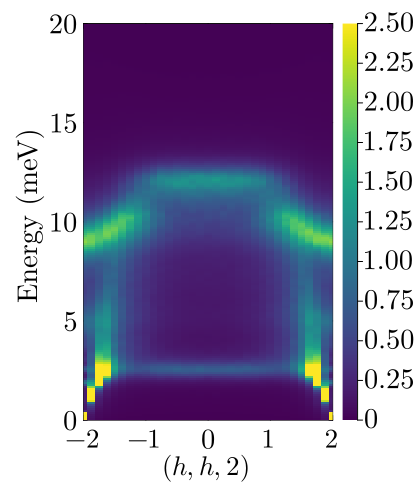}
		\llap{ \parbox[b]{4.5cm}{\textcolor{white}{$\frac{\omega}{k_B T} S_{\sf Q}({\bf q}, \omega)$} \\\rule{0ex}{4.2cm} }}
		}	
	\\[1ex]
	\subfloat[ (e) dCI (dipoles)  \label{fig:4e}]{
  		\includegraphics[width=0.5\columnwidth]{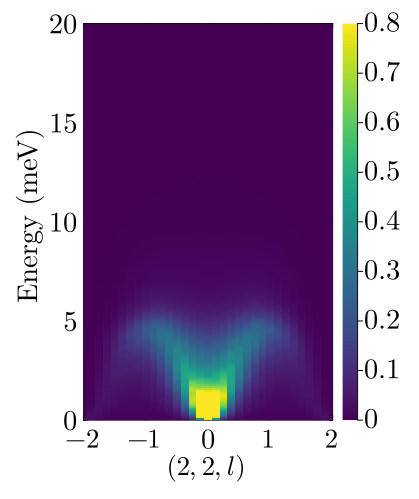}
		\llap{ \parbox[b]{4.5cm}{\textcolor{white}{$\frac{\omega}{k_B T} S_{\sf S}({\bf q}, \omega)$} \\\rule{0ex}{4.2cm} }}
		}
	\subfloat[ (f) dCI (quadrupoles)  \label{fig:4f}]{ 
  		\includegraphics[width=0.5\columnwidth]{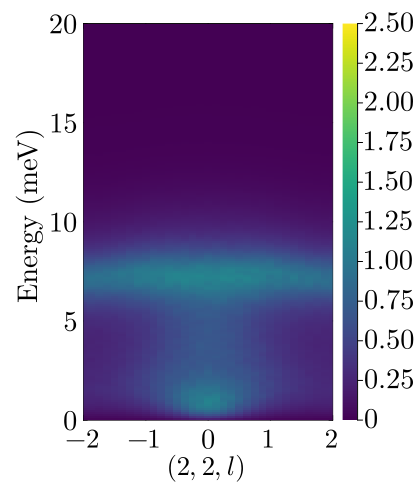}
		\llap{ \parbox[b]{4.5cm}{\textcolor{white}{$\frac{\omega}{k_B T} S_{\sf Q}({\bf q}, \omega)$} \\\rule{0ex}{4.2cm} }}
		}
	\subfloat[ (g) dCI (dipoles) \label{fig:4g}]{
  		\includegraphics[width=0.5\columnwidth]{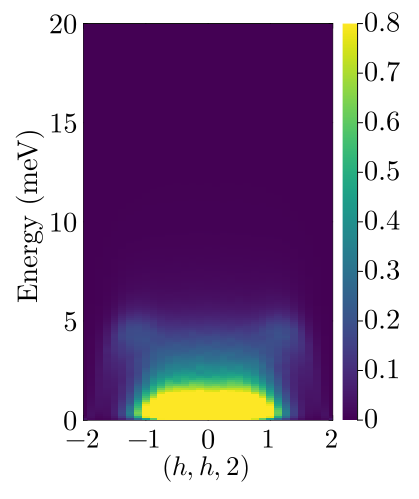}
		\llap{ \parbox[b]{4.5cm}{\textcolor{white}{$\frac{\omega}{k_B T} S_{\sf S}({\bf q}, \omega)$} \\\rule{0ex}{4.2cm} }}
		}
	\subfloat[ (h) dCI (quadrupoles)  \label{fig:4h}]{ 
   		\includegraphics[width=0.5\columnwidth]{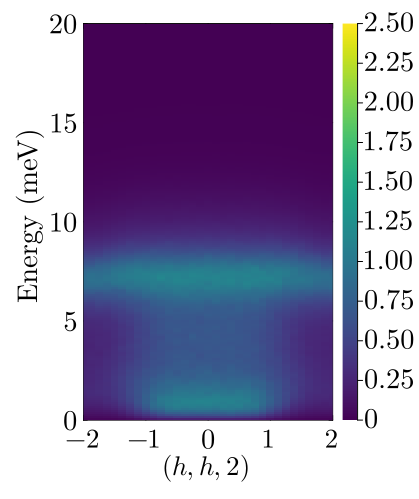}
		\llap{ \parbox[b]{4.5cm}{\textcolor{white}{$\frac{\omega}{k_B T} S_{\sf Q}({\bf q}, \omega)$} \\\rule{0ex}{4.2cm} }}
		}
	\caption{
	Contrasting dynamics of nematic spin liquid and dipolar color ice phases, 
	as revealed by the dynamical structure factors 
	for dipole and quadrupole moments.  
	(a)~Dipolar dynamics of nematic spin liquid [\RPU] 
	for \mbox{$\phi/\pi \approx -0.1$} [\mbox{K < 0}], 
	as found in \mbox{$S_{\sf S}({\bf q}, \omega)$}, on line \mbox{${\bf q} \in (2,2,l)$}.  
	A gap to dipolar excitations is visible at \mbox{${\bf q} = (2,2,0)$}.
	(b)~Equivalent results for quadrupolar structure factor 
	$S_{\sf Q}({\bf q}, \omega)$, showing gapless 
	Goldstone mode associated with spin nematic order.  
	(c)~Corresponding results for \mbox{$S_{\sf S}({\bf q}, \omega)$} 
	on line \mbox{${\bf q} \in (h,h,2)$}.	
	(d)~Equivalent results for \mbox{$S_{\sf Q}({\bf q}, \omega)$}.	
	(e)~Dipolar dynamics of dipolar color ice (dCI)
	for \mbox{$\phi/\pi \approx 0.1$} [\mbox{K > 0}], 
	showing gapless dipoplar excitations, as found in 
	\mbox{$S_{\sf S}({\bf q}, \omega)$} on the line \mbox{${\bf q} \in (2,2,l)$}.  
	(f)~Equivalent results for quadrupolar channel, 
	showing (quasi--)localization of quadrupolar excitations.	
	(g)~Corresponding results for \mbox{$S_{\sf S}({\bf q}, \omega)$} 
	on line \mbox{${\bf q} \in (h,h,2)$}.	
	(h)~Equivalent results for \mbox{$S_{\sf Q}({\bf q}, \omega)$}.	
	Results are taken from semi--classical molecular dynamics (MD) 
	simulation of Eq.~(\ref{eq:H.BBQ}), with parameters 
	[Eq.~(\ref{eq:RP2XU1.parameters}, \ref{eq:dCI.parameters})], 
	as described in the text, and structure factors 
	defined in Eq.~(\ref{eq:Sqw}) of the Methods Section. 
 	}
	\label{fig:dynamics}
\end{figure*}

%%%%%%%%%%%%%%%%%%%%%%%%%%%%%%%%%%%%%

As well as featuring an abundance of spin liquids, the phase diagram shown in 
Fig.~\ref{fig:finite.T.phase.diagram} is also interesting for its topology.
The five different spin liquids found for \mbox{$K>0$}: 
dipolar colour ice, 
the dipolar Coulombic spin liquid, 
quadrupolar colour ice, 
the quadrupolar Coulombic spin liquid, 
and the $su(3)$ spin liquid 
all meet at the zero--temperature $SU(3)$ point.
This reflects the fact that the manifold of ground states found at the 
$SU(3)$ point encompasses the manifold associated with each of 
the other spin liquids.  

%%%%%%%%%%%%%%%%%%%%%%%%%%%%%%%%%%%%%

Meanwhile, simulations for \mbox{$K<0$} suggest that another five phases: 
the ferroquadrupolar phase; the nematic spin liquid; dipolar colour ice; 
the dipolar Coulombic spin liquid; and the high--temperature paramagnet; 
meet at (or near to) a single point \mbox{$(\phi/\pi, T) \approx (-0.2, 0.15)$}.   
These structures, and ferrimagnetic $U(1)$ spin liquid nestled within the FM,  
all merit deeper investigation.  
But in what follows we concentrate on exploring the phenomenology of
the spin--1 pyrochlore AF for \mbox{$J \gg |K| > 0$} [\mbox{$|\phi| \ll 1$}]. 

%%%%%%%%%%%%%%%%%%%%%%%%%%%%%%%%%%%%%
% comparisonf dynamics of RP(2)xU(1) and dCI
%%%%%%%%%%%%%%%%%%%%%%%%%%%%%%%%%%%%%

\subsection{Simulation of dynamics} 
Three different spin liquids: 
the nematic spin liquid;
dipolar colour ice; 
and the dipolar Coulombic spin liqiuid,
are found at or near to
the AF Heisenberg point ($\phi = 0$) [Fig.~\ref{fig:finite.T.phase.diagram}]. 
All of these phases exhibit the same algebraic decay of 
dipolar correlations, and are therefore  difficult to distinguish 
from $S_S({\bf q})$ alone [cf. Fig.~\ref{fig:PinchPoints}(a), 
Fig.~\ref{fig:PinchPoints}(c), 
Fig.~\ref{fig:PinchPoints}(e)].
None the less, these three phases have different dynamical properties, 
which can be probed through the dynamical structure factors 
for dipole and quadrupole moments, $S_{\sf S}({\bf q}, \omega)$ and 
$S_{\sf Q}({\bf q}, \omega)$ [Eq.~(\ref{eq:Sqw})].

%%%%%%%%%%%%%%%%%%%%%%%%%%%%%%%%%%%%%

The dipolar dynamics of the dipolar Coulombic phase found at higher 
temperatures are already well--documented in the context
of the classical $O(3)$ Heisenberg model 
\cite{Moessner1998-PRB58,Conlon2010,Gao2024}.
Here, we focus on the dynamics of the nematic spin liquid and dipolar colour ice, 
found at lower temperatures. 
To this end, we have carried out molecular dynamics (MD) simulations of the \mbox{spin--1} 
BBQ model, Eq.~(\ref{eq:H.BBQ}), using the u3MD method developed in \cite{Remund2022} 
(see Methods).

%%%%%%%%%%%%%%%%%%%%%%%%%%%%%%%%%%%%%

In Fig.~\ref{fig:dynamics}, we compare results for the nematic spin liquid, 
calculated for 
\begin{eqnarray}
J = 2.4\ \text{meV}  \; , \; 
K = -0.8\ \text{meV}  \; , \; 
T \approx 1.8\ \text{K} \; , \; 
\label{eq:RP2XU1.parameters}
\end{eqnarray}
(\mbox{$\phi/\pi \approx -0.1$}), with results for dipolar colour ice, 
calculated for  
\begin{eqnarray}
J = 2.4\ \text{meV}  \; , \; 
K = 0.8\ \text{meV}  \; , \; 
T \approx 1.8\ \text{K} \; , \; 
\label{eq:dCI.parameters}
\end{eqnarray}
(\mbox{$\phi/\pi \approx 0.1$}).  
Predictions for $S_{\sf S}({\bf q}, \omega)$ and $S_{\sf Q}({\bf q}, \omega)$ 
are shown for two perpendicular directions in reciprocal space, 
 \mbox{${\bf q} \in (2,2,l)$} and  \mbox{${\bf q} \in (h,h,2)$}, 
for a cubic cluster of \mbox{$N = 8192$} spins, with linear dimension \mbox{$L=8$}.
A thermal prefactor $\omega/k_B T$ has been included to correct 
structure factors for the classical statistics inherited from 
Monte Carlo simulation \cite{Remund2022}.    

%%%%%%%%%%%%%%%%%%%%%%%%%%%%%%%%%%%%%%%

{\it Dynamics of nematic spin liquid}. 
Dipolar excitations in the nematic spin liquid are gapped, with the lowest lying excitations 
forming a flat band, visible at \mbox{$(2,2,0)$} in Fig.~\ref{fig:4a}, and for a 
finite range of ${\bf q}$ in Fig.~\ref{fig:4c}.
The distribution of spectral weight across this flat band is not uniform, and it 
contributes pinch points to $S_{\sf S}({\bf q}, \omega)$.
At \mbox{$(2,2,0)$} and related zone centers, this flat band has a quadratic 
touching with a dispersing band, visible in Fig.~\ref{fig:4a}.
This dispersing band supports ``half--moon'' features 
in \mbox{$S_S({\bf q}, \omega)$} \cite{Yan2024a}.

%%%%%%%%%%%%%%%%%%%%%%%%%%%%%%%%%%%%%%%

Meanwhile, the quadrupolar channel exhibits a linearly--dispersing Goldstone mode, 
connecting with zone--center Bragg peak, and accompanied by a strongly dispersing band 
of excitations at finite energy, as seen in Fig.~\ref{fig:4b} and Fig.~\ref{fig:4d}.
The existence of gapped dipolar excitations, accompanying gapless quadrupolar 
Goldstone modes, is a general characteristic of spin--nematic order 
\cite{Sato2009,Zhitomirsky2010,Smerald2013}.
However, existence of a flat band is not generic to spin nematics, 
and is instead linked to the $U(1)$ gauge structure of the nematic spin liquid.  
We note that similar results for $S_{\sf S}({\bf q},\omega)$ were 
recently obtained in conventional $O(3)$ MD simulations of a   
closely related ``bond--phonon'' model of Cr spinels \protect{\cite{Gao2024}}, 
for parameters yielding a nematic spin liquid \cite{Shannon2010}.

%%%%%%%%%%%%%%%%%%%%%%%%%%%%%%%%%%%%%%%

{\it Dynamics of dipolar colour ice}. 
Turning to colour ice, we find a radically different phenomenology.
In this case, both dipolar excitations [Fig.~\ref{fig:4e} and Fig.~\ref{fig:4g}]
and quadrupolar excitations [Fig.~\ref{fig:4f} and Fig.~\ref{fig:4h}] are gapless, 
with quadrupoloar excitations showing vanishing spectral weight
for \mbox{$\omega \to 0$}.  
Results for \mbox{$S_S({\bf q}, \omega)$} are qualitatively similar 
to those found in the dipolar Coulombic spin liquid, and by extension 
those found in the classical $O(3)$ Heisenberg AF \cite{Conlon2009,Gao2024}.
Meanwhile results for \mbox{$S_Q({\bf q}, \omega)$} reveal quadrupolar excitations 
which are (nearly) localized, and incoherent in character, with the majority 
of spectral weight found in diffuse, non--dispersing feature at finite energy.
It follows that dynamics in both dipolar and quadrupolar channels sharply distinguish
colour ice from the nematic spin liquid.   

%%%%%%%%%%%%%%%%%%%%%%%%%%%%%%%%%%%%%%%%%%%%%

It is also interesting to contrast the total bandwidth for excitations found in the two spin liquids.
In both cases, quadrupolar excitations have a greater bandwidth than dipolar excitations, 
with the highest energy excitations being found in the quadrupolar channel at 
\mbox{${\bf q} = (2,2,0)$} and related zone centers.  
However, the collinear spin configurations of the nematic spin liquid support a much 
larger dispersion for dipolar excitations than the non--collinear ``tetrahedral'' spin configurations
of colour ice.
This has important implications for the value of exchange constant $J$ needed to 
compare with inelastic neutron scattering, as we will see in the case of \NiF.   

%%%%%%%%%%%%%%%%%%%%%%%%%%%%%%%%%%%%%
\subsection{Implications for experiment}  
%%%%%%%%%%%%%%%%%%%%%%%%%%%%%%%%%%%%%

Semi--classical approximations, based on $SU(3)$ [or $U(3)$] representations 
of spin--1 moments 
\cite{Stoudenmire2009,Remund2015-Thesis,HaoZhang2021,Remund2022,Dahlbom2022a, Dahlbom2022b}, 
have proved a powerful tool for interpreting experiments on 
spin--1 magnets, and related multipolar phenomena. 
Here, we have used this approach to establish a phase diagram for  the spin--1 BBQ model 
on a pyrochlore lattice [Fig.~\ref{fig:finite.T.phase.diagram}], finding an abundance of spin liquids 
[Table~\ref{table:spin.liquids}].
We now consider what this means for experiment.

%%%%%%%%%%%%%%%%%%%%%%%%%%%%%%%%%%%%%

Biquadratic interactions arise automatically in all \mbox{spin--1} 
magnets, with contributions coming from both multi-oribital exchange \cite{Fazekas1999-WorldScientific, Tchernyshyov2011}
and phonons \cite{Yoshimori1981,Yamashita2000,Stoudenmire2009,Hoffmann2020, Soni2022}.
For the majority of \mbox{spin--1} pyrochlore antiferromagnets, we anticipate \mbox{$J \gg |K| > 0$}.    
None the less, at our semi--classical level of approximation,
\mbox{$K$} is a singular perturbation, changing the nature of the  
ground state even for infinitesimal value [cf. Fig.~\ref{fig:zero.T.phase.diagram}].
At finite temperature, this singular behaviour translates into the three competing spin liquids, 
shown for \mbox{$\phi \approx 0$} in  Fig.~\ref{fig:finite.T.phase.diagram}: 
the nematic spin liquid [\RPU]; 
dipolar color ice (dCI); 
and the high--temperature dipolar Coulombic spin liquid (dC).  
These spin liquid phases all have Coulombic character, and exhibit predominantly 
dipolar moments.
And for this reason, they all exhibit the same form of pinch points in the 
equal--time structure factor $S_{\sf S}({\bf q})$ [Fig.~\ref{fig:PinchPoints}(a), (c), (e)].

%%%%%%%%%%%%%%%%%%%%%%%%%%%%%%%%%%%%%%
%%  Fig. 5 - dynamics for NaCaNi$_2$F$_7$
%%%%%%%%%%%%%%%%%%%%%%%%%%%%%%%%%%%%%%
%
\begin{figure*}[t]
	\captionsetup[subfigure]{labelformat=empty, farskip=2pt,captionskip=1pt}
	\centering	
	\subfloat[ (a) INS \label{fig:5a}]{
  		\includegraphics[height=0.242\textheight]{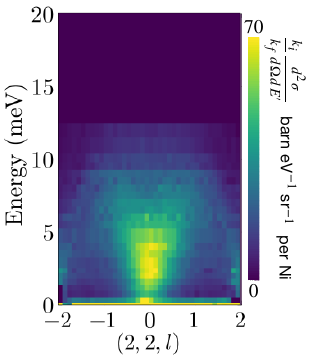}
		}
	\subfloat[ (b)  $\frac{\omega}{k_B T} \tilde{S}_{\sf S}({\bf q}, \omega)$  \label{fig:5b}]{ 
  		\includegraphics[height=0.25\textheight]{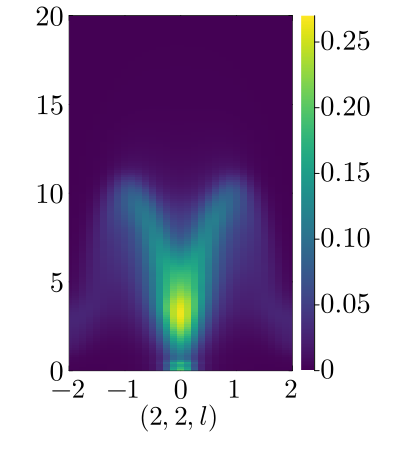}
		}
	\subfloat[ (c) $\frac{\omega}{k_B T} S_{\sf Q}({\bf q}, \omega)$ \label{fig:5c}]{
  		\includegraphics[height=0.25\textheight]{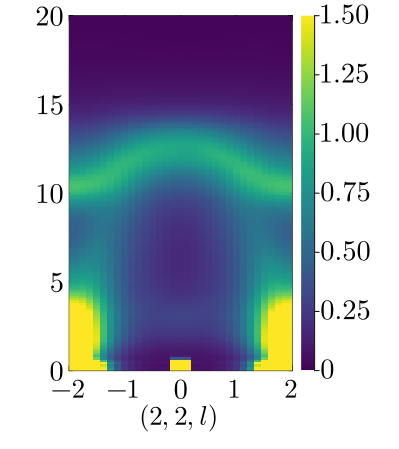}
		}
	\\[1ex]
	\subfloat[ (d) INS  \label{fig:5d}]{
  		\includegraphics[height=0.242\textheight]{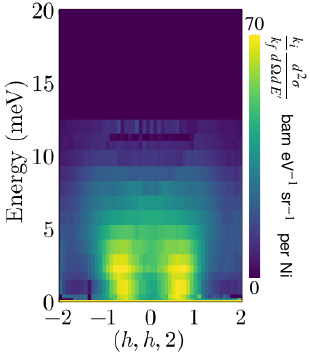}
		}
	\subfloat[ (e) $\frac{\omega}{k_B T} \tilde{S}_{\sf S}({\bf q}, \omega)$  \label{fig:5e}]{ 
  		\includegraphics[height=0.25\textheight]{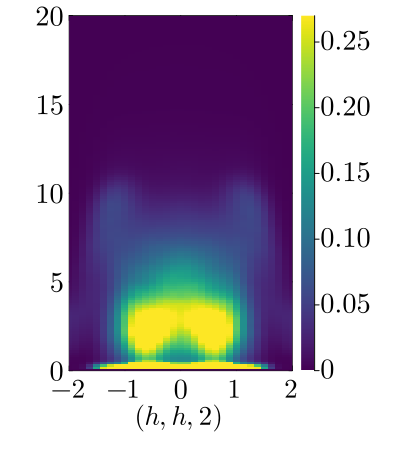}
		}
	\subfloat[ (f) $\frac{\omega}{k_B T} S_{\sf Q}({\bf q}, \omega)$ \label{fig:5f}]{
  		\includegraphics[height=0.25\textheight]{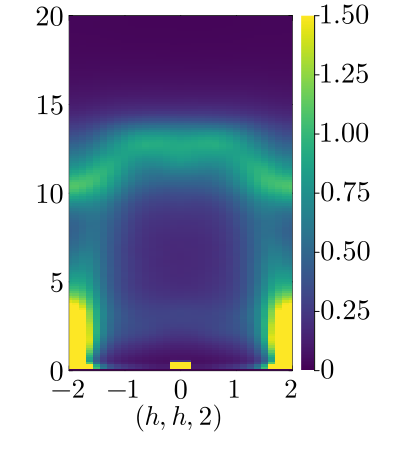}
		}
	\caption{
	Comparison of dynamical structure factors found in simulation of a spin--1 model  
	with biquadratic interactions and experiment on \NiF.
	(a) Inelastic neutron scattering (INS) data for \NiF\ on the line 
	\mbox{${\bf q} \in (2,2,l)$}. 
	(b) Dipolar structure factor \mbox{$\tilde{S}_{\sf S}({\bf q}, \omega)$} 
	found 
	in  simulations of spin--1 model,  
	showing gapped excitations in good agreement with experiment.   
	(c) Corresponding simulation results for the quadrupolar structure factor 
	\mbox{$S_{\sf Q}({\bf q}, \omega)$}, 
	showing dispersing mode at high
	energies, and strong low--energy fluctuations near \mbox{${\bf q} \in (2,2,2)$} 
	and related zone centers.  
	(d) INS data for \NiF\ on the line \mbox{${\bf q} \in (h,h,2)$}. 
	(e) Corresponding predictions for \mbox{$\tilde{S}_{\sf S}({\bf q}, \omega)$}.
	(f) Corresponding predictions for \mbox{$S_{\sf Q}({\bf q}, \omega)$}.   
	Simulation results were obtained for the parameter set in
	Eq.~(\ref{eq:parameters.for.experiment}), 	as described in \cite{SupplementaryMaterials}, 
	with dynamical structure factors
	defined through Eq.~(\ref{eq:Sqw}) and Eq.~(\ref{eq:Sqw.INS}).  
	For purpose of comparison with experiment, simulation results were averaged 
	over $l \pm 0.25$ and $\omega \pm 0.2 {\rm meV}$.
	INS data are taken from experiments described in Ref.~\cite{Zhang2019}.
 	}
	\label{fig:NaCaNi2F7.dynamics}
\end{figure*}
%%%%%%%%%%%%%%%%%%%%%%%%%%%%%%%%%%%%%

This presents a serious challenge for experiment, 
since spin liquids are typically characterized through neutron--scattering 
measurements of $S_{\sf S}({\bf q})$.
In principle, the three competing spin liquids can be cleanly distinguished 
by measurements of quadrupolar correlations, $S_{\sf Q}({\bf q})$, 
with the nematic spin liquid exhibiting Bragg peaks [Fig.~\ref{fig:PinchPoints}(b)], 
the dipolar Coulombic spin liquid
diffuse scattering at zone centers [Fig.~\ref{fig:PinchPoints}(d)], 
and the dipolar colour ice 
a network of diffuse scattering similar to broadened pinch points 
[Fig.~\ref{fig:PinchPoints}(f)].
Information about $S_{\sf Q}({\bf q})$ can be accessed through, e.g., 
resonant inelastic X-ray (RIXS) experiments \cite{Nag2022,Kim2024}.
However, in the absence of knowledge of $S_{\sf Q}({\bf q})$, 
it is necessary to dig deeper into the dynamics of each different spin liquid.
In Fig.~\ref{fig:dynamics}, we have explored the difference in the dynamics of 
the nematic spin liquid and dipolar colour ice. 
In what follows, we consider what this means for inelastic neutron scattering 
experiments on the spin--1 pyrochlore antiferromagnet, NaCaNi$_2$F$_7$.

%%%%%%%%%%%%%%%%%%%%%%%%%%%%%%%%%%%%%
\subsection{Application to NaCaNi$_2$F$_7$}
%%%%%%%%%%%%%%%%%%%%%%%%%%%%%%%%%%%%%
%
The magnetic insulator NaCaNi$_2$F$_7$ contains spin--1 Ni$^{2+}$ ions 
occupying a pyrochlore lattice, which interact 
through antiferromagnetic exchange 
\mbox{$J \approx 36\ \text{K}$}~\cite{Krizan2015, Plumb2019, Zhang2019}.
No magnetic order is observed down to $100\ \text{mK}$; instead 
the system undergoes a spin-glass transition at 
\mbox{$T_{\sf f} \approx 3.6\ \text{K}$}~\cite{Krizan2015,Plumb2019}.
Neutron scattering experiments at $1.8\ \text{K}$ reveal pinch points,  
similar to those expected for a Heisenberg AF,  in the equal-time 
structure factor~\cite{Plumb2019,Zhang2019}.
These pinch points are associated with dipolar excitations that disperse over a 
scale \mbox{$\sim 12\ \text{meV}$}, with the strongest scattering 
occurring at energies \mbox{$\sim 3\ \text{meV}$}~\cite{Zhang2019}.
To date, NaCaNi$_2$F$_7$ has been modeled as an $O(3)$ magnet, 
with small exchange anisotropy and second-neighbour interactions, 
using conventional linear spin-wave, MC, MD and self-consistent Gaussian 
approximation (SCGA) techniques~\cite{Plumb2019,Zhang2019}.
These approaches provide a good account of pinch points, and many other 
aspects of dynamics, for parameters close to an isotropic Heisenberg AF. 
However, they do not explain why the strongest scattering 
occurs at a finite energy, or offer any account of the spin glass transition
seen in experiment.

%%%%%%%%%%%%%%%%%%%%%%%%%%%%%%%%%%%%%%

Given that NaCaNi$_2$F$_7$ is a spin--1 magnet, it must posses some degree
of biquadratic exchange $K$.    
Our results therefore imply that 
its low-temperature Coulombic phase 
should correspond to either the nematic spin liquid [\RPU] or 
dipolar color ice (dCI) \mbox{[cf. Fig.~\ref{fig:finite.T.phase.diagram}]}.
These phases cannot easily be distinguished from the pinch points in 
their equal-time structure factors [Fig.~\ref{fig:PinchPoints} (a), (e)].
However, as we have seen, the low-energy excitations of these two spin liquids are very different.
Spin nematic states support gapless Goldstone modes, which are visible 
 in the quadrupolar channel, while excitations in the dipolar channel 
observed by neutrons remain gapped \cite{Sato2009,Zhitomirsky2010,Smerald2013}.
Meanwhile, states which make up the color ice manifold admit of 
gapless spin-wave excitations, which are dipolar in character, 
and so visible to neutrons.
Inelastic neutron scattering therefore provides a means to distinguish between
the nematic spin liquid and dipolar colour ice.

%%%%%%%%%%%%%%%%%%%%%%%%%%%%%%%%%%%%%%

To make a more quantitative comparison with scattering experiments on
NaCaNi$_2$F$_7$, its is necessary to allow for the effects of exchange 
anisotropy, further-neighbour interactions, and finite experimental resolution. 
A very large number of anisotropic interactions are permitted by the symmetry of the 
pyrochlore lattice, including single-ion terms and anistropic biquadratic 
exchange~\cite{Chung-arXiv}.
Here, for concreteness, we adopt the model used 
by \cite{Plumb2019,Zhang2019}, which considers bilinear exchange 
anisotropy \mbox{$(J_1,J_2,J_3,J_4)$} \cite{Yan2017} 
and second-neighbour interactions $J_{\sf 2nd}$, 
making adjustments to parameters as required to fit general trends 
in the presence of biquadratic exchange $K$.

%%%%%%%%%%%%%%%%%%%%%%%%%%%%%%%%%%%%%%

In Fig.~\ref{fig:NaCaNi2F7.dynamics}, we present MD simulation results obtained for the parameter set
\begin{eqnarray}
& J_1= 2.4\ \text{meV} \; , \; 
J_2= 2.4\ \text{meV} \; , \; 
J_3= 0.15\ \text{meV} \; , \nonumber \\
& J_4=-0.23\  \text{meV} \; , \;
J_{\sf 2nd}=-0.06\ \text{meV}  \; , \; \nonumber \\ 
& K= -0.8\ \text{meV} \; , \; T \approx 1.8\ \text{K} \; .
\label{eq:parameters.for.experiment}
\end{eqnarray}
Simulation results were calculated for an energy 
resolution of~\mbox{$\Delta \omega =0.37\ \text{meV}$}, 
and are 
subject to the same form of post-processing as experimental data, namely averaging 
over \mbox{$l \pm 0.25$} and \mbox{$\omega \pm 0.2\ {\rm meV}$} \cite{Plumb2019}.

%%%%%%%%%%%%%%%%%%%%%%%%%%%%%%%%%%%%%%

The comparison between simulation and experiment is very favourable.   
Predictions for the dynamical strutcure factor measured in experiment, 
\mbox{$\tilde{S}_{\sf S}({\bf q}, \omega)$} [Eq.~(\ref{eq:Sqw.INS})], 
provide an excellent account for of scattering on the line \mbox{${\bf q} \in (2,2,l)$}, reproducing 
both the overall form of dispersion, and the suppression of spectral 
weight at low energies.   
Agreement on line \mbox{${\bf q} \in (h,h,2)$} is also good.   
In \cite{SupplementaryMaterials} 
we provide a further quantitative comparison for pinch points 
and the lineshapes reported at $(2,2,0)$, $(2,2,1/2)$ and $(2,2,1)$. 
These results confirm that it is possible to obtain good fits to experiment 
on NaCaNi$_2$F$_7$, using an extension of the spin--1 BBQ 
model, for parameters associated with the \RPU\ spin liquid.

%%%%%%%%%%%%%%%%%%%%%%%%%%%%%%%%%%%%%%

Once anisotropic terms are included, we find that simulations have 
a strong tendency to fall out of equilibrium, a trend already observed 
in simulations of the $O(3)$ BBQ model in the presence of 
disorder~\cite{Shinaoka2014}.
This spin glass-like transition is accompanied by a rise in fluctuations 
of the quadrupolar order parameter, and for the parameters used 
to model NaCaNi$_2$F$_7$, occurs at \mbox{$T^* \sim 3\ K$}, replacing 
the first-order phase transition found in the parent BBQ model.
This loss of equilibration, in the absence of disorder, reflects the 
fact that states within the \RPU\ manifold are connected by nonlocal 
operations involving the simultaneous exchange of at 
least 6 spins \cite{Shinaoka2010, Benton2012}.
Neither the update used in u3MC simulations, nor the fluctuations 
naturally present in NaCaNi$_2$F$_7$, provide a low-energy mechanism 
for this to happen.
As a consequence, the dynamics calculated in simulation, 
like the dynamics measured in NaCaNi$_2$F$_7$, reflect a nonequilibrium state.

%%%%%%%%%%%%%%%%%%%%%%%%%%%%%%%%%%%%%%

Given the large parameter space, and the large number 
of potential interactions which have been neglected, 
it would be a mistake to over-interpret the numerical values of 
the parameters used to compare with experiment in Fig.~\ref{fig:NaCaNi2F7.dynamics}. 
None the less, we believe it is safe draw two conclusions 
about NaCaNi$_2$F$_7$:
\begin{enumerate}
\item that nematic correlations characteristic of the \RPU\ spin liquid 
can explain the suppression of spectral weight seen at low energy 
near \mbox{${\bf q}=(2,2,0)$}
\item that these nematic correlations enhance the tendency of the system 
to fall out of equilibrium.
\end{enumerate}

%%%%%%%%%%%%%%%%%%%%%%%%%%%%%%%%%%%%%%

It is also interesting to consider whether biquadratic exhange could have 
any other measurable consequence in NaCaNi$_2$F$_7$~?
Substantial biquadratic exhange, \mbox{$K < 0$}, has been identified in Cr spinels, 
whose magnetic Cr$^{3+}$ ions occupy the same pyrochlore lattice as 
Ni$^{2+}$ in NaCaNi$_2$F$_7$.
These interactions arise from coupling to 
phonons \cite{Yamashita2000,Tchernyshyov2002-PRB66,Penc2004,Bergman2006,Gao2024}, 
and have the signal effect of stabilizing a half-magnetization plateau 
in applied magnetic field \cite{Penc2004,Miyata2011-PRL107}.  
No such magnetization plateau is found for \mbox{$K > 0$}, and it might therefore be interesting 
to check for the emergence of a half-magnetization plateau in NaCaNi$_2$F$_7$, 
and other spin--1 pyrochlore magnets.
In the case of NaCaNi$_2$F$_7$, our preliminary Monte Carlo simulations 
for the parameters used to compare with experiment [Eq.~(\ref{eq:parameters.for.experiment})], 
suggest that the onset of such a  plateau would occur for \mbox{$H \sim 70$--$80\ \text{T}$}.
This is comfortably within the reach of modern pulsed-field experiments \cite{Miyata2011-PRL107}.

%%%%%%%%%%%%%%%%%%%%%%%%%%%%%%%%%%%%%%
\section{Discussion} 
%%%%%%%%%%%%%%%%%%%%%%%%%%%%%%%%%%%%%%
%
Spin--1 magnets are not simply ``more classical'' than spin--1/2 magnets, 
but fundamentally different.
In this Article, we have explored the consequences of these differences
for the spin--1 bilinear biquadratic (BBQ) model on a pyrochlore 
lattice,  finding an abundance of spin liquid phases, many of which have 
no possible analogue in classical spin models or spin--1/2 magnets.
Using a combination of numerical techniques built around a $U(3)$ representation
of spin--1 moments, we have documented seven different spin liquids 
\mbox{[Fig.~\ref{fig:finite.T.phase.diagram}, Table~\ref{table:spin.liquids}]}, 
and shown how one of these, a nematic spin liquid, 
offers new insight into the behaviour observed 
in the spin--1 pyrochlore antiferromagnet 
NaCaNi$_2$F$_7$~\cite{Plumb2019,Zhang2019}.   
In many of these spin liquids, quadrupolar fluctuations predominate, suggesting 
that measurements of \mbox{$S_{\sf Q}({\bf q},\omega)$} 
have an important role to play~\cite{Mitsuru2017, Nag2022}.   
Extending the model to include the effect of anisotropic exchange between dipoles 
\cite{Yan2017}, and quadrupoles \cite{Chung-arXiv}, will lead to a still richer range
of possible spin liquid phases.

These results open a number of exciting new vistas.
From the experimental perspective, they resolve key open questions about NaCaNi$_2$F$_7$, 
explaining the anomalous suppression of spectral weight at low--energies, 
and establishing it as a potential model system for the study of intrinsic spin glass behaviour.
And from a theoretical perspective, the identification of a spin nematic phase, 
and of spin liquids built of quadrupole moments, offers connections 
to both liquid crystals \cite{Penc2011-Springer}, 
and the theory of gravity \cite{Chojnacki2024}.
Meanwhile, the presence of a spin liquid with an underlying $SU(3)$ symmetry raises 
intriguing parallels with the colour physics of quarks.

Taken together, these results open a tantalizing landscape for the exploration of new phenomena 
in theoretical models of \mbox{spin--1} magnets, and suggest that the search
for spin liquids in \mbox{spin--1} pyrochlore materials may be every bit 
as rewarding as the search for spin liquids in their (pseudo-)spin--1/2 counterparts 
\cite{Smith2024}.  

%%%%%%%%%%%%%%%%%%%%%%%%%%%%%%%%%%%%%%
% \appendix
%%%%%%%%%%%%%%%%%%%%%%%%%%%%%%%%%%%%%%

%%%%%%%%%%%%%%%%%%%%%%%%%%%%%%%%%%%%%%%%%%%%%%%
\section{Materials and Methods}
%%%%%%%%%%%%%%%%%%%%%%%%%%%%%%%%%%%%%%%%%%%%%%%
\label{section.methods}

%%%%%%%%%%%%%%%%%%%%%%%%%%%%%%%%%%%%%
\subsection{Representation of spin--1 moments}
\label{sec:A.matrix}
%%%%%%%%%%%%%%%%%%%%%%%%%%%%%%%%%%%%%

All numerical simulation were carried out within a semi--classical approach which treats  
spin--1 moments exactly at the level of a single site~\cite{,Remund2022}.
This is built on a representation of spin--1 moments in terms of ``${\mathcal A}$--matrices'', 
rank--2 tensors satisfying the algebra $U(3)$, introduced in \cite{Papanicolaou1988}.

Within this approach, the Hamiltonian ${\mathscr H}_{ {\sf BBQ} }$ [Eq.~\eqref{eq:H.BBQ}] can 
be written 
\begin{equation}
	{\mathscr H}_{\sf BBQ} = 
		\sum_{\langle ij \rangle} \left[ J \mathcal{A}^{\alpha}_{i  \beta} \mathcal{A}^{\beta}_{j  \alpha} 
		+ (K - J) \mathcal{A}^{\alpha}_{i \beta} \mathcal{A}^{\alpha}_{j  \beta}  
		+ K \mathcal{A}^{\alpha}_{i  \alpha} \mathcal{A}^{\beta}_{j \beta} \right] \; , 
\label{eq:HBBQ.A}
\end{equation}
where $\mathcal{A}^{\alpha}_{i \beta}$ are $3\times 3$ matrices satisfying 
\begin{eqnarray}
	\begin{aligned}
		[\mathcal{A}^{\alpha}_{i \beta}, \mathcal{A}^{\gamma}_{i \eta}] &=& 
		\delta^{\gamma}_{\beta} \mathcal{A}^{\alpha}_{i \eta} - 
		\delta^{\alpha}_{\eta} \mathcal{A}^{\gamma}_{i \beta}		\, ,   \\
		[\mathcal{A}^{\alpha}_{i \beta}, \mathcal{A}^{\gamma}_{j \eta}] &=& 0	\, , \quad
		\rm{Tr}\ \mathcal{A}^{\alpha}_{i \beta} = 1 \; , 
	\end{aligned}	
\label{eq:A.com.rel}
\end{eqnarray}
and magnetic moments are characterized through 
\begin{eqnarray}
	S_{i}^{\alpha} &=& -i \epsilon^{\alpha \ \gamma}_{\ \beta} \mathcal{A}^{ \beta}_{i  \gamma}	\, , 
	\label{eq:A.to.dipole} \\
\	Q^{\alpha}_{i  \beta} &=& - \mathcal{A}^{\alpha}_{i \beta} - \mathcal{A}^{\beta}_{i  \alpha} 	
					+ \frac{2}{3}\delta^{\alpha \beta}  \mathcal{A}^{\gamma}_{i  \gamma}	\, ,\label{eq:A.to.quadrupole}
\end{eqnarray}
where $\epsilon^{\alpha \ \gamma}_{\ \beta}$ is the totally antisymmetric tensor.

For purposes of simulation it is convenient to charcaterize ${\mathcal A}$--matrices 
in terms of a complex vector of unit length
\begin{eqnarray}
     		{\bf d} = \begin{pmatrix} 
			x_1 + i \ x_2		\\ 
			x_3 + i \ x_4		\\ 
			x_5 + i \ x_6		
		\end{pmatrix}  \; ,  \quad {\bf d}^*{\bf d} = 1 \; , 
\label{eq:d.vector}
\end{eqnarray}
such that 
\begin{equation}
	\mathcal{A}^{\alpha}_{\beta} = ({\rm d}^{\alpha})^* \  {\rm d}_{\beta} 	\; .
\label{eq:A-matrix_Director}
\end{equation}
${\mathcal A}$--matrices can then be generated through a generalized 
Marsaglia construction
\begin{equation}
	\begin{aligned}
			x_1 &= \theta_2^{1/4} \ \theta_1^{1/2} \ \sin{\phi_1}		\ , \\ 
			x_2 &= \theta_2^{1/4} \ \theta_1^{1/2} \ \cos{\phi_1}		\ , \\ 
			x_3 &= \theta_2^{1/4} \ (1-\theta_1)^{1/2} \ \sin{\phi_2}		\ , \\ 
			x_4 &= \theta_2^{1/4} \ (1-\theta_1)^{1/2} \ \cos{\phi_2}	\ , \\ 
			x_5 &= (1-\theta_2^{1/2})^{1/2} 			\ , \\ 
			x_6 &= 0			\ ,
	\end{aligned}
	\label{eq:sampling.d}
\end{equation}
where the variables $(\theta_1, \theta_2, \phi_1, \phi_2)$ are drawn from the uniform distribution
\begin{subequations}
\begin{eqnarray}
&&0 \leq \theta_1 , \theta_2 \leq 1 \; , \\
&& 0 \leq \phi_1 , \phi_2  < 2\pi \; .
\end{eqnarray} \; 
\label{eq:uniform.distribution}
\end{subequations}

Further details of the mathematics of spin--1 moments are given in 
Supplementary Materials~\cite{SupplementaryMaterials}.

%%%%%%%%%%%%%%%%%%%%%%%%%%%%%%%%%%%%%
\subsection{Variational energy minimization}
\label{sec:JAX}
%%%%%%%%%%%%%%%%%%%%%%%%%%%%%%%%%%%%%

To obtain the \mbox{$T=0$} ground-state properties, we perform large-scale variational 
energy minimizations of the BBQ Hamiltonian, ${\mathscr H}_{\sf BBQ}$ [Eq.~\eqref{eq:HBBQ.A}], 
following the approach introduced in \cite{Pohle2023}.
These calculations make use of the machine learning library JAX~\cite{Jax2018}, accelerated by a 
NVIDIA\textsuperscript{\textregistered} Jetson Orin Nano\textsuperscript{\texttrademark} 8GB 
GPU compute module.

Ground states are parameterized in terms of the product wave function 
\begin{eqnarray}
\mid \Psi_0 \rangle_{\sf var}   = \prod_i \mid \mathcal{A}^{\alpha}_{i \beta} \rangle 
\end{eqnarray}
where $\mathcal{A}^{\alpha}_{i \beta}$ is parameterized through Eq.~(\ref{eq:A-matrix_Director}), 
and a minimum--energy solution is found by optimizing the parameters 
$(\theta_1, \theta_2, \phi_1, \phi_2)$ [Eq.~(\ref{eq:sampling.d})] at each site of the lattice.

This is accomplished using the optimizer ``Adam'' within the gradient processing and optimization 
library ``Optax''~\cite{Optax2020}.
Simulations are initialized from relevant ordered and disordered states, as well 
as random configurations, for finite-size clusters of up to $N=128,00$ sites, 
with cubic symmetry and periodic boundary conditions.
The optimization process for each set of model parameter involved about $10{,}000$
optimization steps.

%%%%%%%%%%%%%%%%%%%%%%%%%%%%%%%%%%%%%
\subsection{$U(3)$ Monte Carlo simulation [u3MC]}
%%%%%%%%%%%%%%%%%%%%%%%%%%%%%%%%%%%%%
\label{sec:u3MC}

Monte Carlo simulations of thermodynamic properties at finite temperature 
were carried out using the (semi-)classical 
formalism  introduced in~\cite{Remund2022}, 
which is based on the algebra $u(3)$ [u3MC].
Within this approach, the matrices which define the configurations 
of individual spin--1 moments,  
$\mathcal{A}^{\alpha\beta}_i$~[Eq.~(\ref{eq:A-matrix_Director})], 
are sampled by generating random instances of the complex vector ${\bf d}_i$ [Eq.~(\ref{eq:d.vector})]. 
This is accomplished through the generalized Marsaglia construction Eq.~(\ref{eq:sampling.d}), 
with variables $(\theta_1, \theta_2, \phi_1, \phi_2)$ drawn from a uniform distribution, 
Eq.~(\ref{eq:uniform.distribution}).  
After calculating the corresponding change in energy through ${\mathscr H}_{\sf BBQ}$ 
[Eq.~\eqref{eq:HBBQ.A}], updates to $\mathcal{A}^{\alpha\beta}_i$ are accepted or 
rejected according to a standard local-update Metropolis algorithm.

Simulations were carried out for regular cubic clusters of $N$ spins with periodic boundary 
conditions, with a single u3MC step consisting of $N$ attempted local updates
at randomly chosen sites.
To reduce the autocorrelation time, we typically used the replica-exchange method (parallel 
tempering)~\cite{Earl2005},
initiated every $100$ MC steps. 
Thermodynamic quantities were averaged over $1 \times 10^5$ statistically independent 
samples, after using $5 \times 10^5$ steps of simulated annealing from high temperature, 
followed by an additional  $5 \times 10^5$ MC steps for thermalization at fixed $T$.

%%%%%%%%%%%%%%%%%%%%%%%%%%%%%%%%%%%%%
\subsection{$U(3)$ molecular dynamics simulation [u3MD]}
\label{sec:u3MD}
%%%%%%%%%%%%%%%%%%%%%%%%%%%%%%%%%%%%%

To obtain dynamical structure factors for the spin--1 bilinear biquadratic model 
on the pyrochlore lattice we time evolve product wave functions written in terms of 
$\mathcal{A}^{\alpha}_{i \beta}$ by numerically integrating the equation of motion 
\begin{eqnarray}
	\begin{aligned}
		\frac{d}{dt}  \mathcal{A}^{\gamma}_{i \eta}
		= &-i \left[\mathcal{A}^{\gamma}_{i \eta}, \mathscr{H}_{ {\sf BBQ}} \right] \\
		= &-i \sum_{\delta} \Big[ J \left(\mathcal{A}^{\gamma}_{i \alpha}  \mathcal{A}^{\alpha}_{i+{\delta}, \eta}
						-    \mathcal{A}^{\alpha}_{i \eta}  \mathcal{A}^{\gamma}_{i+{\delta}, \alpha} \right)		\\
		  &+ (K - J) \left(\mathcal{A}^{\gamma}_{i \alpha}  \mathcal{A}^{\eta}_{i+{\delta}, \alpha}	
						-    \mathcal{A}^{\alpha}_{i \eta}  \mathcal{A}^{\alpha}_{i+{\delta}, \gamma} \right) \Big] \, ,
	\label{eq:EOM}	
	\end{aligned}
\end{eqnarray}
where $\sum_\delta$ sums over the first neighbors of site $i$. 

We start by generating statistically independent and sufficiently thermalized
$\mathcal{A}$-matrix configurations using u3MC simulations, 
as outlined in Section~\ref{sec:u3MC}.  
For each $\mathcal{A}$ matrix configuration, we generate a time series
$\{  \mathcal{A}^{\alpha}_{i \beta} (t) \}$, by numerically integrating the the equations of motion 
given in Eqs.~\eqref{eq:EOM} using a fourth-order 
Runge-Kutta method~\cite{NumericalRecipes2007}.
The Fourier transform of this time series is then computed with
% 
%%%%%%%%%%%%%%%%%%%%%%
\begin{equation}
	{m_{\sf A}}^{\alpha}_{~\beta}({\bf q}, \omega) 
		= \frac{1}{\sqrt{N_t N}} \sum_{i}^{N} \sum_{t}^{N_t}  
			e^{i {\bf q}\cdot{\bf r}_i} e^{i \omega t}  \mathcal{A}^{\alpha}_{i \beta}(t)  \, ,
\end{equation}
%%%%%%%%%%%%%%%%%%%%%%
%
using a Fast Fourier Transform (FFT)~\cite{FFTW05}. 
To minimize numerical artifacts such as the Gibbs phenomenon, and to mimic 
frequency resolution from experiment, we convolute numerical data  
with a Gaussian envelope.

Since ${m_{\sf A}}^{\alpha}_{~\beta}({\bf q}, \omega)$ contains the information of 
all fluctuations allowed for a spin--1 moment, we separate the contributions from  
dipole and quadrupole moments via 
% 
%%%%%%%%%%%%%%%%%%%%%%
\begin{align}
	&{m_{\sf S}}^{\alpha}_{~\alpha}({\bf q}, \omega)  
		= -i \sum_{\beta,\gamma} \epsilon^{\alpha \ \gamma}_{\ \beta} {m_{\sf A}}^{\beta}_{~\gamma}({\bf q}, \omega)  \, , 
	\label{eq:defn.m.S}	 \\
	&\begin{aligned}
	{m_{\sf Q}}^{\alpha}_{~\beta}({\bf q}, \omega)  
		= -{m_{\sf A}}^{\alpha}_{~\beta}({\bf q}, \omega) - &{m_{\sf A}}^{\beta}_{~\alpha}({\bf q}, \omega)  \\
			+\frac{2}{3}  \delta^{\alpha \beta}
			\sum_\gamma &{m_{\sf A}}^{\gamma}_{~\gamma}({\bf q}, \omega) \; .
	\end{aligned}	
	\label{eq:defn.m.Q}
\end{align}
%%%%%%%%%%%%%%%%%%%%%%
%
The dynamical structure factors for the dipole and quadrupole channels can
then be  calculated as
%
%%%%%%%%%%%%%%%%%%%%%%
\begin{eqnarray}
	S_{\rm{\mu}}({\bf q},\omega) = \left\langle \sum_{\alpha\beta}
		 	| {m_{\mu}}^{\alpha}_{~\beta}({\bf q}, \omega) |^2 \ \right\rangle &&\,  , \nonumber\\
	\mu \in \{\sf S, Q, {\sf A}\} && \; , 
\label{eq:Sqw}
\end{eqnarray}
%%%%%%%%%%%%%%%%%%%%%%
%
where $\langle \ldots  \rangle$ denotes an average over
statistically-independent samples from the time series obtained 
in numerical integration of the equations of motion.

%%%%%%%%%%%%%%%%%%%%%%%%%%%%%%%%%%%%%
\subsection{Comparison with experiment}
\label{sec:NaCaNi2F7}
%%%%%%%%%%%%%%%%%%%%%%%%%%%%%%%%%%%%%
 
When comparing with inelastic neutron scattering (INS) experiments on \NiF, we work  
with the extended model 
\begin{equation}
	\begin{aligned}
	{\mathscr H}_{\sf NaCaNi_2F_7} &= 
		\sum_{\langle ij \rangle}  \left( J^{\alpha \mu}_{\beta \nu} \right)_{ij}  \mathcal{A}^{\alpha}_{i  \beta} \mathcal{A}^{\mu}_{j  \nu} 	\\
		&+ \ J_{\sf 2nd}  \sum_{\langle ij \rangle_2}  
		\mathcal{A}^{\alpha}_{i  \beta} (\mathcal{A}^{\beta}_{j \alpha} - \mathcal{A}^{\alpha}_{j  \beta}) \, ,
	\end{aligned}
\label{eq:HNaCaNi2F.A}
\end{equation}
where $\left(J^{\alpha \mu}_{\beta \nu} \right)_{ij}$ is a bond-dependent tensor, describing 
spin--anisotropic interactions on first-neighbour bonds $\langle ij \rangle$, together with 
an appropriate generalization of the equation of motion, Eq.~(\ref{eq:EOM}), described in \cite{SupplementaryMaterials}.  

Following \cite{Plumb2019,Zhang2019}, we parameterize the dipolar contribution to the interaction between spins on 
neighbouring sites as 
\begin{eqnarray}
\hat{J}^{\sf S}_{01} 
	= \begin{pmatrix}
		J_2		& 	J_4	&	J_4	\\
		-J_4		& 	J_1	&	J_3	\\
		-J_4		& 	J_3	&	J_1	
	\end{pmatrix}
\end{eqnarray}
where the interactions on other five bonds within a tetrahedron can be found by symmetry \cite{SupplementaryMaterials}.

Interactions between quadrupole moments on first neighbour bonds 
are assumed to be spin--isotropic, and therefore fully characterized by biquadratic exchange $K$.   
Following \cite{Plumb2019,Zhang2019}, interactions on second-neighbour bonds $\langle ij \rangle_2$ are also 
assumed to be spin--isotropic, and purely dipolar in character, and so fully characterized 
by a Heisenberg interaction $J_{\sf 2nd}$.
The resulting model has total of 6 parameters, 
\begin{eqnarray}
(J_1, J_2, J_3, J_4, K, J_{\sf 2nd}) \nonumber
\end{eqnarray}
as listed in Eq.~(\ref{eq:parameters.for.experiment}).

To compare with experiment, we evaluate 
\begin{align}
	\begin{aligned}
		&\tilde{S}_{\rm S}({\bf q},\omega) = \\
		& \left\langle
		\sum_{\alpha \beta} \left( \delta^{\alpha \beta} - \frac{q^{\alpha} q^{\beta}}{|{\bf q}|^2} \right)
		\Big[{m_{\sf S}}^{\alpha}_{~\alpha}({\bf q}, \omega)   \Big]^*  \  {m_{\sf S}}^{\beta}_{~\beta} ({\bf q}, \omega)  
		\right\rangle 	\, ,
	\end{aligned}
\label{eq:Sqw.INS}
\end{align}
which takes into account the dipolar projection coming from the interaction between 
the neutron and electron spins.

In the dynamical simulations, we evolve replicas over approximately 
\mbox{$N_t = 6000$} time steps 
with a time increment of 
\mbox{ $\delta t = 0.008 $ meV$^{-1}$}.
For the Fourier transform in time, unless stated otherwise, we use 
every $20^{th}$ time snapshot and apply 
a Gaussian envelope to the signal to reduce numerical artifacts, resulting in a 
frequency resolution with a full width at half maximum (FWHM)
of \mbox{$0.2\ \text{meV}$} for the simulations of ${\mathscr H}_{\sf BBQ}$ [Eq.~(\ref{eq:HBBQ.A})]
shown in Fig.~4 of the main text.
For purposes of comparison with experiment (Fig. 5 of the main text, 
and Figs.~S6, S7 and S8 of \cite{SupplementaryMaterials}), 
simulations of ${\mathscr H}_{\sf NaCaNi_2F_7}$ [Eq.~(\ref{eq:HNaCaNi2F.A})]
were carried out with a resolution of  \mbox{$0.37\ \text{meV}$}.
Results were averaged over $500-1000$ statistically independent time series for 
cluster sizes with a linear dimension of $ L=8$ ($N=8192$).

%%%%%%%%%%%%%%%%%%%%%%%%%%%%%%%%%%%%%%
\section{Acknowledgments}
%%%%%%%%%%%%%%%%%%%%%%%%%%%%%%%%%%%%%%

The authors are pleased to acknowledge helpful discussions with 
Colin Broholm, 
Shintaro Hoshino,
Ludovic Jaubert,
Yukitoshi Motome, 
and Karlo Penc, 
and are indebted to Kemp Plumb for sharing many details of  
experiment, including the INS data reproduced 
from Ref.~\cite{Plumb2019,Zhang2019}.

%%%%%%%%%%%%%%%%%%%%%%%%%%%%%%%%%%%%%%
\subsection{Funding}
%%%%%%%%%%%%%%%%%%%%%%%%%%%%%%%%%%%%%%

This work was supported by Japan Society for the Promotion of Science (JSPS) KAKENHI Grants No. 
JP19H05822,  
JP19H05825,  
JP25K17335,  
MEXT as ``Program for Promoting Researches on the Supercomputer Fugaku'' (Grant No. JPMXP1020230411), 
and by the Theory of Quantum Matter Unit, OIST.
Numerical calculations were carried out using HPC facilities provided by
the Supercomputer Center of the Institute of Solid State Physics, the University of Tokyo.

%%%%%%%%%%%%%%%%%%%%%%%%%%%%%%%%%%%%%%
\subsection{Author Contributions}
%%%%%%%%%%%%%%%%%%%%%%%%%%%%%%%%%%%%%%

R.P. carried out all numerical simulations and analysis of experimental data.
N.S. conceived of the project, and wrote the manuscript.
Both authors contributed equally to the interpretation of simulation results.

%%%%%%%%%%%%%%%%%%%%%%%%%%%%%%%%%%%%%%
\subsection{Competing interests}
%%%%%%%%%%%%%%%%%%%%%%%%%%%%%%%%%%%%%%
 The authors declare that they have no competing interests.

%%%%%%%%%%%%%%%%%%%%%%%%%%%%%%%%%%%%%%
\subsection{Data and materials availability}
%%%%%%%%%%%%%%%%%%%%%%%%%%%%%%%%%%%%%%
Numerical data that support the findings of this study are available from 
the authors upon request.

%%%%%%%%%%%%%%%%%%%%%%%%%%%%%%%%%%%%%
\bibliography{paper}
%%%%%%%%%%%%%%%%%%%%%%%%%%%%%%%%%%%%%

%%%%%%%%%%%%%%%%%%%%%%%%%%%%%%%%%%%%%
% Supplementary Materials 
%%%%%%%%%%%%%%%%%%%%%%%%%%%%%%%%%%%%%

\clearpage
\newpage

%%%%%%%%%%%%%%%%%%%%%%%%%%%%%%%%%%%%%
\title{ {\it Supplementary Materials for} \\
Nematic spin liquid in a \mbox{spin--1} pyrochlore magnet and its realization 
in $\mathrm{NaCaNi}_2\mathrm{F}_7$}
%%%%%%%%%%%%%%%%%%%%%%%%%%%%%%%%%%%%%

%%%%%%%%%%%%%%%%%%%%%%%%%%%%%%%%%%%%%
\author{Rico Pohle}
%%%%%%%%%%%%%%%%%%%%%%%%%%%%%%%%%%%%%
\affiliation{Creative Science Unit, Faculty of Science, 
Shizuoka University, Shizuoka 422-8529, Japan}
\affiliation{Institute for Materials Research,
Tohoku University, Sendai, Miyagi 980-8577, Japan}

%%%%%%%%%%%%%%%%%%%%%%%%%%%%%%%%%%%%%
\author{Nic Shannon}
%%%%%%%%%%%%%%%%%%%%%%%%%%%%%%%%%%%%%
\affiliation{Theory of Quantum Matter Unit, Okinawa Institute of Science and 
Technology Graduate University, Onna-son, Okinawa 904-0412, Japan} 
%%%%%%%%%%%%%%%%%%%%%%%%%%%%%%%%%%%%%

%%%%%%%%%%%%%%%%%%%%%%%%%%%%%%%%%%%%%
\date{\today}
%%%%%%%%%%%%%%%%%%%%%%%%%%%%%%%%%%%%%

%%%%%%%%%%%%%%%%%%%%%%%%%%%%%%%%%%%%%%%%%%
\maketitle
%%%%%%%%%%%%%%%%%%%%%%%%%%%%%%%%%%%%%%%%%%

%%%%%%%%%%%%%%%%%%%%%%%%%%%%%%%%%%%%%%%%%%
\section{Spin--$1$ magnets within a $U(3)$ formalism}	
%%%%%%%%%%%%%%%%%%%%%%%%%%%%%%%%%%%%%%%%%%
\label{section:U3.formalism}

Spin--$1$ moments encompass both magnetic dipole and quadrupole components
on a single site.
To describe these, we adopt a formalism based on the algebra $u(3)$, introduced
in \cite{Papanicolaou1988}, and developed in detail in \cite{Remund2022}.
In the Methods Section of the main text, we outline how this approach was used
to simulate the spin--1 bilinear biquadratic (BBQ) model.
Here we provide further details of the mathematical structures which underpin these simulations.

%%%%%%%%%%%%%%%%%%%%%%%%%%%%%%%%%%%%%%%%%%
\subsection{BBQ model}	
%%%%%%%%%%%%%%%%%%%%%%%%%%%%%%%%%%%%%%%%%%

We take as a starting point the bilinear biquadratic model
\begin{eqnarray}
	{\mathscr H}_{\sf BBQ}  &=&  J \sum_{\langle ij \rangle}  {\bf S}_i \cdot {\bf S}_j 
						+ \ K \ \sum_{\langle ij \rangle} ( {\bf S}_i \cdot {\bf S}_j )^2   \; , 
\label{eq:H.BBQ}
\end{eqnarray}
[Eq.~(1) of the main text], parameterized as 
\begin{equation}
	(J, K) = (\cos{\phi}, \sin{\phi} ) 	\; ,
\label{eq:H.BBQ.parametrization}  
\end{equation}
[Eq.~(4) of the main text].
This Hamiltonian can be written in terms of dipole and quadrupole moments as 
\begin{eqnarray}				
{\mathscr H}_{\sf BBQ} &=& \sum_{\langle ij \rangle} 
		\left[\left( J - \frac{K}{2} \right) {\bf S}_i \cdot {\bf S}_j
		+ \frac{K}{2} \left. \bf{Q}_i \cdot \bf{Q}_j \right.  +  \frac{4}{3} K 
		\right]  \; , 
\nonumber \\
\label{eq:H.BBQ.multipole}
\end{eqnarray}
[Eq.~(2) of the main text], where 
\begin{eqnarray}
	{\bf S}_i &=& \begin{pmatrix}
		S_i^x\\
		S_i^y\\
		S_i^z
	\end{pmatrix} \; ,
\label{eq:dipole}
\end{eqnarray}
and
\begin{eqnarray}
	{\textbf {\textit Q}}_i &=& \begin{pmatrix}
	Q_i^{x^2-y^2}	\\
	Q_i^{3z^2-r^2}	\\
	Q_i^{xy}		\\
	Q_i^{xz}		\\
	Q_i^{yz}				
\end{pmatrix}  =  \begin{pmatrix}
	(S^x_i)^2 - (S^y_i)^2  	\\
	\frac{2}{\sqrt{3}} \left( 
	   S^z_i S^z_i - \frac{1}{2} \left(S^x_i S^x_i + S^y_i S^y_i \right) 
	 \right) 	\\
	S^x_i S^y_i  + S^y_i S^x_i\\
	S^x_i S^z_i  + S^z_i S^x_i\\
	S^y_i S^z_i  + S^z_i S^y_i	\\
\end{pmatrix}  \, . \nonumber\\
\label{eq:quadrupole}
\end{eqnarray}
While it is possible to completely describe a spin--1 moment in terms of 
its dipole moment ${\textbf {\textit S}}$ and quadrupole moment 
${\textbf {\textit Q}}$ \cite{Balla2014-Thesis,Remund2015-Thesis,Dahlbom2022a,Dahlbom2022b}, 
here we take a different approach. 

The first step is to introduce a basis of time-reversal invariant states 
\begin{subequations}
\begin{eqnarray}
 	| x \rangle &=& \frac{i}{\sqrt{2}} ( | 1 \rangle -  | \overline{1} \rangle )  \\
	| y \rangle &=& \frac{1}{\sqrt{2}}(  | 1 \rangle + | \overline{1} \rangle )  \\
	| z \rangle &=& -i | 0 \rangle \; .
\end{eqnarray}
\label{eq:TRstates}
\end{subequations}
where 
\begin{eqnarray}
\{  | 1 \rangle \; , \;  | 0 \rangle \; , \;  | \overline{1} \rangle \}
\end{eqnarray}
are eigenstates of $S^z$.  
All possible quantum states of a spin--1 moment can the.be written as 
\begin{eqnarray}
 | {\bf d} \rangle
 	= \sum_{\alpha=x,y,z} d_\alpha^* | \alpha \rangle  
\label{eq:Defd} 
\end{eqnarray}
where 
\begin{eqnarray}
{\bf d} = (d_x, d_y, d_z)  \quad , \quad d_\alpha \in \mathbb{C} \; ,
\end{eqnarray}
is a complex vector of unit length 
\begin{eqnarray}
 	{\bf d}^* {\bf d}	= |{\bf d}|^2 = 1  \; . 
\end{eqnarray}

The vector ${\bf d}$ is sometimes referred to as a 
director~\cite{Lauchli2006, Tsunetsugu2006, Stoudenmire2009, Remund2022}, 
and can be resolved through a generalized Marsaglia construction \cite{Marsaglia1972} as 
\begin{eqnarray}
     		{\bf d} = \begin{pmatrix} 
			x_1 + i \ x_2		\\ 
			x_3 + i \ x_4		\\ 
			x_5 + i \ x_6		
		\end{pmatrix}  \; , 
\end{eqnarray}
where
\begin{equation}
	\begin{aligned}
			x_1 &= \theta_2^{1/4} \ \theta_1^{1/2} \ \sin{\phi_1}		\ , \\ 
			x_2 &= \theta_2^{1/4} \ \theta_1^{1/2} \ \cos{\phi_1}		\ , \\ 
			x_3 &= \theta_2^{1/4} \ (1-\theta_1)^{1/2} \ \sin{\phi_2}		\ , \\ 
			x_4 &= \theta_2^{1/4} \ (1-\theta_1)^{1/2} \ \cos{\phi_2}	\ , \\ 
			x_5 &= (1-\theta_2^{1/2})^{1/2} \ \sin{\phi_3}			\ , \\ 
			x_6 &= (1-\theta_2^{1/2})^{1/2} \ \cos{\phi_3}			\ ,
	\end{aligned}
	\label{eq:sampling.d}
\end{equation}
and
\begin{subequations}
\begin{eqnarray}
&&0 \leq \theta_1 , \theta_2 \leq 1 \\
&& 0 \leq \phi_1 , \phi_2, \phi_3 < 2\pi 
\end{eqnarray}
\end{subequations}
This parameterization includes an unphysical gauge degree of freedom (undetermined overall phase) 
which can eliminated by restricting $d^z$ to purely real values, i.e. setting 
\begin{eqnarray}
	\phi_3 = \pi/2 \; .
\end{eqnarray}
This leaves a total of four degrees of 
freedom, as expected for a spin--$1$ moment~\cite{Zhang2021, Remund2022}.

From this starting point, we introduce the $3\times 3$ Hermitian matrix formed
by the outer product of ${\bf d}$ vectors 
\begin{equation}
	\mathcal{A}^{\alpha}_{\beta} = ({\rm d}^{\alpha})^* \  {\rm d}_{\beta} 	\; ,
\label{eq:A-matrix_Director}
\end{equation}
which we refer to as an $\mathcal{A}$--matrix.
This can conveniently be decomposed in terms of the basis 
 \begin{eqnarray}
 \mathcal{A}^1_1 &=&
 \begin{pmatrix}
 	1 & 0 & 0 \\
 	0 & 0 & 0 \\
 	0 & 0 & 0
 \end{pmatrix} \; , \; 
\mathcal{A}^1_2  =
 \begin{pmatrix}
 	0 & 1 & 0 \\
 	0 & 0 & 0 \\
 	0 & 0 & 0
 \end{pmatrix} \; , \; 
\mathcal{A}^1_3  =
 \begin{pmatrix}
 	0 & 0 & 1 \\
 	0 & 0 & 0 \\
 	0 & 0 & 0
 \end{pmatrix} \; , \; \nonumber\\
\mathcal{A}^2_1 &=&
 \begin{pmatrix}
 	0 & 0 & 0 \\
 	1 & 0 & 0 \\
 	0 & 0 & 0
 \end{pmatrix} \; , \; 
\mathcal{A}^2_2  =
 \begin{pmatrix}
 	0 & 0 & 0 \\
 	0 & 1 & 0 \\
 	0 & 0 & 0
 \end{pmatrix} \; , \; 
\mathcal{A}^2_3  =
 \begin{pmatrix}
 	0 & 0 & 0 \\
 	0 & 0 & 1 \\
 	0 & 0 & 0
 \end{pmatrix} \; , \; \nonumber\\
\mathcal{A}^3_1   &=&
 \begin{pmatrix}
 	0 & 0 & 0 \\
 	0 & 0 & 0 \\
 	1 & 0 & 0
 \end{pmatrix} \; , \; 
\mathcal{A}^3_2   =
 \begin{pmatrix}
 	0 & 0 & 0 \\
 	0 & 0 & 0 \\
 	0 & 1 & 0
 \end{pmatrix} \; , \; 
\mathcal{A}^3_3  =
 \begin{pmatrix}
 	0 & 0 & 0 \\
 	0 & 0 & 0 \\
 	0 & 0 & 1
 \end{pmatrix} \; , \;  \nonumber\\
 \label{eq:u3.A.matrix.rep}
 \end{eqnarray}
which satisfies the closed algebra $u(3)$ 
\begin{equation}
	\begin{aligned}
		[\mathcal{A}^{\alpha}_{i \beta}, \mathcal{A}^{\gamma}_{i \eta}] &= 
		\delta^{\gamma}_{\beta} \mathcal{A}^{\alpha}_{i \eta} - 
		\delta^{\alpha}_{\eta} \mathcal{A}^{\gamma}_{i \beta}		\, ,   \\
		[\mathcal{A}^{\alpha}_{i \beta}, \mathcal{A}^{\gamma}_{j \eta}] &= 0	\, ,
	\end{aligned}	
\label{eq:A.com.rel}
\end{equation}
where $\delta^\alpha_\beta$ is the Kronecker delta. 
It follows from Eq.~(\ref{eq:Defd}) that  $\mathcal{A}^\alpha_\beta$ encodes information 
about both dipole moments 
\begin{eqnarray}
	S_{i}^{\alpha} = -i \epsilon^{\alpha \ \gamma}_{\ \beta} \mathcal{A}^{ \beta}_{i  \gamma}	\, , 
	\label{eq:A.to.dipole}
\end{eqnarray}
and quadrupole moments 
\begin{align}
	Q^{\alpha}_{i  \beta} = - \mathcal{A}^{\alpha}_{i \beta} - \mathcal{A}^{\beta}_{i  \alpha} 	
					+ \frac{2}{3}\delta^{\alpha \beta}  \mathcal{A}^{\gamma}_{i  \gamma}	\, ,\label{eq:A.to.quadrupole}
\end{align}
where $\epsilon^{\alpha \ \gamma}_{\ \beta}$ is the totally antisymmetric tensor 
(Levi-Civita symbol).

Since both Eq.~(\ref{eq:A.to.dipole}) and Eq.~(\ref{eq:A.to.quadrupole}) are linear 
in $\mathcal{A}^{\alpha}_{\beta}$, it follows that the spin--1 BBQ model 
[Eq.~(\ref{eq:H.BBQ.multipole})] can be rewritten in a purely bilinear form
\begin{equation}
	{\mathscr H}_{\sf BBQ} = 
		\sum_{\langle ij \rangle} \left[ J \mathcal{A}^{\alpha}_{i  \beta} \mathcal{A}^{\beta}_{j  \alpha} 
		+ (K - J) \mathcal{A}^{\alpha}_{i \beta} \mathcal{A}^{\alpha}_{j  \beta}  
		+ K \mathcal{A}^{\alpha}_{i  \alpha} \mathcal{A}^{\beta}_{j \beta} \right] \; , 
\label{eq:HBBQ.A}
\end{equation}
where we adopt the Einstein convention of summing over repeated indices.  
This version of the Hamiltonian is well suited to Monte Carlo simulation (u3MC), 
since random instances of $\mathcal{A}^{\alpha}_{\beta}$ [Eq.~(\ref{eq:A-matrix_Director})], 
can be generated efficiently using Eq.~(\ref{eq:sampling.d}), as described in Methods Section
of the main text.

Meanwhile, the very simple structure of the algebra $u(3)$ [Eq.~(\ref{eq:A.com.rel})], 
leads to an equally simple equation of motion 
\begin{equation}
	\begin{aligned}
		\frac{d}{dt}  \mathcal{A}^{\gamma}_{i \eta}
		= &-i \left[\mathcal{A}^{\gamma}_{i \eta}, \mathscr{H}_{ {\sf BBQ}} \right] \\
		= &-i \sum_{\delta} \Big[ J \left(\mathcal{A}^{\gamma}_{i \alpha}  \mathcal{A}^{\alpha}_{i+{\delta}, \eta}
						-    \mathcal{A}^{\alpha}_{i \eta}  \mathcal{A}^{\gamma}_{i+{\delta}, \alpha} \right)		\\
		  &+ (K - J) \left(\mathcal{A}^{\gamma}_{i \alpha}  \mathcal{A}^{\eta}_{i+{\delta}, \alpha}	
						-    \mathcal{A}^{\alpha}_{i \eta}  \mathcal{A}^{\alpha}_{i+{\delta}, \gamma} \right) \Big] \, ,
	\label{eq:EOM}	
	\end{aligned}
\end{equation}
where $\sum_\delta$ sums over the first neighbors of site $i$. 
This provides an efficient basis for ``Molecular Dynamics'' (u3MD) simulations of spin dynamics, 
as described in the Methods Section of the main text.
Results for dynamical structure factors were calculated as 
\begin{eqnarray}
	S_{\rm{\mu}}({\bf q},\omega) = \left\langle \sum_{\alpha\beta}
		 	| {m_{\mu}}^{\alpha}_{~\beta}({\bf q}, \omega) |^2 \ \right\rangle &&\,  , \nonumber\\
	\mu \in \{\sf S, Q, {\sf A}\} && \; , 
\label{eq:Sqw}
\end{eqnarray}
where 
\begin{align}
	&{m_{\sf S}}^{\alpha}_{~\alpha}({\bf q}, \omega)  
		= -i \sum_{\beta,\gamma} \epsilon^{\alpha \ \gamma}_{\ \beta} {m_{\sf A}}^{\beta}_{~\gamma}({\bf q}, \omega)  \, , 
	\label{eq:defn.m.S}	 \\
	&\begin{aligned}
	{m_{\sf Q}}^{\alpha}_{~\beta}({\bf q}, \omega)  
		= -{m_{\sf A}}^{\alpha}_{~\beta}({\bf q}, \omega) - &{m_{\sf A}}^{\beta}_{~\alpha}({\bf q}, \omega)  \\
			+\frac{2}{3}  \delta^{\alpha \beta}
			\sum_\gamma &{m_{\sf A}}^{\gamma}_{~\gamma}({\bf q}, \omega) \; .
	\end{aligned}	
	\label{eq:defn.m.Q}
\end{align}

We conclude by returning to the original formulation of the model, 
Eq.~(\ref{eq:H.BBQ.multipole}), and noting that dipole and quadrupole moments 
of spin only supply a total of eight operators $\{ {\bf S}, {\bf Q}  \}$, 
while the algebra $u(3)$ has 9 generators [cf. Eq~(\ref{eq:u3.A.matrix.rep})].
The ``missing'' operator is 
\begin{eqnarray}
    {\bf S}^2 = S(S+1) = 2 \; , 
\end{eqnarray}
and once this is included, it is possible to connect
the two formulations of the problem 
\begin{equation}
  \begin{pmatrix}
  	{\bf S}^2 	\\
  	S^x 	\\
  	S^y	\\
	S^z 	\\
  	Q^{x^2-y^2	}\\ 
  	Q^{3z^2-r^2}	\\
  	Q^{xy}	\\
  	Q^{xz}	\\
  	Q^{yz}	\\
  \end{pmatrix}	=	C
  \begin{pmatrix}
  	\mathcal{A}^{1}_{~ 1}	\\
  	\mathcal{A}^{1}_{~ 2}	\\
  	\mathcal{A}^{1}_{~ 3}	\\
	\mathcal{A}^{2}_{~ 1}	\\
	\mathcal{A}^{2}_{~ 2}	\\
	\mathcal{A}^{2}_{~ 3}	\\
	\mathcal{A}^{3}_{~ 1}	\\
	\mathcal{A}^{3}_{~ 2}	\\
	\mathcal{A}^{3}_{~ 3}	\\
  \end{pmatrix}
  \label{eq:su3_u3}	\; ,
\end{equation}
through the linear transformation 
\begin{equation}
  C=
  \begin{pmatrix}
  	2 & 0 & 0 & 0 & 2 & 0 & 0 & 0 & 2  \\
  	0 & 0 & 0 & 0 & 0 & -i & 0 & i & 0 \\
  	0 & 0 & i & 0 & 0 & 0 & -i & 0 & 0 \\
  	0 & -i & 0 & i & 0 & 0 & 0 & 0 & 0 \\
    	-1 & 0 & 0 & 0 & 1 & 0 & 0 & 0 & 0 \\
  	\frac{1}{\sqrt{3}} & 0 & 0 & 0 & \frac{1}{\sqrt{3}} & 0 & 0 & 0 & -\frac{2}{\sqrt{3}} \\
  	0 & -1 & 0 & -1 & 0 & 0 & 0 & 0 & 0 \\
  	0 & 0 & -1 & 0 & 0 & 0 & -1 & 0 & 0 \\
  	0 & 0 & 0 & 0 & 0 & -1 & 0 & -1 & 0 \\
  \end{pmatrix} \; .
\label{eq:C}
\end{equation}
This transformation will prove extremely useful when discussing spin anisotropic 
interactions, below.

%%%%%%%%%%%%%%%%%%%%%%%%%%%%%%%%%%%%%
\subsection{Extension to anisotropic exchange interactions}
%%%%%%%%%%%%%%%%%%%%%%%%%%%%%%%%%%%%%
\label{section:simulation.of.experiment}

%%%%%%%%%%%%%%%%%%%%%%%%%%%%%%%%%%%%%
%  Fig. - def. single tetrahedron
%%%%%%%%%%%%%%%%%%%%%%%%%%%%%%%%%%%%%
%
\begin{figure}[t]
	\centering
  	\includegraphics[width=0.35\textwidth]{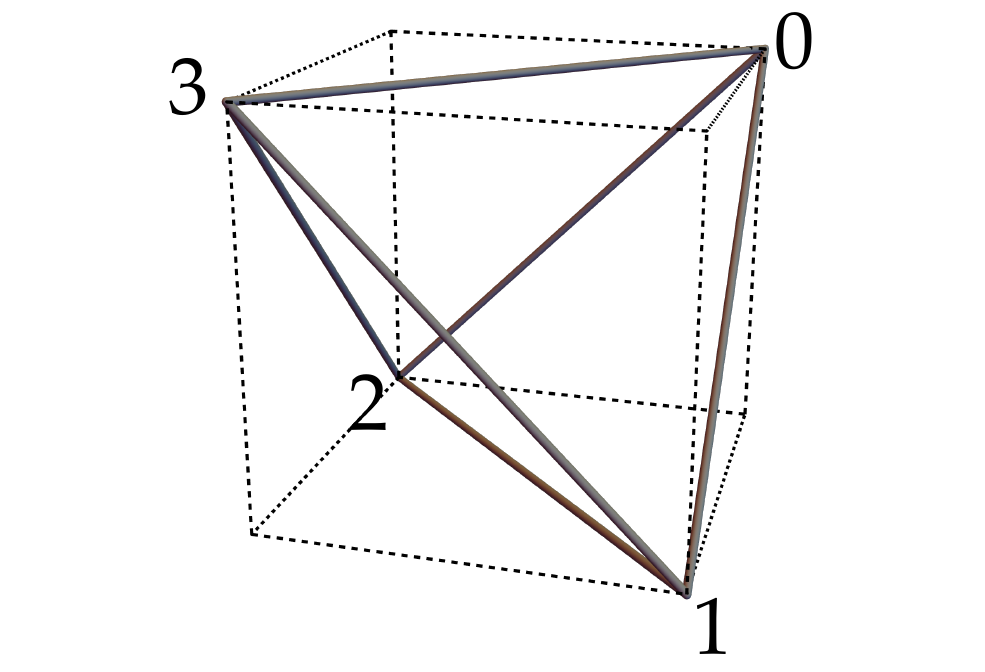}
	\caption{ 
	 Site indices for a single tetrahedron in the pyrochlore lattice, 
	 consistent with Eq.~\eqref{eq:Aniso.Exchange}.
	}
\label{fig:tetra.def}
\end{figure}
%%%%%%%%%%%%%%%%%%%%%%%%%%%%%%%%%%%%%

In order to simulate the dynamical properties of \NiF, we generalize ${\mathscr H}_{\sf BBQ}$, 
in Eq.~\eqref{eq:HBBQ.A} to encompass anisotropic exchange interactions
\begin{equation}
	\begin{aligned}
	{\mathscr H}_{\sf NaCaNi_2F_7} &= 
		\sum_{\langle ij \rangle}  \left( J^{\alpha \mu}_{\beta \nu} \right)_{ij}  \mathcal{A}^{\alpha}_{i  \beta} \mathcal{A}^{\mu}_{j  \nu} 	\\
		&+ \ J_{\sf 2nd}  \sum_{\langle ij \rangle_2}  
		\mathcal{A}^{\alpha}_{i  \beta} (\mathcal{A}^{\beta}_{j \alpha} - \mathcal{A}^{\alpha}_{j  \beta}) \, ,
	\end{aligned}
\label{eq:HNaCaNi2F.A}
\end{equation}
where $\left(J^{\alpha \mu}_{\beta \nu} \right)_{ij}$ is a bond-dependent tensor, describing 
interactions on first-neighbour bonds $\langle ij \rangle$.
Interactions on second-neighbour bonds $\langle ij \rangle_2$ are assumed to be spin-isotropic.

The interaction tensor, $\left(J^{\alpha \mu}_{\beta \nu} \right)_{ij}$, 
can be constructed using the transformation $C$ defined 
in Eq.~\eqref{eq:C}
\begin{equation}
	\left(J^{\alpha \mu}_{\beta \nu}\right)_{ij} = 
	C^{-1} \left(\tilde{J}^{\alpha \mu}_{\beta \nu} \right)_{ij}  C	\, ,
\end{equation}
where
\begin{align}
	\left(\tilde{J}^{\alpha \mu}_{\beta \nu}\right)_{ij} = \begin{pmatrix}
	\left[ \mathbf{0}	\right]_{1 \times 1}	& 	\mathbf{0}					 		&	\mathbf{0}	\\
	\mathbf{0}						& 	\left[ \hat{J}^{\sf S}_{ij}\right]_{3 \times 3}	&	\mathbf{0}	\\
	\mathbf{0}						& 	\mathbf{0}							&	\left[ \hat{J}_{ij}^{\sf Q} \right]_{5 \times 5}
	\end{pmatrix} 	\, .
\end{align}
encompasses both interactions between spin-dipole moments $\hat{J}^{\sf S}_{ij}$
and spin-quadrupole moments $ \hat{J}_{ij}^{\sf Q}$.

Following \cite{Plumb2019, Zhang2019}, we consider anisotropies in interactions 
between spin-dipole moments on first-neighbour bonds, and ``borrow'' 
from the analysis of interactions between \mbox{(pseudo-)spin--1/2} moments 
on the pyrochlore lattice \cite{Curnoe2007,McClarty2009,Ross2011,Yan2017}.
Within that analysis, the most general form of interactions can be written
%
%%%%%%%%%%%%%%%%%%%%%%
\begin{equation}
	\begin{aligned}
		\hat{J}^{\sf S}_{01} = \begin{pmatrix}
		J_2		& 	J_4	&	J_4	\\
		-J_4		& 	J_1	&	J_3	\\
		-J_4		& 	J_3	&	J_1	
		\end{pmatrix} , \
		\hat{J}^{\sf S}_{23} = \begin{pmatrix}
		J_2		& 	-J_4	&	J_4	\\
		J_4		& 	J_1	&	-J_3	\\
		-J_4		& 	-J_3	&	J_1	
		\end{pmatrix}	,		\\
%%%%%%%%%%%%%%%%%%%%%%
%
%%%%%%%%%%%%%%%%%%%%%%
		\hat{J}^{\sf S}_{02} = \begin{pmatrix}
		J_1		& 	-J_4	&	J_3	\\
		J_4		& 	J_2	&	J_4	\\
		J_3		& 	-J_4	&	J_1	
		\end{pmatrix}	 , \
		\hat{J}^{\sf S}_{13} = \begin{pmatrix}
		J_1		& 	J_4	&	-J_3	\\
		-J_4		& 	J_2	&	J_4	\\
		-J_3		& 	-J_4	&	J_1	
		\end{pmatrix}	,		\\
%%%%%%%%%%%%%%%%%%%%%%
%
%%%%%%%%%%%%%%%%%%%%%%
		\hat{J}^{\sf S}_{03} = \begin{pmatrix}
		J_1		& 	J_3	&	-J_4	\\
		J_3		& 	J_1	&	-J_4	\\
		J_4		& 	J_4	&	J_2	
		\end{pmatrix}	, \
		\hat{J}^{\sf S}_{12} = \begin{pmatrix}
		J_1		& 	-J_3	&	J_4	\\
		-J_3		& 	J_1	&	-J_4	\\
		-J_4		& 	J_4	&	J_2	
		\end{pmatrix} ,		\\
	\end{aligned}
\label{eq:Aniso.Exchange}
\end{equation}
%%%%%%%%%%%%%%%%%%%%%%
%
where 
\begin{eqnarray}
\{ J_1, J_2, J_3, J_4\} \nonumber
\end{eqnarray}
are independent parameters, and the convention for labelling interactions  on inequivalent bonds 
is defined through Fig.~\ref{fig:tetra.def}, and follows \cite{Yan2017,Ross2011,Plumb2019,Zhang2019}.
For simplicity, interactions between spin-quadrupoles are assumed to be isotropic,  
and limited to first-neighbour bonds, such that
%
%%%%%%%%%%%%%%%%%%%%%%
\begin{equation}
	\hat{J}^{\sf Q}_{ij} = K \cdot \mathds{1}_{5 \times 5}	\quad , \quad  ij \in \langle ij \rangle \; .
\end{equation}

Following these steps, we extend the equations of motion in Eq.~\eqref{eq:EOM} to 
%
%%%%%%%%%%%%%%%%%%%%%%
\begin{widetext}
\begin{equation}
	\begin{aligned}
		\frac{d}{dt}  \mathcal{A}^{\gamma}_{i \eta}
		= &-i \left[\mathcal{A}^{\gamma}_{i \eta}, \mathscr{H}_{ {\sf NaCaNi_2F_7}} \right] \\
		= &-i \sum_{\delta \in \{ \delta_1 \}} \Big[ \left({J}^{\eta \mu}_{\beta \nu}\right)_{i, i+\delta} 
		\mathcal{A}^{\gamma}_{i \beta}  	
		- \left({J}^{\alpha \mu}_{\gamma \nu}\right)_{i, i+\delta} 	
		\mathcal{A}^{\alpha}_{i \eta}  \Big] 	\mathcal{A}^{\mu}_{{i+\delta}, \nu}  \\
		 &- i \ J_{\sf 2nd}  \sum_{\delta' \in \{ \delta_2 \} } \Big[ \mathcal{A}^{\gamma}_{i \beta}  
		 \left( \mathcal{A}^{\beta}_{{i+\delta'}, \eta}	 - \mathcal{A}^{\eta}_{i+\delta', \beta} \right)
		 + \mathcal{A}^{\beta}_{i \eta} 
		 \left( \mathcal{A}^{\beta}_{{i+\delta'}, \gamma} - \mathcal{A}^{\gamma}_{{i+\delta'}, \beta}  \right) \Big] \, .
	\label{eq:EOM.NaCaNi2F}	
	\end{aligned}
\end{equation}
\end{widetext}
where $\{ \delta_1 \}$ is the set of first-neighbour bonds, 
and $\{ \delta_2 \}$ is the set of second-neighbour bonds.
The results for dynamics shown in Fig.~5 of the main text, 
and Fig.~\ref{fig:NaCaNi2F7.planes}, 
Fig.~\ref{fig:NaCaNi2F7.line.cuts.frequency} and 
Fig.~\ref{fig:NaCaNi2F7.line.cuts.momentum}
below, were obtained by numerically integrating Eq.~(\ref{eq:EOM.NaCaNi2F}), 
using the methods described in the main text.

For purposes of comparison with experiment, the dynamical structure factor is calculated as 
\begin{align}
	\begin{aligned}
		&\tilde{S}_{\rm S}({\bf q},\omega) = \\
		& \left\langle
		\sum_{\alpha \beta} \left( \delta^{\alpha \beta} - \frac{q^{\alpha} q^{\beta}}{|{\bf q}|^2} \right)
		\Big[{m_{\sf S}}^{\alpha}_{~\alpha}({\bf q}, \omega)   \Big]^*  \  {m_{\sf S}}^{\beta}_{~\beta} ({\bf q}, \omega)  
		\right\rangle 	\, ,
	\end{aligned}
\label{eq:Sqw.INS}
\end{align}
where ${m_{\sf S}}^{\alpha}_{~\alpha}({\bf q}, \omega)$ was calculated through Eq.~(\ref{eq:defn.m.S}), 
and the additional prefactor, relative to Eq.~(\ref{eq:Sqw}), originates in the dipolar interaction 
between neutron and electron spins.

%%%%%%%%%%%%%%%%%%%%%%%%%%%%%%%%%%%%%%%%%%
\section{Characterisation of Spin liquids}
%%%%%%%%%%%%%%%%%%%%%%%%%%%%%%%%%%%%%%%%%%

The BBQ model defined in Eq.~\eqref{eq:HBBQ.A} exhibits a total of seven spin liquid phases, 
five of which also occur in the ground state, as summarized in Fig.~1 of the main text.
Spin liquids are typically challenging to observe and distinguish using conventional 
magnetic order parameters or thermodynamic quantities such as the specific heat alone. 
Thus, to better characterize these phases, we introduce 
bond correlation functions, following Refs.~\cite{Penc2004, Penc2007, Shannon2010,Yan2020},
which distinguish spin liquids based on the irreducible representations (irreps) 
associated with their local correlations (constraint).

%%%%%%%%%%%%%%%%%%%%%%%%%%%%%%%%%%%%%%%%%%
\subsection{Correlations on bonds} 
%%%%%%%%%%%%%%%%%%%%%%%%%%%%%%%%%%%%%%%%%%

We consider the primitive four-site unit cell of the pyrochlore lattice 
(see Fig.~\ref{fig:tetra.def}), and introduce bond variables 
\begin{eqnarray}
	\rho^{\sf S}_{ij} = {\bf S}_i \cdot {\bf S}_j  
	\quad , \quad 
	\rho^{\sf Q}_{ij} = {\bf Q}_i \cdot {\bf Q}_j  \; .  
\end{eqnarray}
The relevant fluctuations transform with irreps 
\begin{eqnarray}
	\nu \in \{{\sf A_{1}}, {\sf E}, {\sf T_{2}} \}  \; ,
\end{eqnarray}
of the tetrahedral symmetry group $T_d$, defined through
\begin{eqnarray}
	\begin{pmatrix}
   		\Lambda_{\sf A_{1}}		\\
		\Lambda_{\sf E,1}		\\
		\Lambda_{\sf E,2}		\\
		\Lambda_{\sf T_{2},1}	\\
		\Lambda_{\sf T_{2},2}	\\
		\Lambda_{\sf T_{2},3}	
	\end{pmatrix}^{\mu} = 
	\begin{pmatrix}
   		\frac{1}{\sqrt{6}}		&  \frac{1}{\sqrt{6}}	& \frac{1}{\sqrt{6}}	&  \frac{1}{\sqrt{6}}	&  \frac{1}{\sqrt{6}}	&  \frac{1}{\sqrt{6}}	 \\
		\frac{1}{\sqrt{3}}		&  \frac{-1}{2\sqrt{3}}	& \frac{-1}{2\sqrt{3}}	&  \frac{-1}{2\sqrt{3}}	&  \frac{-1}{2\sqrt{3}}	&  \frac{1}{\sqrt{3}}	 \\
		0				&  \frac{1}{2}		& - \frac{1}{2}		&  - \frac{1}{2}		&  \frac{1}{2}		&  0				 \\
		0				&  0				& \frac{- 1}{\sqrt{2}}	&  \frac{1}{\sqrt{2}}	&  0				&  0				 \\
		0				&  \frac{-1}{\sqrt{2}}	& 0				&  0				&  \frac{1}{\sqrt{2}}	&  0				 \\
		 \frac{-1}{\sqrt{2}}	& 0				& 0				&  0				& 0				&   \frac{1}{\sqrt{2}}	 
	\end{pmatrix} 
	\begin{pmatrix}
		\rho^\mu_{0,1}	\\
		\rho^\mu_{0,2}	\\
		\rho^\mu_{0,3}	\\
		\rho^\mu_{1,2}	\\
		\rho^\mu_{1,3}	\\
		\rho^\mu_{2,3}	
	\end{pmatrix} \, , \nonumber \\
\label{eq:OP.S}
\end{eqnarray}
and can have dipolar or quadrupolar character
\begin{eqnarray}
	\mu \in  \{\sf S, Q\} \; .
\end{eqnarray}
To resolve fluctuations in a given symmetry channel, we introduce the operator 
\begin{equation}
	\lambda_{\nu}^{\mu} = \frac{4}{N} \sum_{\tetrahedron} (\boldsymbol\Lambda_{\nu}^{\mu})^2 \; , 
\label{eq:bond.order.parameter}
\end{equation}
where $\sum_{\tetrahedron}$ denotes a sum over all tetrahedra, 
and we collect higher dimensional irreps as  
\begin{eqnarray}
\boldsymbol\Lambda_E &=& (\Lambda_{\sf E,1}, \Lambda_{\sf E,2}) \; , \\
\boldsymbol\Lambda_{\sf T_{2}} &=& (\Lambda_{\sf T_{2},1}, \Lambda_{\sf T_{2},2}, \Lambda_{\sf T_{2},3}) \; .
\end{eqnarray}
This operator has the associated susceptibility 
\begin{equation}
	\chi[\lambda_{\nu}^{\mu}] = \frac{N}{T} \left[ \left \langle (\lambda_{\nu}^{\mu})^{2} \right\rangle 
						 - \left\langle \lambda^{\mu}_{\nu} \right\rangle^2 \right]	\, ,
\label{eq:bond.order.parameter.sus}
\end{equation}
where $\langle \ldots  \rangle$ represents an average over statistically-independent 
Monte Carlo measurements.

The location of maxima in $\chi[\lambda_\nu^\mu] $ as a function of $(T, \phi)$ can 
then be used to track the onset of different types of fluctuations, and thereby 
crossovers into spin liquid phases.  
Table~\ref{table:critarium.spin.liquids} provides information about which 
$\lambda_\nu^\mu$ is associated with a given spin liquid phase.

%%%%%%%%%%%%%%%%%%%%%%%%%%%%%%%%%%%%%%%%%
%. Table S1 
%%%%%%%%%%%%%%%%%%%%%%%%%%%%%%%%%%%%%%%%%%

\begin{table}[t]
	\centering
	\begin{tabular}[t]{|c|c|c|c|}
	\hline
	spin liquid 	& spin moments	&	correlators and 		 	&	Eq.	\\
			 	& 				&	local constraints		&		\\
	\hline
	nematic	 [\RPU] 		& dipolar		& $\lambda_{\sf E}^{\sf S}$	& \eqref{eq:bond.order.parameter}		\\
	dipolar Coulombic	[dC]			& dipolar		& $\Gamma^{\sf S}$ 			& \eqref{eq:OP.Coulomb.S}		\\
 	dipolarcolorice		[dCI] 		& dipolar		& $\lambda_{\sf A_1}^{\sf S}$	& \eqref{eq:bond.order.parameter}		\\
	$SU(3)$ coulombic	 [su3C] 		& mixed		& $\Gamma^{\sf A}$ 			& \eqref{eq:OP.Coulomb.A}		\\
	quadrupolar color ice	 [qCl]		& quadrupolar	& $\lambda_{\sf A_1}^{\sf Q}$	& \eqref{eq:bond.order.parameter}	\\
	quadrupolar Coulombic	[qC] 		& quadrupolar	& $\Gamma^{\sf Q}$ 		& \eqref{eq:OP.Coulomb.Q}		\\
	ferrimagnetic	 [Ferri $U(1)$]	& mixed		&  $\lambda_{\sf T_2}^{\sf S}$,  $\lambda_{\sf T_2}^{\sf Q}$ & \eqref{eq:bond.order.parameter} \\
	\hline
	\end{tabular}
	\caption{
	Bond-based operators and local constraints used in constructing the 
	finite-temperature phase diagram, Fig.~1 of the main text.
	The onset of each spin liquid phase is identified through the 
	maximum in the susceptibility associated with the relevant 
	bond-based operator $\chi[\lambda_{\nu}^{\mu}]$ [Eq.~(\ref{eq:bond.order.parameter.sus})], 
	or local constraint $\chi[\Gamma^{\mu}]$ [Eq.~(\ref{eq:OP.Coulomb.Susceptibility})].
	While all spin liquids exhibit crossovers from their high-temperature state, 
	only the \RPU{} spin liquid undergoes a phase transition.	
	}
\label{table:critarium.spin.liquids}
\end{table}

%%%%%%%%%%%%%%%%%%%%%%%%
\subsection{Local Constraints}
%%%%%%%%%%%%%%%%%%%%%%%%

Figure~2 of the main text and Fig.~\ref{fig:Pinch.Points.hk0} in this Supplemental Material 
clearly show characteristic ``pinch-point'' singularities in the equal-time 
structure factors for both dipole and quadrupole spin moments.
These signatures provide direct evidence of an underlying Coulombic gauge structure within
a given spin liquid manifold, which corresponds to a local divergence-free 
condition~\cite{Moessner1998-PRB58, Moessner1998-PRL80, Henley2005}. 
To capture the crossover from a high-temperature paramagnet
to a low-temperature 
Coulombic spin liquid,
we evaluate local constraints
that explicitly measure 
the divergence-free condition~\cite{Greitemann2019}.

The operator characterizing the divergence-free condition for spin-dipoles is
%
%%%%%%%%%%%%%%%%%%%%%%%%%%%%%%%%%%%%%%%%%%
\begin{equation}
\Gamma^{\sf S} = \frac{1}{N} \sum_{\tetrahedron} \left( \sum_i^4 \left\lVert {\textbf {\textit S}}_i \right\rVert^2 - 
					\frac{1}{4}  \left\lVert \sum_i^4 {\textbf {\textit S}}_i \right\rVert^2 \right) 	\, ,
\label{eq:OP.Coulomb.S}
\end{equation}
%%%%%%%%%%%%%%%	%%%%%%%%%%%%%%%%%%%%%%%%%%%
%
while for spin-quadrupoles, it is given by
%
%%%%%%%%%%%%%%%%%%%%%%%%%%%%%%%%%%%%%%%%%%
\begin{equation}
	\Gamma^{\sf Q} = \frac{1}{N} \sum_{\tetrahedron} \left( \sum_i^4 \left\lVert {\textbf {\textit Q}}_i \right\rVert^2 - 
					\frac{1}{4}  \left\lVert \sum_i^4 {\textbf {\textit Q}}_i \right\rVert^2 \right) 	\, .
\label{eq:OP.Coulomb.Q}
\end{equation}
%%%%%%%%%%%%%%%%%%%%%%%%%%%%%%%%%%%%%%%%%%
%

The BBQ model not only supports Coulombic spin liquids 
with dominant fluctuations in either the dipolar or quadrupolar channels, 
but also allows for a spin liquid where dipoles and quadrupoles contribute equally.
Due to the symmetry of the Hamiltonian, such fluctuations are particularly prominent 
near the antiferromagnetic $SU(3)$ point at $\phi/\pi = 1/4$.
To evaluate the Coulomb
law for these fluctuations, we define 
an operator characterizing the local constraint
for the $\mathcal{A}$ matrices, which, following Eq.~\eqref{eq:su3_u3},
is constructed by a linear combination of Eqs.~\eqref{eq:OP.Coulomb.S} and 
\eqref{eq:OP.Coulomb.Q}:
%
%
%%%%%%%%%%%%%%%%%%%%%%
\begin{equation}
	\begin{aligned}
		\Gamma^{\sf A} = \frac{1}{N}  \sum_{\tetrahedron} \Bigg( 
		\sum_i^4 &\left[ \left\lVert {\textbf {\textit S}}_i \right\rVert^2 + \left\lVert {\textbf {\textit Q}}_i \right\rVert^2 \right] - 	\\
		\frac{1}{4} &\left[ \left\lVert \sum_i^4 {\textbf {\textit S}}_i \right\rVert^2 +  \left\lVert \sum_i^4 {\textbf {\textit Q}}_i \right\rVert^2 \right]  \Bigg) 	\, .
	\label{eq:OP.Coulomb.A}
	\end{aligned}
\end{equation}
%%%%%%%%%%%%%%%%%%%%%%
%

Finite-temperature crossovers into spin liquid phases characterized by 
one of these local constraints can be identified through maxima in 
the associated susceptibility 
\begin{equation}
	\chi[\Gamma^{\mu}] = \frac{N}{T} \left[ \langle \left(\Gamma^{\mu}\right)^2\rangle - 
					\langle \Gamma^{\mu}\rangle^2  \right]
\label{eq:OP.Coulomb.Susceptibility}
\end{equation}
where 
\begin{eqnarray}
	\mu \in \{\sf S, Q, \mathcal{A}\}  \; . 
\end{eqnarray}

A summary of all relevant bond-based operators and local constraints
used in constructing the finite-temperature phase diagram, 
Fig.~1 of the main text, is provided in Table~\ref{table:critarium.spin.liquids}.

%%%%%%%%%%%%%%%%%%%%%%%%%%%%%%%%%%%%%
%  Fig. - JAX T=0 phase diagram 
%%%%%%%%%%%%%%%%%%%%%%%%%%%%%%%%%%%%%
%
\begin{figure*}[htb]
	\centering
  	\includegraphics[width=0.9\textwidth]{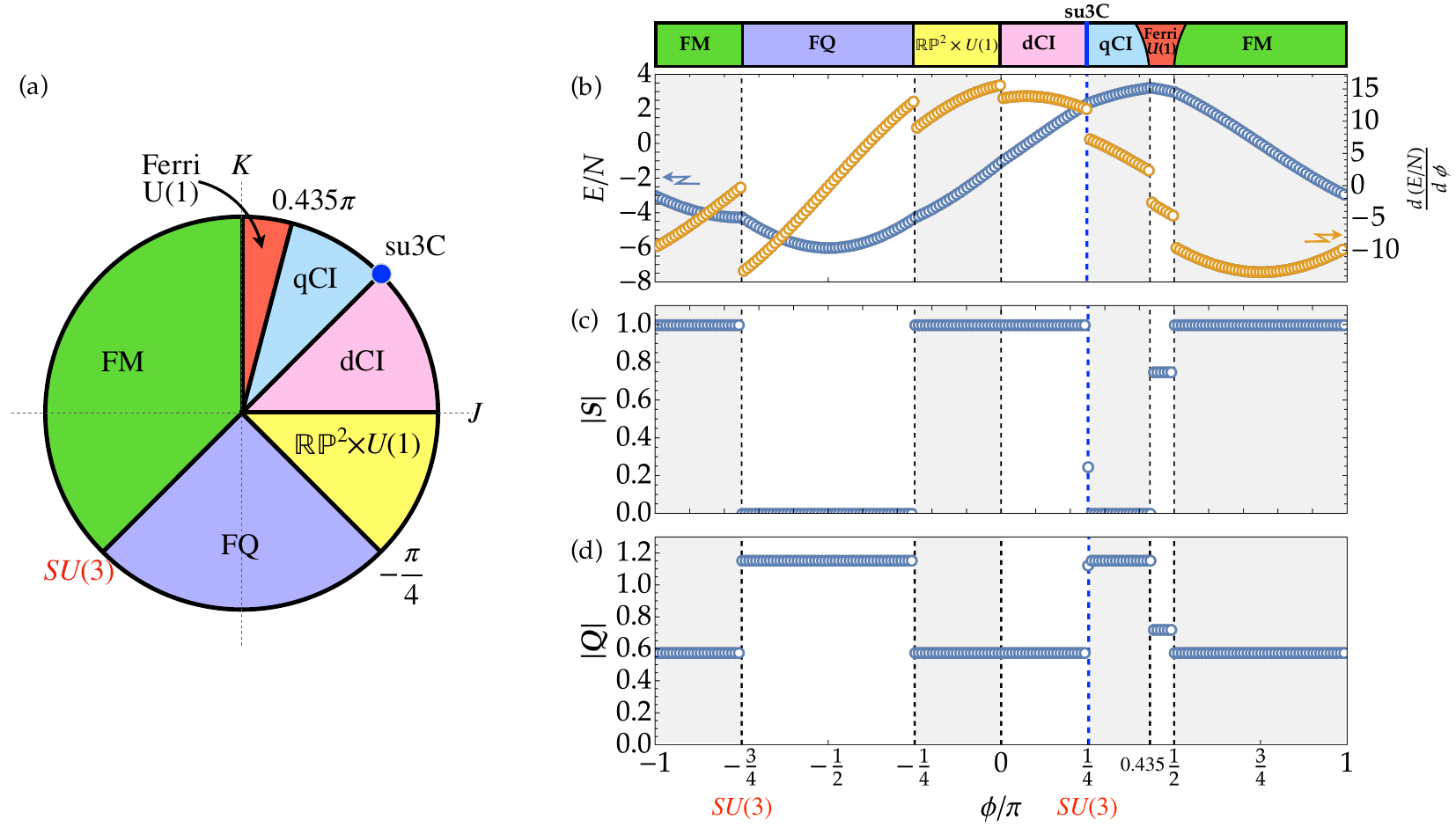}	
	\caption{ 
	Ground states of the bilinear biquadratic (BBQ) model as function 
	of model parameters, and character of associated spin moments.
	(a)~Ground-state phase diagram of ${\mathscr H}_{ {\sf BBQ}}$ [Eq.~\eqref{eq:HBBQ.A}], 
	as a function of the parameter $\phi$ 
	 [Eq.~\eqref{eq:H.BBQ.parametrization}].
	(b)~Energy per site, $E/N$, and its derivative, $d(E/N)/d\phi$.
	(c)~Averaged spin-dipole norm, $|{\textbf {\textit S}}|$ [Eq.~\eqref{eq:S.norm}]. 
	(d)~Averaged spin-quadrupole norm, $|{\textbf {\textit Q}}|$  [Eq.~\eqref{eq:Q.norm}].
	Discontinuities in $d(E/N)/d\phi$ indicate that all phases are separated by
	first-order phase transitions. 
	Calculations were performed using variational energy minimization  
	based on JAX libraries, for finite-size clusters of linear dimension
	$L=20$ and total number of sites $N = 128,000$, as described in the main text.
	}
	\label{fig:PhaseDiagram.JAX}
\end{figure*}

%%%%%%%%%%%%%%%%%%%%%%%%%%%%%%%%%%%%%%%%%%
\section{Variational ground state phase diagram}	
%%%%%%%%%%%%%%%%%%%%%%%%%%%%%%%%%%%%%%%%%%

To determine the ground-state phase diagram in Fig.~2 of the main text, 
we perform variational energy minimization of ${\mathscr H}_{\sf BBQ}$ 
[Eq.~\eqref{eq:HBBQ.A}], as discussed in the Methods Section of the main text.
The coupling constants are parametrized according to Eq.~(\ref{eq:H.BBQ.parametrization}),   
and finite-size clusters are simulated under periodic boundary conditions, 
preserving the cubic symmetry of the pyrochlore lattice. 
Simulations are carried out for clusters with a linear system size of 
$L=20$, containing a total number of sites 
$N = 16 L^3 = 128,000$.

In Fig.~\ref{fig:PhaseDiagram.JAX}(a) we present the ground-state phase 
diagram as function of normalized model parameters $J$ and $K$.
The model stabilizes magnetically ordered, ferromagnetic (FM) and 
ferroquadrupolar (FQ) spin nematic phases over a wide range of model parameters. 
In the region where $J$ and $K$ frustrate the model, we observe in total 
five different spin liquid phases, namely the 
\RPU{} nematic spin liquid,  
dipolar color ice (dIC), 
the $SU(3)$ Coulombic spin liquid (su3C) at the singular $SU(3)$ point ($\phi / \pi = 1/4$),
the quadrupolar color ice (qCI), and the 
ferrimagnetic (ferri) $U(1)$ spin liquid. 
Phase boundaries are determined from the normalized energy, $E/N$, 
and its derivative, $d(E/N)/d\phi$, as shown in panel \ref{fig:PhaseDiagram.JAX}(b).
Correspondingly, panels \ref{fig:PhaseDiagram.JAX}(c) and \ref{fig:PhaseDiagram.JAX}(d), 
display the average spin-dipole norm, $|{\textbf {\textit S}}|$, and the average 
spin-quadrupole norm, $|{\textbf {\textit Q}}|$, calculated as
%
%%%%%%%%%%%%%%%%%%%%%%
\begin{align}
	|{\textbf {\textit S}}| & = \frac{1}{N} \sum_i |{\textbf {\textit S}}_i | \, ,  \label{eq:S.norm}  \\
	|{\textbf {\textit Q}}| &= \frac{1}{N} \sum_i |{\textbf {\textit Q}}_i | \, ,  \label{eq:Q.norm}
\end{align}
%%%%%%%%%%%%%%%%%%%%%%
%
with the definitions of spin-dipole and spin-quadrupole components 
provided in Eqs.~\eqref{eq:dipole} and \eqref{eq:quadrupole}.

Discontinuities in $d(E/N)/d\phi$ confirm that all 
phases in this model are separated by first-order phase transitions.
Each phase exhibits a constant value of the spin-dipole norm 
$|{\textbf {\textit S}}|$, indicating that the spin length remains unchanged with each phase
upon varying model parameters. 
Phases such as FM, \RPU{} and dCI are purely dipolar, whereas FQ, 
and qCI are purely quadrupolar. 
The Ferri $U(1)$ phase exhibits \mbox{$|{\textbf {\textit S}}| = 0.75$}, which arises 
because one spin in each tetrahedron supports a quadrupole moment, 
while the other three support dipole moments [cf. Fig.~2(h) of the main text].

%%%%%%%%%%%%%%%%%%%%%%%%%%%%%%%%%%%%%%%%%%
\section{Spin liquid phases and their Pinch-points in the $(h,k,0)$ plane}	
%%%%%%%%%%%%%%%%%%%%%%%%%%%%%%%%%%%%%%%%%%
%

%%%%%%%%%%%%%%%%%%%%%%%%%%%%%%%%%%%%%
%  Fig. - pinch-points in (h,k,0) plane 
%%%%%%%%%%%%%%%%%%%%%%%%%%%%%%%%%%%%%
%
\begin{figure}[htb!]
	\centering
  	\includegraphics[width=0.49\textwidth]{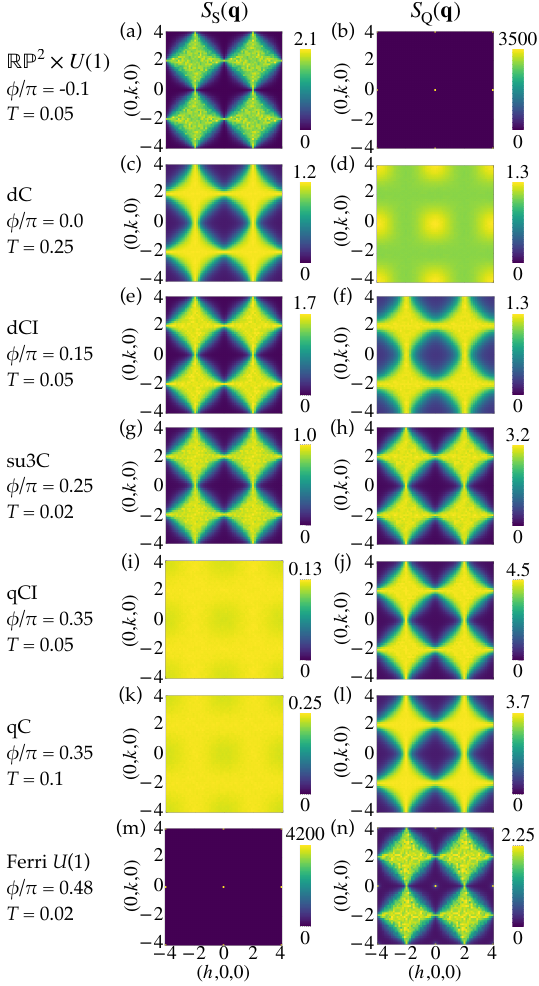}	
	\caption{ 	
	Equal-time structure factors
	within the $(h,k,0)$ plane for the seven spin liquid phases found in the 
	spin--1 bilinear biquadratic model on the pyrochlore lattice
	(see Fig.~3 of the main text and Table~\ref{table:critarium.spin.liquids}).
	{\it Left column}: spin-dipoles, $S_{\sf S}({\bf q})$.
	{\it Right column}: spin-quadrupoles, $S_{\sf Q}({\bf q})$.
	All spin-liquids exhibit pinch-point singularities at Brillouin-zone 
	centers, directly indicating the presence of local divergence-free conditions. 
	Notably, the Coulombic liquid at the antiferromagnetic $SU(3)$ point (su3C) is 
	unique, displaying pinch-points in both channels.
	The dipolar color ice (dCI) and \RPU \ phases exhibit pinch-points in the 
	spin-dipole channel, while the quadrupolar color ice (qCI) and ferrimagnetic 
	Coulombic [Ferri $U(1)$] phases show pinch-points in 
	the spin-quadrupole channel. 
	Results were obtained from $U(3)$ Monte Carlo (u3MC) simulations, for cubic clusters 
	with periodic boundary conditions and 
	linear dimension $L=8$ ($N=8192$).
	}
	\label{fig:Pinch.Points.hk0}
\end{figure}
%%%%%%%%%%%%%%%%%%%%%%%%%%%%%%%%%%%%%
%

In this Section, we supplement the energy-integrated structure factors 
shown in Fig.~3 of the main text for the $(h,h,l)$ 
with results for the $(h,k,0)$ plane.
Figure~\ref{fig:Pinch.Points.hk0} shows the corresponding structure factors 
for the same model parameters as used in Fig.~3. 
In every spin liquid phase, characteristic pinch-point singularities appear in either 
the spin-dipole or spin-quadrupole channel, directly indicating the presence of an
underlying divergence-free condition
in the corresponding Coulomb gauge field, 
namely
%
%%%%%%%%%%%%%%%%%%%%%%%%%%%%%%%%%%%%%%%%%
\begin{align}
	\nabla {\bf m}_{\sf S} &= \sum_i^4 {\bf S}_i = 0	\quad \forall \  \tetrahedron	\, ,  \\
	\nabla {\bf m}_{\sf Q} &= \sum_i^4 {\bf Q}_i = 0	\quad \forall \  \tetrahedron	\, .
\label{eq:Div.Free}
\end{align}
%%%%%%%%%%%%%%%%%%%%%%%%%%%%%%%%%%%%%%%%%
%
for spin-dipoles and spin-quadrupoles, respectively.

A particularly interesting case is the su3C spin liquid, where the $SU(3)$ symmetry 
of the Hamiltonian makes dipoles and quadrupoles equivalent.  
As a result, this spin liquid exhibits pinch points in both channels simultaneously.
The distribution of pinch points in the dipole and quadrupole channels follows a 
somewhat symmetric pattern around this $SU(3)$ point. 
Specifically, the dCI, dC, and \RPU{} phases display pinch points in the dipole 
channel, while the qCI, qC, and Ferri $U(1)$ phases exhibit pinch points in the 
quadrupole channel. 
Interestingly, in the \RPU{} and Ferri $U(1)$ phases, one channel features pinch 
points while the other shows Bragg peaks, providing valuable examples of 
(semi-) classical Coulombic spin liquids which coexist with 
{a spontaneously broken symmetry.

%%%%%%%%%%%%%%%%%%%%%%%%%%%%%%%%%%%%%
%  Fig. - Thermodynamics for RP2xU1
%%%%%%%%%%%%%%%%%%%%%%%%%%%%%%%%%%%%%
%
\begin{figure}[tb!]
	\centering
  	\includegraphics[width=0.39\textwidth]{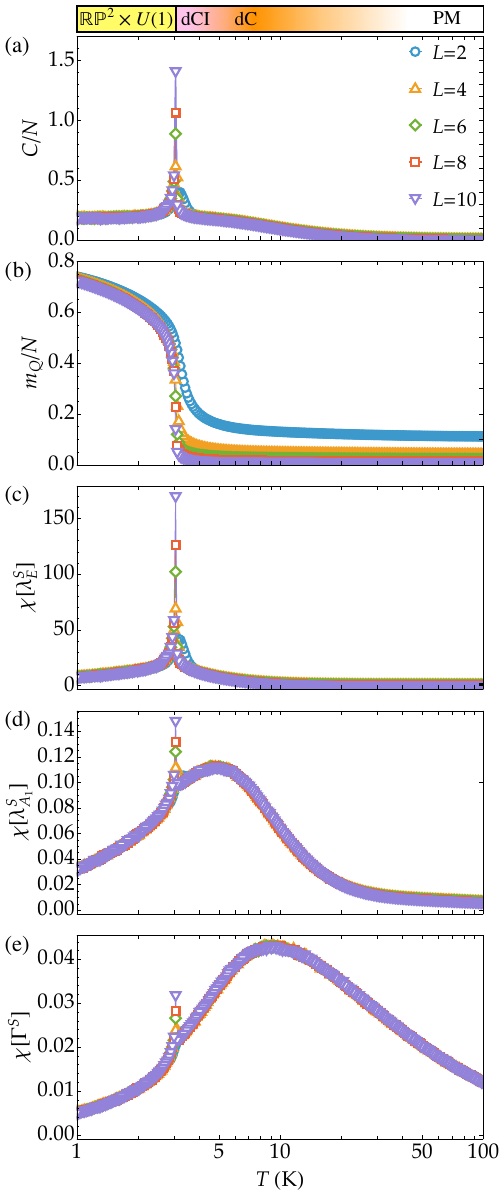}	
	\caption{ 	
	Thermodynamic observables for the BBQ model, ${\mathscr H}_{ {\sf BBQ} }$
	[Eq.~\eqref{eq:HBBQ.A}], at model parameters in Eq.~\eqref{eq:RP2XU1.parameters} 
	that stabilize the \RPU{} spin liquid at low temperatures. 
	Shown are
	(a) specific heat,
	(b) nematic order parameter [see Eq.~\eqref{eq:nematic.OP}], 
	(c) susceptibility of the dipolar bond-order parameter with irrep ${\sf E}$,  
	(d) susceptibility of the dipolar bond-order parameter with irrep ${\sf A_1}$ 
	[see Eq.~\eqref{eq:bond.order.parameter.sus}],  and 
	(e) susceptibility of the local constraints for spin-dipoles 
	[see Eq.~\eqref{eq:OP.Coulomb.Susceptibility}].
	All observables exhibit a singular behaviour at the phase 
	transition into the \RPU{} spin liquid at \mbox{$T_c = 3.05(5)$ K}.
	Additionally, (d) and (e) reveal broad crossovers spanning multiple regimes, 
	from the high-temperature paramagnet (PM) to the   
	dipolar Coulombic (dC) spin liquid, and subsequently to dipolar color ice (dCI).
	}
	\label{fig:Thermo.RP2xU1}
\end{figure}
%%%%%%%%%%%%%%%%%%%%%%%%%%%%%%%%%%%%%
%

%%%%%%%%%%%%%%%%%%%%%%%%%%%%%%%%%%%%%%%%%%
\section{Thermodynamics for \RPU{} and dipolar color ice}	
%%%%%%%%%%%%%%%%%%%%%%%%%%%%%%%%%%%%%%%%%%
%

%%%%%%%%%%%%%%%%%%%%%%%%%%%%%%%%%%%%%
%  Fig. - Thermodynamics for dipolar color ice 
%%%%%%%%%%%%%%%%%%%%%%%%%%%%%%%%%%%%%
%
\begin{figure}[t]
	\centering
  	\includegraphics[width=0.39\textwidth]{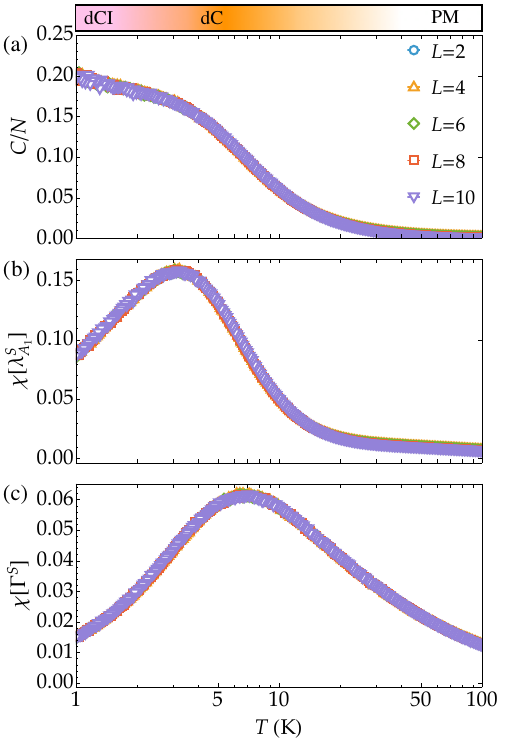}	
	\caption{ 	
	Thermodynamic observables for the BBQ model, ${\mathscr H}_{ {\sf BBQ} }$
	[Eq.~\eqref{eq:HBBQ.A}], at model parameters in Eq.~\eqref{eq:dCI.parameters} 
	that stabilize the dipolar color ice (dCI) spin liquid at low temperatures. 
	Shown are
	(a) specific heat,  
	(b) the susceptibility of the dipolar bond-order parameter with 
	irrep ${\sf A_1}$ [see Eq.~\eqref{eq:bond.order.parameter.sus}],  and 
	(c) the susceptibility of the local constraints for spin-dipoles 
	[see Eq.~\eqref{eq:OP.Coulomb.Susceptibility}].
	Observables show no signatures of a phase transition, only 
	crossovers from the high-temperature paramagnet (PM) to   
	dipolar Coulombic (dC) spin liquid, and subsequently to dipolar color ice (dCI).
	}
	\label{fig:Thermo.colorIce}
\end{figure}
%%%%%%%%%%%%%%%%%%%%%%%%%%%%%%%%%%%%%

\NiF{} is a spin--1 magnet in which magnetic dipole and quadrupole moments 
are unavoidable and appear to play a crucial role for its physical properties.
The simplest and most general model incorporating interactions between these moments 
is the BBQ model, commonly expressed in Eq.~(1) of the main text and 
formulated within the $U(3)$ formalism 
in Eq.~\eqref{eq:HBBQ.A} of this Supplemental Material.
As discussed in Fig.~1 of the main text, finite quadrupolar interactions  
immediately drive the system into one of two distinct spin liquid phases: 
the \RPU{} nematic spin liquid for $K<0$ and the dipolar color ice for $K>0$.
In Figs.~4 and 5 of the main text, we compare the dynamical signatures of these two 
spin liquid phases and propose
that the characteristically gapped spin structure factor observed in \NiF{} may originate
from nonzero negative biquadratic interactions ($K<0$).
To provide context for these results, here we review the 
 thermodynamic properties of the \RPU\ and dCI phases, as found 
 in u3MC simulations for the same sets of model parameters.

%%%%%%%%%%%%%%%%%%%%%%%%%%%%%%%%%%%%%%%%
\subsection{Thermodynamics of nematic spin liquid [\RPU]}
%%%%%%%%%%%%%%%%%%%%%%%%%%%%%%%%%%%%%%%%
\label{section.themodynamics.RPU}

Figure~\ref{fig:Thermo.RP2xU1} presents thermodynamic observables at 
\mbox{$\phi/\pi \approx -0.1$}, with
%
%%%%%%%%%%%%%%%%%%%%%%%%%%%
\begin{eqnarray}
	J = 2.4\ \text{meV}  \; , \; 
	K = -0.8\ \text{meV}  \; , \; 
\label{eq:RP2XU1.parameters}
\end{eqnarray}
%%%%%%%%%%%%%%%%%%%%%%%%%%%
%
which stabilizes the \RPU{} nematic spin liquid at low temperatures.
The specific heat, shown in Fig.~\ref{fig:Thermo.RP2xU1}(a), exhibits a sharp 
first-order phase transition into the \RPU{} nematic spin liquid at 
\begin{eqnarray}
	T_N = 3.05(5)~\text{K} \; .
\end{eqnarray}
As shown in Fig.~\ref{fig:Thermo.RP2xU1}(b), this transition is accompanied 
by the sharp rise in the nematic order parameter,
%
% %%%%%%%%%%%%%%%%%%%%%%%%%%%%%%%%%%%%%%%%%
\begin{equation}
	m_{\sf Q} = \sqrt{ \sum_{\alpha \beta} \Big(\sum_i Q^{\alpha}_{i  \beta} \Big)^2  }		\, ,
\label{eq:nematic.OP}
\end{equation}
%%%%%%%%%%%%%%%%%%%%%%%%%%%%%%%%%%%%%%%%%
%
with $Q^{\alpha}_{i  \beta}$ defined in Eq.~\eqref{eq:A.to.quadrupole}.

%%%%%%%%%%%%%%%%%%%%%%%%%%%%%%%%%%%%%
%  Fig. - dynamics in (h,h,l) and (h,k,0) planes 
%%%%%%%%%%%%%%%%%%%%%%%%%%%%%%%%%%%%%
%
\begin{figure*}[tbh]
	\centering
  	\includegraphics[width=0.98\textwidth]{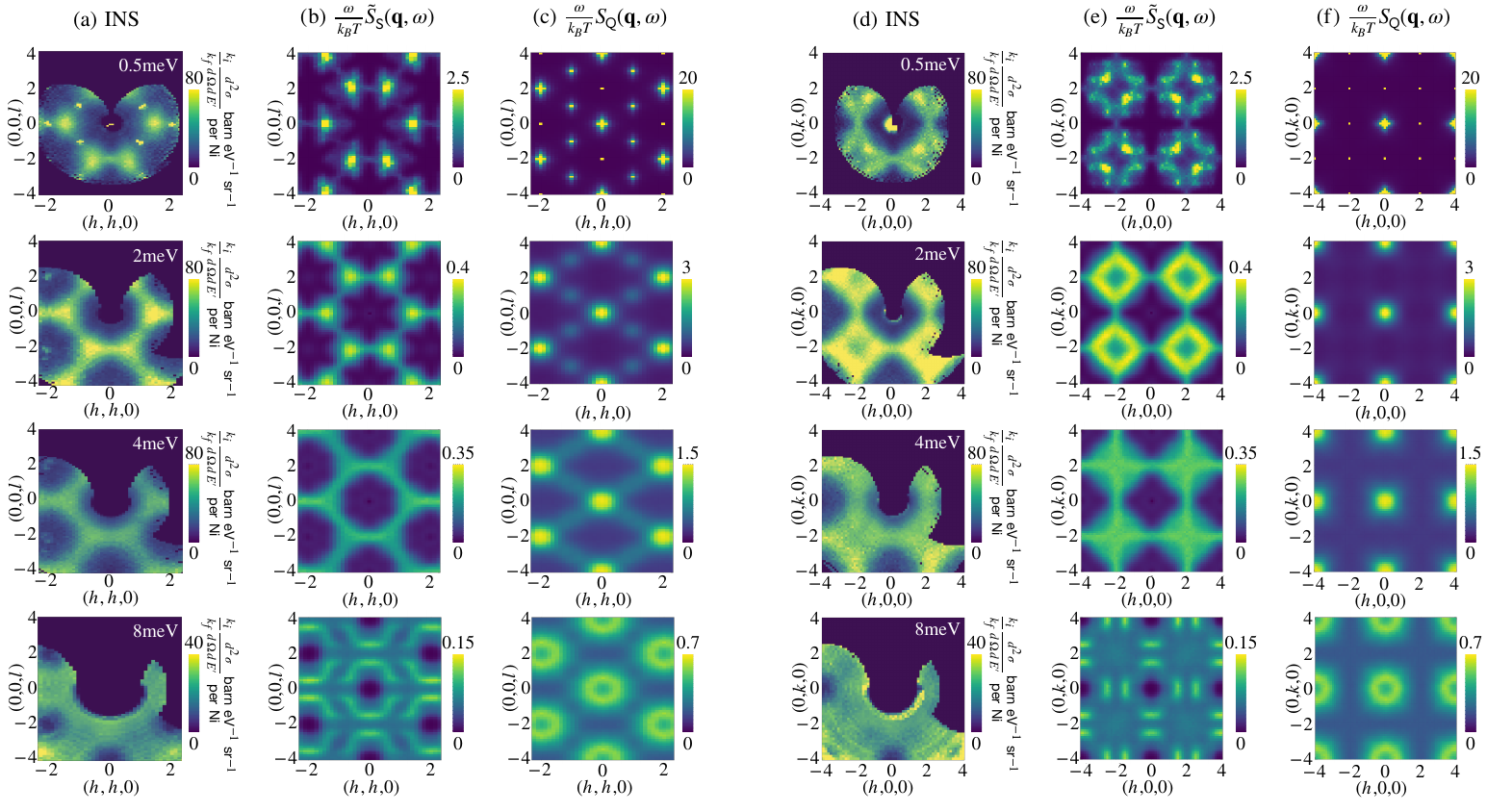}	
	\caption{ 	
	Comparison of dipolar structure factors found in experiment and simulation, 
	together with corresponding theoretical predictions for quadrupolar structure factor.
	Results are shown for both the $(h,h,l)$ plane [columns on left] 
	and $(h,k,0)$ plane [columns on right]. 
	(a), (d) Inelastic neutron scattering (INS) data for \NiF{}, 
	at energies of $0.5~\text{meV}$, $2~\text{meV}$, $4~\text{meV}$ and $8~\text{meV}$, 
	taken from experiments described in~\cite{Plumb2019}.
	(b), (e) Dipolar structure factor, $\tilde{S}_{\rm S}({\bf q},\omega)$ [Eq.~\eqref{eq:Sqw.INS}], 
	found in simulations of a minimal model of \NiF{}.  
	(c), (f) Corresponding results for quadrupolar structure factor, 
	$S_{\rm{Q}}({\bf q},\omega)$ [Eq.~\eqref{eq:Sqw}].
	Numerical simulations of the model Eq.~\eqref{eq:HNaCaNi2F.A}, 
	were carried out using the methods described in Section~\ref{section:U3.formalism} 
	and the main text, for parameters Eq.~\eqref{eq:NaCaNi2F7.model}, 
	at $T \approx 1.8\ \text{K}$.
	Simulation results were obtained with a 
	resolution of \mbox{$0.37\ \text{meV}$} (FWHM), 
	and subsequently averaged over $l \pm 0.25$ and 
	$\omega \pm 0.2~\text{meV}$ to replicate processing 
	of experimental data~\cite{Plumb2019}.   
	}
	\label{fig:NaCaNi2F7.planes}
\end{figure*}
%%%%%%%%%%%%%%%%%%%%%%%%%%%%%%%%%%%%%

The \RPU{} spin liquid is symmetric under the bond irrep ${\sf E}$~\cite{Shannon2010}.
Accordingly, we present the susceptibility of the bond correlation function,
$\chi[\lambda^{\rm S}_E]$ [see Eq.~\eqref{eq:bond.order.parameter.sus}], 
in Fig.~\ref{fig:Thermo.RP2xU1}(c), which exhibits a sharp singularity at 
the ordering temperature $T_N$.
A weaker signature of this transition is also observed in the susceptibilities 
of the bond correlation function
with irrep ${\sf A_1}$,
$\chi[\lambda^{\rm S}_{A_1}]$,
and the local constraints for spin-dipoles, 
$\chi[\Gamma^{\rm S}]$ [see Eq.~\eqref{eq:OP.Coulomb.Susceptibility}],
as shown in Figs.~\ref{fig:Thermo.RP2xU1}(d) and \ref{fig:Thermo.RP2xU1}(e),
respectively. 
However, these correlators
primarily capture system-size-independent 
crossovers over a broad temperature range,  
with peak maxima at approximately 
\mbox{$T_1 \approx 9$ K}, and 
\mbox{$T_2 \approx 5$ K}.
This suggests a three-step evolution: an initial crossover from the 
high-temperature paramagnetic (PM) state into the dC spin liquid, followed 
by a second crossover into the dCI phase, and finally a first-order phase 
transition into the \RPU{} spin liquid.
%

%%%%%%%%%%%%%%%%%%%%%%%%%%%%%%%%%%%%%%%%
\subsection{Thermodynamics of dipolar colour ice [dCI]}
%%%%%%%%%%%%%%%%%%%%%%%%%%%%%%%%%%%%%%%

In Fig.~\ref{fig:Thermo.colorIce}, we present thermodynamic observables for 
positive biquadratic exchange at $\phi / \pi \approx 0.1$, with
%
%%%%%%%%%%%%%%%%%%%%%%%%%%%
\begin{eqnarray}
	J = 2.4\ \text{meV}  \; , \; 
	K = 0.8\ \text{meV}  \; , \; 
\label{eq:dCI.parameters}
\end{eqnarray}
%%%%%%%%%%%%%%%%%%%%%%%%%%%
% 
which stabilize the dCI spin liquid at low temperatures.
The specific heat in Fig.~\ref{fig:Thermo.colorIce}(a) shows no indication of a 
phase transition.
Meanwhile, the susceptibilities of the dipolar 
bond correlation function,
$\chi[\lambda^{\rm S}_{A_1}]$, and the local constraints for spin-dipoles, 
$\chi[\Gamma^{\rm S}]$,
exhibit distinct crossovers over a broad temperature range.
This suggests a gradual evolution from the high-temperature PM into the 
dC and eventually the dCI spin liquid phases as the temperature decreases.
It is important to note that the dipolar color ice state has been associated with 
the development of octupolar correlations in the O(3) version of this 
model~\cite{Cerkauskas-Thesis}.
While our model differs slightly, a similar tendency may arise at low 
temperatures.
A detailed investigation of octupolar correlations in this setting is left for future work.

The temperature evolution described in Figs.~\ref{fig:Thermo.RP2xU1} and 
Figs.~\ref{fig:Thermo.colorIce} is reflected in the finite-temperature phase diagram 
shown in Fig.~1 of the main text.
We note however that this phase diagram was plotted using a 
relative temperature scale, consistent with Eq.~\eqref{eq:H.BBQ.parametrization}, 
rather than an absolute temperature scale in Kelvin.

%%%%%%%%%%%%%%%%%%%%%%%%%%%%%%%%%%%%%%%%
\subsection{Effect of anisotropic exchange}
%%%%%%%%%%%%%%%%%%%%%%%%%%%%%%%%%%%%%%%

We conclude with a few comments on thermodynamics in the presence of anisotropic 
exchange interactions, as used to compare with experiment on \NiF.
Here we adopt a model Hamiltonian ${\mathscr H}_{ {\sf NaCaNi_2F_7} }$ [Eq.~\eqref{eq:HNaCaNi2F.A}], 
with parameters 
\begin{align}
	\begin{aligned}	
		J_1 &= \phantom{-} 2.4\ \text{meV} \, ,  		\\
		J_2 &= \phantom{-} 2.4\ \text{meV} \, ,			\\
		J_3 &= \phantom{-} 0.15\ \text{meV} \, ,		\\
		J_4 &=- 0.23\  \text{meV} \, ,		\\
		J_{\sf 2nd} &=-0.06\ \text{meV} \, ,	\\
		K &= -0.8\ \text{meV} \, .
	\end{aligned}	
\label{eq:NaCaNi2F7.model}
\end{align}
building on the pattern of earlier attempts to fit neutron scattering 
experiment \cite{Plumb2019,Zhang2019},  but including biquadratic exchange $K$.
The choice of these parameters is justified by fits to inelastic neutron scattering, 
discussed below.

In terms of the magnitude of interactions, this model is very close 
to the parameter set used to chracterize the nematic spin liquid 
in Section~\ref{section.themodynamics.RPU} [Eq.~(\ref{eq:H.BBQ}), Eq.~(\ref{eq:RP2XU1.parameters})],
differing only in the addition of small Dzyaloshinskii-Moriya (DM) interaction, $J_4$;
pseudodipolar interaction $J_3$; 
and second-neighbour Heisenberg interaction $J_{\sf 2nd}$.
The presence of second-neighbour exchange of magnitude $|J_{\sf 2nd}/J_1| = 0.025$ does not 
substantially modify thermodynamic properties at temperatures relevant to experiment.
However the presence of even small anisotropic interactions have a profound affect on 
thermodynamic properties, 
causing the system to fall out of equilibrium below a spin glass temperature $T* \lesssim T_N$.  
In simulation, this failure to equilibrate is signalled by qualitative changes in autocorrelation functions, 
and a corresponding breakdown in the convergence of estimators for thermodynamic quantities.

The nonequilibrium properties of the model describing \NiF, 
and characterization of its spin-glass phase at low temperatures 
represent a substantial topic in their own right, 
and will be reported elsewhere. 
None the less, the thermodynamic properties below $T^*$ show many 
echos of the spin glass found in \NiF.
And on that basis it makes sense to 
compare MD results within the spin-glass phase of the model, 
with INS results measured in \NiF\ below its spin glass temperature of
$3.6~\text{K}$ \cite{Krizan2015}.

\newpage
%%%%%%%%%%%%%%%%%%%%%%%%%%%%%%%%%%%%%%%%%%
\section{Dynamics of \NiF }	
%%%%%%%%%%%%%%%%%%%%%%%%%%%%%%%%%%%%%%%%%%
\label{sec:dynamics.in.experiment}

%%%%%%%%%%%%%%%%%%%%%%%%%%%%%%%%%%%%%%
%   Fig. - line cuts in \omega 
%%%%%%%%%%%%%%%%%%%%%%%%%%%%%%%%%%%%%%
%
\begin{figure}[t]
	\captionsetup[subfigure]{farskip=10pt,captionskip=1pt}
	\centering	
	\subfloat[ linear scale \label{fig:S7.linear.scale}]{ \includegraphics[width=0.95\columnwidth]{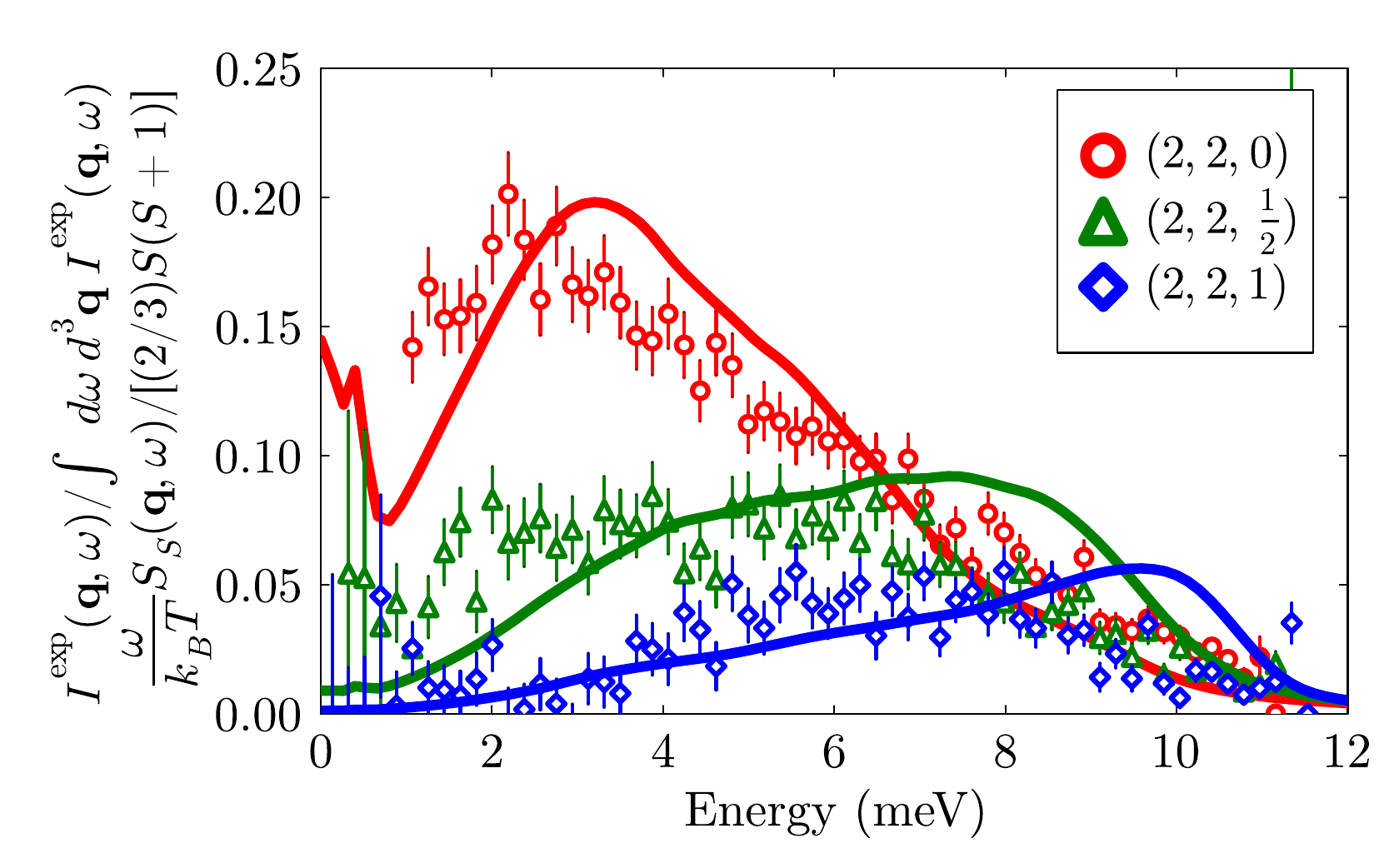} }  \\
	\subfloat[ semi-log scale \label{fig:S7.semi-log.scale}]{\includegraphics[width=0.95\columnwidth]{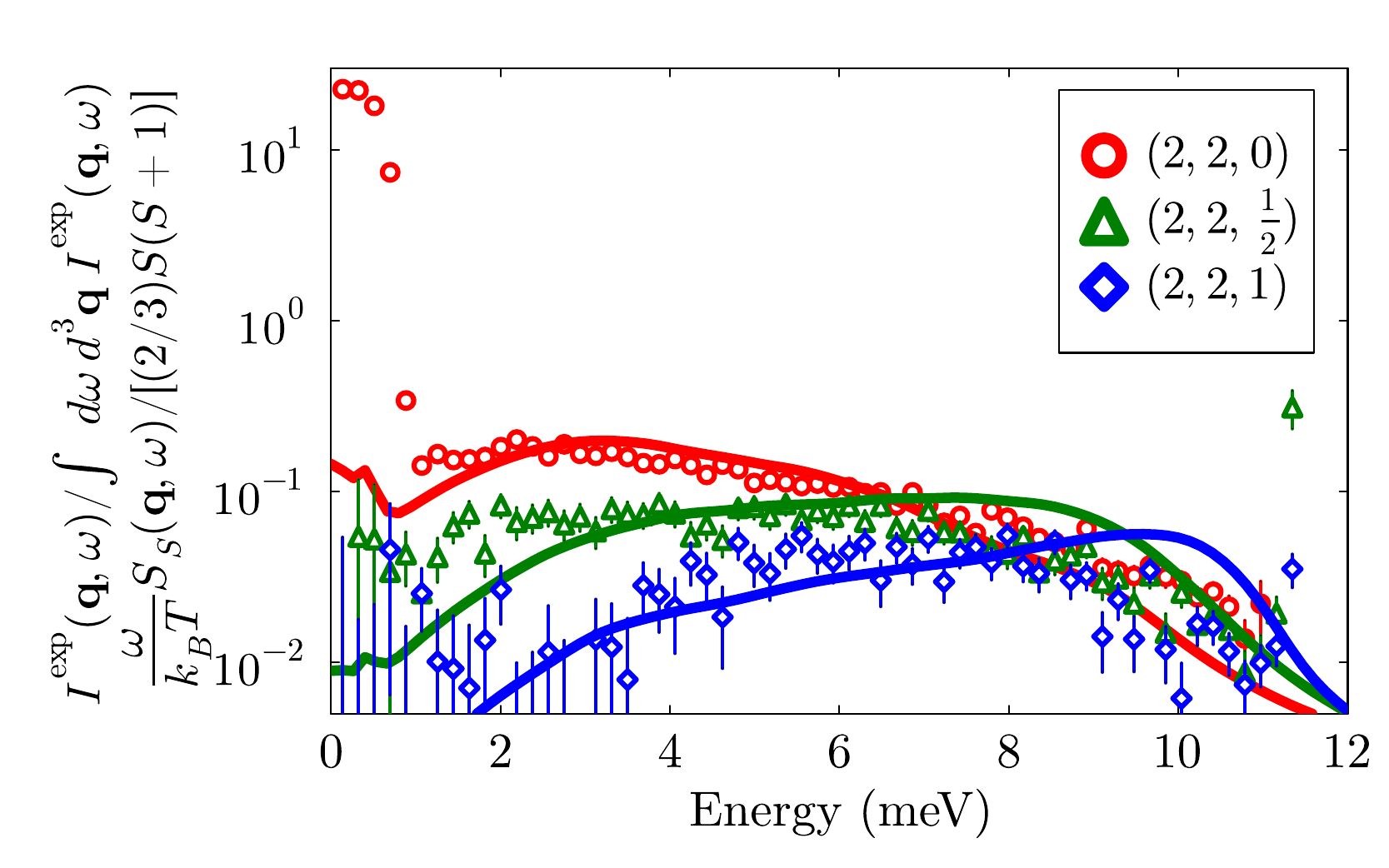}	} 
	\caption{ 
	Dynamical structure factor of \NiF\ as a function of energy, 
	showing good agreement between experiment and simulation.
	(a) Results for wavevector 
	${\bf q} = (2,2,0)$, $(2,2,1/2)$, $(2,2,1)$, plotted on a linear scale.
	Inelastic neutron scattering (INS) data for \mbox{$T=1.8\ \text{K}$} 
	are plotted as open symbols, 
	while equivalent theoeretical predictions for $\tilde{S}({\bf q},\omega)$ [Eq.~(\ref{eq:Sqw.INS})] 
	are shown as solid lines.
	(b) 	Identical results plotted on a semi-log scale, to facilitate comparison 
	with Fig.~4 of \cite{Zhang2019}.
	Experimental data is taken from experiment described in~\cite{Zhang2019}, 
	while theoretical results reflect molecular dynamics (u3MD) simulations 
	of Eq.~\eqref{eq:HNaCaNi2F.A}, with parameters Eq.~(\ref{eq:NaCaNi2F7.model}), 
	as described in the text.	
	Following \cite{Zhang2019}, results for both experiment and simulation have been 
	normalized to total scattering.    
        }
        \label{fig:NaCaNi2F7.line.cuts.frequency}
\end{figure}
%%%%%%%%%%%%%%%%%%%%%%%%%%%%%%%%%%%%%
%

%%%%%%%%%%%%%%%%%%%%%%%%%%%%%%%%%%%%%%
%   Fig. - line cuts in q
%%%%%%%%%%%%%%%%%%%%%%%%%%%%%%%%%%%%%%
%
\begin{figure}[b]
	\captionsetup[subfigure]{farskip=2pt,captionskip=1pt}
	\centering	
  	\includegraphics[width=0.48\textwidth]{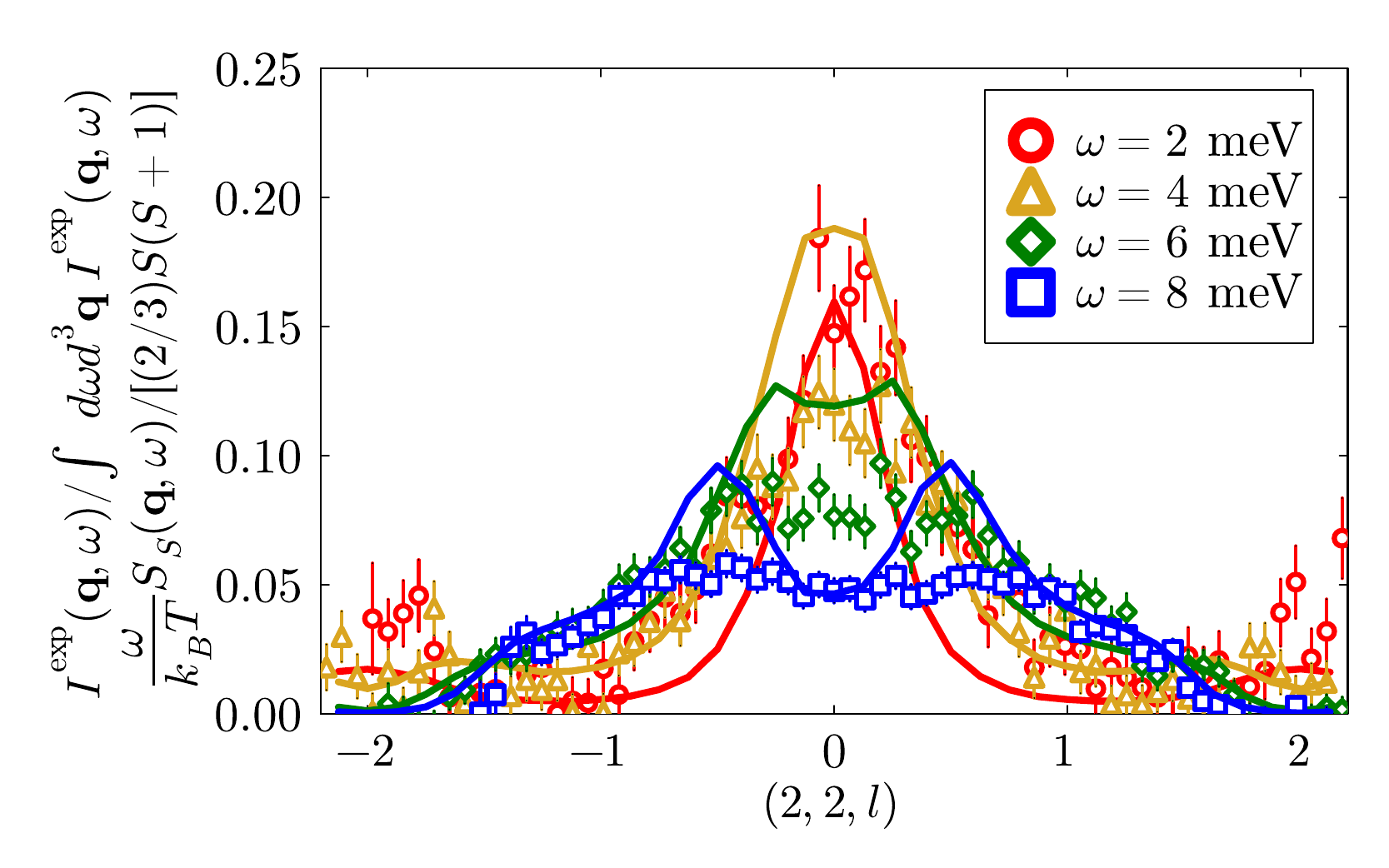}	
	\caption{ 
	Dynamical structure factor of \NiF\ as a function of wavevector, 
	showing agreement between experiment and simulation.
	Results are shown for the line ${\bf q} \in (2,2,l)$, 
	and are taken from simulation in the space of $U(3)$ matrices, 
	and experiments reported in \cite{Zhang2019}, as described 
	in the caption for Fig.~\ref{fig:NaCaNi2F7.line.cuts.frequency}.
         }
        \label{fig:NaCaNi2F7.line.cuts.momentum}
\end{figure}
%%%%%%%%%%%%%%%%%%%%%%%%%%%%%%%%%%%%%
%

In Fig.~5 of the main text, we compare the dynamical structure factor obtained 
from u3MD simulations to inelastic neutron scattering (INS) results for \NiF.
Simulations were performed for the BBQ model with anisotropic interactions,
${\mathscr H}_{ {\sf NaCaNi_2F_7} } $ [Eq.~\eqref{eq:HNaCaNi2F.A}], 
using the  model parameters Eq.~(\ref{eq:NaCaNi2F7.model}), 
at $T \approx 1.8\ \text{K}$.
Dynamical properties are computed by integrating the equations of motion 
given in Eq.~\eqref{eq:EOM.NaCaNi2F}, followed by Fourier transformation of the 
real-space and time-domain signals into momentum and energy, as described in 
the methods section of the main text. 
For comparison with experimental INS data, dipolar correlations are evaluated 
using the directionally resolved structure factor $\tilde{S}_{\rm S}({\bf q},\omega)$ 
[Eq.~\eqref{eq:Sqw.INS}], while quadrupolar structure factors are 
computed via Eq.~\eqref{eq:Sqw}.
To ``correct'' the effects of classical statistics, we apply the 
prefactor $\omega / (k_B T)$ to the simulated 
signal~\cite{Zhang2019, Remund2022}.

In Fig.~\ref{fig:NaCaNi2F7.planes}, we complement the $\tilde{S}(\mathbf{q},\omega)$
signal shown in the main text by presenting energy cuts through 
the $(h,h,l)$ plane in panels (a)--(c), and through the $(h,k,0)$ 
plane in panels (d)--(f).
We fit the model parameters to qualitatively reproduce the dominant 
features observed in the INS data from Ref.~\cite{Plumb2019} [shown in 
Figs.~\ref{fig:NaCaNi2F7.planes}(a) and \ref{fig:NaCaNi2F7.planes}(d)] with 
the corresponding features seen in the  spin-dipole channel of our u3MD simulations
[shown in Figs.~\ref{fig:NaCaNi2F7.planes}(b) and \ref{fig:NaCaNi2F7.planes}(e)].
The signatures from quadrupolar excitations, which predominantly represent 
the gapless Goldstone mode excitations, are shown in 
Figs.~\ref{fig:NaCaNi2F7.planes}(c) and \ref{fig:NaCaNi2F7.planes}(f).

We emphasize that our model is underconstrained in the sense that we consider 
only anisotropic interactions between dipoles while adjusting parameters to capture 
general trends in the presence of biquadratic exchange $K$.
Future developments, such as self-consistent Gaussian approximations within the 
$U(3)$ formalism~\cite{Remund-unpub}, may allow for a more 
systematic treatment of all allowed 
anisotropies (including those between quadrupolar terms) to improve the quantitative 
agreement between simulation and experiment.

In Fig.~\ref{fig:NaCaNi2F7.line.cuts.frequency}, we present a quantitative 
comparison between u3MD simulation results [cf. Fig.~\ref{fig:NaCaNi2F7.planes}] 
and INS data taken from \cite{Zhang2019}, measured at 
\begin{eqnarray}
	{\bf q} \in \{ (2, 2, 0) \; , (2, 2, 1/2) \; , (2, 2, 1) \} \; ,
\end{eqnarray}
over a range of energies 
\begin{eqnarray}
	 \hbar\omega  \in [0, 12]\ \text{meV} \; .   
\end{eqnarray}
Results are shown on a linear scale in Fig.~\ref{fig:S7.linear.scale}, 
and on a semi-log scale in Fig.~\ref{fig:S7.semi-log.scale}, 
to facilitate easier comparison with fits published in \cite{Zhang2019}.

At higher energy transfers, the agreement between experiment and simulation 
is good enough for differences to be difficult to observe on a semi-logarithmic scale 
[Fig.~\ref{fig:S7.semi-log.scale}].   
Both experiment and simulation show a suppression 
of spectral weight below $\hbar\omega \sim 2\ \text{meV}$.
This contrasts with published results for an $O(3)$ model without biquadratic interactions, 
which found a sharp rise in spectral weight at low energies [cf. Fig.~4(a) of Ref.~\onlinecite{Zhang2019}].

In Fig.~\ref{fig:NaCaNi2F7.line.cuts.momentum}, we show an equivalent
comparison of simulation and experimental results at fixed energy, 
for a line in reciprocal space  
\begin{eqnarray}
	{\bf q} \in (2,2,l)
\end{eqnarray}
with results plotted for 
\begin{eqnarray}
      \hbar \omega  \in \{ 2 , 4 , 6 , 8 \} \; \text{meV} \; . 
\end{eqnarray}
Here simulations give an excellent account of experiment for 
$ |l| \in (1, 2)$,  
with qualitative agreement, but some quantitative differences for 
$l \in (-1,1)$.
These results can be compared with $O(3)$ simulations presented Fig.~4(b) of Ref.~\onlinecite{Zhang2019}, 
which show broadly similar trends, but bigger quantative differences from experiment, 
particularly at low energy.

%%%%%%%%%%%%%%%%%%%%%%%%%%%%%%%%%%%%%
\end{document}